\documentclass[aps,prd,showpacs,eqsecnum,twocolumn,superscriptaddress]{revtex4-1}
\usepackage{amsmath,amssymb,graphicx,color,ulem,multirow}

\begin{document}

\title{Global simulations of strongly magnetized remnant massive neutron stars formed in binary neutron star mergers}

\author{Kenta Kiuchi}
\affiliation{Center for Gravitational Physics, Yukawa Institute for Theoretical Physics, 
Kyoto University, Kyoto, 606-8502, Japan} 

\author{Koutarou Kyutoku} 
\affiliation{Theory Center, Institute of Particle and Nuclear Studies, KEK, Tsukuba 305-0801, Japan}
\affiliation{Department of Particle and Nuclear Physics, the Graduate University
for Advanced Studies (Sokendai), Tsukuba 305-0801, Japan}
\affiliation{Interdisciplinary Theoretical Science (iTHES) Research Group, RIKEN, Wako, Saitama 351-0198, Japan}
\affiliation{Center for Gravitational Physics, Yukawa Institute for Theoretical Physics, Kyoto University, Kyoto, 606-8502, Japan} 

\author{Yuichiro Sekiguchi}
\affiliation{Department of Physics, Toho University, Funabashi, Chiba 274-8510, Japan}
\affiliation{Center for Gravitational Physics, Yukawa Institute for Theoretical Physics, Kyoto University, Kyoto, 606-8502, Japan} 

\author{Masaru Shibata}
\affiliation{Center for Gravitational Physics, Yukawa Institute for Theoretical Physics, 
Kyoto University, Kyoto, 606-8502, Japan}

\date{\today}

\begin{abstract}

We perform a general-relativistic magnetohydrodynamics simulation for 
$\approx 30$ ms from merger of a binary neutron star throughout the formation of a remnant massive neutron star (RMNS) 
with a high spatial resolution of the finest grid resolution $12.5$ m. 
First, we show that the Kelvin-Helmholtz instability at merger could amplify the magnetic-field energy at least up to $\sim 1\%$ of 
the thermal energy. Then, we show that the magnetorotational instability in the RMNS envelope and torus with 
$\rho < 10^{13}~{\rm g~cm^{-3}}$ sustains magneto-turbulent state and 
the effective viscous parameter in these regions is likely to converge to $\approx 0.01$--$0.02$ with respect to the grid resolution. 
We also point out that the current grid resolution is not still fine enough to sustain magneto-turbulent state in 
the RMNS with $\rho \ge 10^{13}~{\rm g~cm^{-3}}$. 

\end{abstract}


\maketitle


\section{Introduction}
On August 17, 2017, the first direct detection of gravitational waves from a binary neutron star (BNS) merger GW170817 
was achieved by Advanced LIGO and Advanced VIRGO~\cite{TheLIGOScientific:2017qsa}. 
Half a day after the gravitational wave event, electromagnetic emissions were observed in the UV-Optical-NIR bands
~\cite{Evans:2017mmy,Drout:2017ijr,Kilpatrick:2017mhz,Kasliwal:2017ngb,Nicholl:2017ahq,Utsumi:2017cti,Tominaga:2017cgo,Chornock:2017sdf,Arcavi:2017vbi,Diaz:2017uch,Shappee:2017zly,Coulter:2017wya,Soares-Santos:2017lru,Valenti:2017ngx,Pian:2017gtc,Smartt:2017fuw}, and 
it was concluded that these emissions named as AT2017gfo were associated with GW170817. 
Furthermore, a long-term monitoring of AT2017gfo in the X-ray and radio bands is being continued up to about 200 days after GW170817, and 
the observed emissions can be explained by synchrotron radiation associated with blast waves between the ejecta and interstellar medium~\cite{Alexander:2017aly,Hallinan:2017woc,Margutti:2018xqd,Dobie:2018zno,Mooley:2017enz,Haggard:2017qne,Margutti:2017cjl,Troja:2017nqp}.

The observed emissions in the UV-Optical-NIR bands are very consistent with the so-called kilonova/macronova model
~\cite{Li:1998bw,Kulkarni:2005jw,Metzger:2010sy}, and a significant amount of neutron-rich matter is likely to have been ejected 
during the merger. 
Recent numerical relativity simulations of the BNS mergers suggest that the mass ejection can be classified into two components: 
dynamical ejecta at the merger~\cite{Hotokezaka:2012ze,Dietrich:2015pxa,Bauswein:2013yna,Bernuzzi:2015opx,Radice:2016dwd,Palenzuela:2013hu,Foucart:2015gaa,Sekiguchi:2015dma,Sekiguchi:2016bjd}; and post-merger ejecta from a merger remnant~\cite{Shibata:2017xdx,Fujibayashi:2017xsz,Fujibayashi:2017puw,Just:2015dba,Siegel:2017nub,Fernandez:2013tya,Fernandez:2014cna,Shibata:2011fj,Dessart:2008zd,Shibata:2017jyf,Radice:2017zta,Kiuchi:2012,Perego:2014fma}. 
The dynamical ejecta are driven by tidal stripping of NSs and/or shock heating at a contact interface of two NSs. 
The post-merger ejecta from a merger remnant is driven by the angular momentum transport and the viscous heating due to effective turbulent viscosity
~\cite{Shibata:2017jyf,Fujibayashi:2017puw} (see also Refs.~\cite{Fujibayashi:2017xsz,Perego:2014fma} for the neutrino driven wind from the merger remnant). 
The effective viscosity is generated by magneto-turbulence inside the merger remnant, 
and the magneto-turbulent state is realized by several magnetohydrodynamical instabilities~\cite{Rasio,Balbus:1998ja,Balbus:1991ay,Balbus:1999fk,Kiuchi:2014hja,Kiuchi:2015sga,Price:2006fi,Liu:2008xy,Neilsen:2014hha,Anderson:2008zp,Duez:2005cj,Shibata:2005mz,Siegel:2013nrw}.

In particular, in the case of formation of a long-lived remnant massive neutron star (RMNS), the post-merger ejecta could be a dominant component of the 
mass ejection from a BNS merger. The lifetime of the RMNS is determined by the total mass of BNSs, the equation of state (EOS), 
and the angular momentum transport process inside the RMNS. The observed total mass of GW170817 is between 
$\approx 2.73M_\odot$ and $\approx 2.78M_\odot$~\cite{TheLIGOScientific:2017qsa} and this is consistent with the observed mass of binary pulsars
~\cite{Lorimer:2008se}. In this event, the merger remnant is likely to collapse to a black hole within O(10) s after the merger
~\cite{Shibata:2017xdx,Margalit:2017dij,Metzger:2018uni} because the formation of a permanently stable or very long-lived RMNS implies that 
additional energy injection due to the magnetic dipole radiation could occur and this is unlikely to be consistent with 
the electromagnetic emissions AT2017gfo~\cite{Evans:2017mmy,Drout:2017ijr,Kilpatrick:2017mhz,Kasliwal:2017ngb,Nicholl:2017ahq,Utsumi:2017cti,Tominaga:2017cgo,Chornock:2017sdf,Arcavi:2017vbi,Diaz:2017uch,Shappee:2017zly,Coulter:2017wya,Soares-Santos:2017lru,Valenti:2017ngx,Pian:2017gtc,Smartt:2017fuw}. 
On the other hand, if the total mass of BNSs is smaller than that of GW170817, a very long-lived RMNS could be formed. 
Indeed, very recently, a new BNS system PSR J1946+2052 was discovered, and its total mass is estimated as 
$2.50\pm 0.04M_\odot$~\cite{Stovall:2018ouw}.
Numerical relativity simulations in conjunction with these observational facts suggest that 
a bright electromagnetic counterpart to a gravitational-wave event which indicates the existence of a very long-lived RMNS could be observed 
in the near future~\cite{Hotokezaka:2013iia,Foucart:2015gaa}.   

Thus, it is an urgent issue to investigate the mass ejection from very long-lived RMNSs for future observation. 
Because the post-merger mass ejection is driven primarily by the effective turbulent viscosity as mentioned above, general relativistic 
magnetohydrodynamics (GRMHD) simulation is an essential tool to explore the fate of the very long-lived RMNSs. In particular, 
the required grid resolution is high because the magneto-turbulence is easily killed by the large numerical diffusion for an insufficient grid 
resolution. 
However, it is computationally challenging to simulate entire evolution of the RMNS for the viscous timescale 
while keeping such a high grid resolution. 

We will tackle this problem step by step. As a first step, we read off the effective 
viscous parameter from a high-resolution GRMHD simulation of a BNS merger. As a second step, we will perform a long-term viscous hydrodynamics simulation 
with a hypothetical value of the viscous parameter which is suggested by the GRMHD simulation to explore the mass ejection and electromagnetic emission 
from the very long-lived RMNSs~\cite{Shibata:2017jyf,Radice:2017zta,Fujibayashi:2017puw}.

In this paper, we perform a high-resolution GRMHD simulation for a BNS merger and investigate to what extent the effective turbulent viscosity is 
generated inside the RMNSs. Specifically, we estimate the Shakura-Sunyaev $\alpha$ parameter and the convergence metrics 
which measure the sustainability of the magneto-turbulent state~\cite{Hawley:2011tq,Hawley:2013lga}. 
In particular, we investigate the dependence of these quantities on the grid resolution by performing several simulations with different grid resolution. 
Finally, we discuss an implication to the value of the effective viscous parameter which is necessary for viscous hydrodynamics simulations. 

This paper is organized as follows. In Sec.~II, we describe our numerical method, grid setup, and initial condition. 
Section~III presents simulation results. 
We provide discussions in Sec.~IV and a summary in Sec.~V. 

\section{Method, grid setup, and initial models}

Our simulations are performed using a GRMHD code developed in Refs.~\cite{Kiuchi:2012,Kiuchi:2014hja}. 
Einstein's equation is formulated in the framework of the puncture-Baumgarte-Shapiro-Shibata-Nakamura
method~\cite{SN,BS,Capaneli,Baker}. Fourth-order finite differencing and lop-sided finite differencing for the 
advection terms are employed to discretize the field equations. 
GRMHD equations are formulated in a conservative form and solved by a high-resolution 
shock capturing scheme with a third-order reconstruction scheme~\cite{Kurganov}. 
We implement a fixed mesh refinement (FMR) algorithm together with the Balsara's method~\cite{Balsara:2011} 
to guarantee the divergence-free property of the magnetic field and magnetic flux conservation simultaneously. 

In our FMR implementation, a simulation domain consists of several Cartesian boxes which have a common coordinate origin. 
The domain of each Cartesian refinement box is $x_{l},y_{l}\in[-N\Delta x_{l},N\Delta x_{l}]$, and $z_{l}\in[0,N\Delta x_{l}]$ 
where $N$ is an integer and $\Delta x_{l}$ is the grid spacing for the $l$-th refinement level. 
The relation between the grid spacing of coarser and finer refinement boxes is $\Delta x_{l-1}=2\Delta x_{l}$ with $l=2,3,\cdots l_{\rm max}$. We impose 
the reflection symmetry across the orbital plane $(z=0)$. 
For the highest resolution run, we set $N=702$, $l_{\rm max}=10$, and $\Delta x_{10}=50~{\rm m}$ until $5$ ms before merger. 
With this setup, the volume of the largest refinement box is $\approx (35,900~{\rm km})^3/2$. 
Subsequently, we apply a prescription in Ref.~\cite{Kiuchi:2015sga} to improve the grid resolution in the central region. Specifically, we generate 
two ${\it new}$ finer FMR boxes of $\Delta x_{11}=25~{\rm m}$ and $\Delta x_{12}=12.5~{\rm m}$ while keeping the grid number $N$. 
With this setting, we performed simulations up to about $30$ ms after merger using $32,000$ cores on the Japanese K computer. 
The simulation cost is about 40 million core hours. 
To investigate numerical convergence, we also performed a middle resolution run with $N=482$, $l_{\rm max}=10$, and $\Delta x_{10}=70$ m and a 
low resolution run with $N=312$, $l_{\rm max}=10$, and $\Delta x_{10}=110$ m. During these simulations, we did not improve the resolution in
the central region. 

We employ a BNS in a quasi-circular orbit with mass $1.25M_{\odot}$--$1.25M_{\odot}$ as initial data. 
This NS mass is close to the lower end of the observed NS mass in the BNS systems, PSR J1946+2052~\cite{Stovall:2018ouw}. 
The initial orbital angular velocity is $G m_0\Omega/c^3=0.0221,$ where $m_0=2.5M_\odot$, 
$G$ is the gravitational constant, and $c$ is the speed of light. 
We adopt the H4 EOS~\cite{H4} to model the NS with which the maximum mass of a cold 
spherical NS is $\approx 2.03M_\odot$. For the numerical evolution of the system, a piecewise polytrope prescription~\cite{rlof2009} 
is employed to model the cold part of the EOS. The thermal part of the EOS is written in a $\Gamma$-law form with $\Gamma=1.8$
~\cite{Hotokezaka:2012ze}. 

Following Ref.~\cite{Kiuchi:2014hja}, we set the vector potential of the initial magnetic field in the form
\begin{align}
A_i = [ - (y-y_{{\rm NS}}) \delta^x_i + (x-x_{{\rm NS}}) \delta^y_i ]A_{\rm b}\max(P-P_c,0)^2, 
\end{align}
where $x_{\rm NS}$ and $y_{\rm NS}$ denote the coordinate center of the NS. $P$ is the pressure 
and $P_c$ is a cutoff value, which we set to be the value of the pressure at 4\% of the maximum rest-mass density. 
$A_{\rm b}$ is a constant which determines the amplitude of the magnetic field and we set the initial maximum magnetic-field 
strength to be $10^{15}$ G. 
This initial magnetic-field strength is justified by our recent study~\cite{Kiuchi:2015sga}: 
We have already found that a moderately weak 
initial magnetic field of $10^{13}$ G is amplified to $\gtrsim 10^{15.5}$--$10^{16}$ G by the Kelvin-Helmholtz instability 
only within a few milliseconds after the onset of merger, and thus, the final value of the magnetic-field strength depends very weakly on the initial
value~\cite{footnote3}.

\section{Results}

\subsection{Dynamics}\label{subsec:dyn}
We start the simulation from an inspiral part of about 5 orbits before the onset of merger. 
After merger, a RMNS is formed.
Employing the nuclear-theory-based neutron-star EOS, 
the maximum mass of cold rigidly-rotating NSs is by $\approx 20\%$ larger than that of cold spherical NSs~\cite{CST}, which is 2.03$M_\odot$ for
the present model. 
During the merger, the material is shock-heated and the temperature of the RMNS is increased to several tens of MeV. 
Then, the resultant thermal pressure provides additional force to support the self-gravity of the RMNS~\cite{Kaplan:2013wra}. 
This effect increases the maximum mass by several percents~\cite{Sekiguchi:2011zd,Kaplan:2013wra}. 
With these effects, the maximum mass of the hot and rigidly rotating NS with the H4 EOS is greater than $\approx 2.44 M_\odot$~\cite{footnote2}. 
Because the gravitational wave energy emitted during the inspiral and merger phases is $\approx 2.2\%$ 
of the initial gravitational mass, the gravitational mass of the RMNS is smaller than $2.44M_\odot$. 
Thus, such a RMNS should survive for a timescale of neutrino cooling 
or for that of dissipation of angular momentum by, e.g., magnetic dipole radiation. 

Figure~\ref{fig1} plots profiles of the rest-mass density (panels a1--a4), the magnetic-field strength (panels b1--b4), 
the plasma beta defined by $\beta\equiv P/P_{\rm mag}$ (panels c1--c4), and the angular velocity (panels d1--d4) 
on a meridional plane at different time slices after merger. Here $P_{\rm mag}$ is magnetic pressure. 
The merger time $t_{\rm merger}$ is defined as the time at which the amplitude of gravitational waves achieves its maximum 
(see also the visualization in Ref.~\cite{viz}). 
The merger remnant is composed of a dense RMNS surrounded by a massive torus. We define the RMNS and its core by fluid elements with 
rest-mass density $\rho \ge 10^{13} {\rm g~cm^{-3}}$ and $\rho \ge 10^{14} {\rm g~cm^{-3}}$, respectively. 
The RMNS has a highly flattened structure due to the rapid and differential rotation, as shown in 
Fig.~\ref{fig1} (panel a1). The matter with $\rho \le 10^{13} {\rm g~cm^{-3}}$ constitutes a torus and envelope. 
Thermal pressure and centrifugal force push the fluid elements outward. 
Due to torque exerted by the non-axisymmetric structure of the rest-mass density of the RMNS, the angular momentum is transported outward. 
Consequently, the torus gradually expands quasi-radially as shown in Fig.~\ref{fig1} (panels a2--a4). 

In the early stage of merger, the magnetic field is steeply amplified 
and a strongly magnetized RMNS is formed as shown in Fig.~\ref{fig1} (panel b1).
The magnetic-field amplification is caused primarily by the Kelvin-Helmholtz instability that is developed in shear layers. 
The shear layer first emerges when the two NSs come into contact. It is reinforced whenever the two dense cores formed after merger
collide until they settle to a single core~\cite{Kiuchi:2014hja,Kiuchi:2015sga}. 
The magnetic field is also amplified in the outer envelope by MRI (panels b2--b4). 
Note that the growth rate of the Kelvin-Helmholtz instability is proportional to the wavenumber, i.e., small-scale vortices 
grow faster than large-scale vortices. Therefore, even the $12.5$ m run does not fully capture the growth of the magnetic field. 
We analyze the magnetic-field amplification due to the Kelvin-Helmholtz instability and MRI in Secs.~\ref{subsec:bamp} and \ref{subsec:mri}.

Figure~\ref{fig1} (c1--c4) plots the plasma beta on a meridional plane. These panels show that 
the matter pressure dominates the magnetic-field pressure in both RMNS and its envelope. 
This indicates that the force-free magnetic field is not developed in the RMNS envelope at this moment. 
Note that the plasma beta in the RMNS core may be smaller in reality than that found in this study because the Kelvin-Helmholtz instability could 
further amplify the magnetic field. 

The RMNS settles to a quasi-stationary state at $13$--$15$ ms after merger (panels d1--d2 of Fig.~\ref{fig1} and see also Fig.~\ref{fig2}). 
The angular velocity deep inside the RMNS core is smaller than that of the RMNS core surface in our numerical result~\cite{Shibata:2005ss}. 
Thus, the radial profile of the angular velocity has an off-center peak (panels d3--d4 of Fig.~\ref{fig1}).
The reason for this is that at the collision of two NSs, the kinetic energy is dissipated at the contact interface (but see a discussion below). 
Figure~\ref{fig2} shows spacetime diagrams of the rest-mass density and the angular velocity on the orbital plane for three different grid resolutions.
We average both profiles along the azimuthal direction. 
The off-center peak of the angular velocity profile appears at $r \approx 10$ km for $t - t_{\rm merger} \gtrsim 13$--$15$ ms. 
This figure also shows that the angular velocity around the center is damped for $t-t_{\rm merger} \lesssim 12$--$14~{\rm ms}$ 
for the $12.5~{\rm m}$ and $70~{\rm m}$ runs. The damping is seen for $t-t_{\rm merger} \lesssim 5$--$6~{\rm ms}$ for the $110~{\rm m}$ run. 
Note that the quick damping of the angular velocity around the center may not be conclusive 
because our simulation is not convergent for resolving the Kelvin-Helmholtz instability 
(see also Fig.~\ref{fig8} for the convergence in the power spectrum.) However, we may conclude that irrespective of the grid resolution the region with $\rho \lesssim 10^{14-14.5}~{\rm g~cm^{-3}}$ inside 
the RMNS is subject to the MRI~\cite{Balbus:1991ay,Balbus:1998ja}, because of its extremely rapid and strongly-differential rotation 
with negative radial gradient of the angular velocity. 
On the other hand, the angular velocity profile inside the RMNS core depends significantly on the grid resolution. If we believe the results in the highest-resolution run,
the region with $\rho \gtrsim 10^{14-14.5}~{\rm g~cm^{-3}}$ is not likely to be subject to the MRI. 
However in the presence of a highly developed poloidal magnetic field in a differentially rotating medium, 
an efficient angular momentum transport could work by magnetic winding and associated magnetic braking~\cite{Shapiro:2000zh}. 

Figure~\ref{fig3} plots the radial profiles of the rest-mass density and the angular velocity on the orbital plane for all the runs 
at $t-t_{\rm merger}=15$ ms and $30$ ms. As in Fig.~\ref{fig2}, we average the profiles along the azimuthal direction. 
The rest-mass density profiles depend weakly on the grid resolution. 
The off-center peak of the angular velocity is located at $R=9$--$10$ km and its position does not change significantly during the simulation.
However, this result is not still conclusive because in the central region, 
the angular velocity exhibits dependence on the grid resolution. 
The profiles with $R \gtrsim 15$ km are not likely to depend significantly on the grid resolution. 

Figure~\ref{fig4} plots the evolution of the magnetic-field energy, the rotational kinetic energy and the internal energy. 
The solid and dashed curves in the left panel correspond to poloidal and 
toroidal components, respectively. Both components are amplified exponentially in the early stage of merger and 
saturate eventually. Here, the time of the saturation depends on the grid resolution.
This exponential growth is initiated by the 
Kelvin-Helmholtz instability~\cite{Kiuchi:2015sga}. 
There is no prominent growth of the magnetic-field energy after the saturation. 
For the $70$ m and $110$ m runs, the rapid growth of the magnetic-field energy due to the Kelvin-Helmholtz instability
becomes less prominent and it is found that the toroidal magnetic-field energy is amplified for $t-t_{\rm merger} \gtrsim 10$ ms due to the magnetic winding 
and MRI. 
This feature is obviously unphysical because in reality, the magnetic-field energy is steeply increased by the Kelvin-Helmholtz instability
until the saturation is reached. 

The middle and right panels of Fig.~\ref{fig4} 
show that the rotational kinetic energy is $\approx 10^{53}~{\rm erg}$ and the internal energy is $\approx 2.6 \times 10^{53}~{\rm erg}$ 
at $t-t_{\rm merger} \approx 30$ ms for the highest resolution run. 
Because both energies are larger than the magnetic-field energy, the saturation energy of the 
magnetic field could be larger than that found in the current work. 
We expect that the magnetic-field energy could increase up to $\sim 10^{51}~{\rm erg}$ in reality as discussed in Sec.~\ref{sec:dis}.

\subsection{Tomography of magnetic-field amplification}\label{subsec:bamp}

Because of a highly dynamical situation, it is not trivial to disentangle the amplification due to the Kelvin-Helmholtz instability, the MRI, 
and magnetic winding. 
Therefore, we perform a detailed analysis for the magnetic-field amplification in this section. 
First, we foliate the RMNS and its envelope in terms of the rest-mass density and estimate a volume average defined by
\begin{align}
\langle q \rangle_a = \frac{\displaystyle \int_{V_a} q {\rm d}^3x}{\displaystyle \int_{V_a} {\rm d}^3x} \label{eq:v-ave}
\end{align}
where $V_a$ denotes a region with $a \le \log_{10}[\rho~({\rm g~cm^{-3}})] < a+1$ and $q$ is any physical quantity such as the magnetic-field component
and the rest-mass density. In this subsection, we choose $q=b_i$ with $i=R,\varphi$ where $b_i$ is a spatial component of magnetic field measured in the fluid 
rest frame. Figure~\ref{fig5} plots $\langle b_R \rangle$ and $\langle b_\varphi \rangle$ as functions of time for all the runs 
with $a=10$--$14$. 

We first describe our finding for the results of the $12.5~{\rm m}$ run. 
Irrespective of choice of the density range, both components exhibit a prominent growth in the early stage of merger due to 
the Kelvin-Helmholtz instability for $t-t_{\rm merger} \lesssim 3$--4 ms at which the exponential growth of the magnetic-field energy is
saturated in the high-density range with $a=12$--$14$.
After the saturation is reached, there is no prominent growth of the magnetic field in these density ranges. 
On the other hand, in the low-density region with $a=10$ and 11
the exponential growth is still seen for $t-t_{\rm merger}\lesssim6$--7 ms. The growth rate of the 
poloidal magnetic-field strength is $\approx 1400~{\rm s}^{-1}$ for $4 \lesssim t-t_{\rm merge} \lesssim 5$ ms with $a=11$.
For the relatively low-density region with $\rho \alt 10^{12}~{\rm g~cm^{-3}}$,
the fastest growing mode of the MRI is covered by more than $10$ grid points (see Fig.~\ref{fig6}). 

For the $70~{\rm m}$ and $110~{\rm m}$ runs, the magnetic-field amplification for $t-t_{\rm merger} \lesssim 4$--$5~{\rm ms}$ 
is less prominent compared to that for the $12.5~{\rm m}$ run irrespective of the density range. 
After this early amplification phase, the toroidal component is amplified for the $70~{\rm m}$ and $110~{\rm m}$ runs in the density range with $a=11$--$14$. 
This is due to the magnetic winding and the non-axisymmetric MRI~\cite{Kiuchi:2014hja}. Note that the fastest growing mode of the non-axisymmetric 
MRI is covered by more than $10$ grid points in both $70~{\rm m}$ and $110~{\rm m}$ runs for $\rho < 10^{13}~{\rm g~cm^{-3}}$ 
as we discuss in the next subsection (see also Table~\ref{tab1}). 
However, the amplification due to the winding and non-axisymmetric MRI found in the low resolution runs is unphysical
because this tomography suggests that the magnetic-field energy would saturate within a short timescale after merger in all the density ranges due to 
the Kelvin-Helmholtz instability and the MRI. 

Because the MRI in combination with the Kelvin-Helmholtz instability develops in the RMNS and its envelope and subsequently it drives magneto-turbulence, 
the resultant effective turbulent viscosity should transport angular momentum outward~\cite{Balbus:1998ja,Balbus:1991ay}. 
In Sec.~\ref{sec:ang_transport}, we analyze angular momentum transport due to the MRI-driven 
turbulence in detail. 

\subsection{MRI-driven turbulence}\label{subsec:mri}

After the saturation of the magnetic-field growth, the MHD-driven turbulence is likely to be developed. 
In the presence of a region of $\partial \Omega / \partial R < 0$, MRI plays a role for sustaining the MHD-driven turbulence. 
Following Refs.~\cite{Hawley:2011tq,Hawley:2013lga}, we here evaluate the convergence 
metrics to investigate the sustainability of the MRI-driven turbulence:
\begin{align}
&Q_z = \frac{\lambda^z_{\rm MRI}}{\Delta x_l}~,\\
&Q_\varphi = \frac{\lambda^\varphi_{\rm MRI}}{\Delta x_l}~,\\
&{\cal R}=\frac{b_R^2}{b_\varphi^2}.
\end{align}
and 
\begin{align}
\alpha_{\rm mag} = \frac{W_{R\varphi}}{b^2/8\pi}. 
\end{align}
We estimate the wavelength of the fastest growing mode of 
the MRI as
\begin{align}
\lambda^i_{\rm MRI} = \frac{b_i}{\sqrt{4\pi\rho h + b^2}} \frac{2\pi c}{\Omega}~(i=z~{\rm or}~\varphi)~,
\end{align}
where $h$ is the relativistic specific enthalpy and $b^2=b^\mu b_\mu$. $b^\mu$ is the magnetic field 
measured in the fluid rest frame. 
$W_{R\varphi}$ is the Maxwell stress defined by 
\begin{align}
W_{R\varphi}=-\left\langle\frac{b_Rb_\varphi}{4\pi}\right\rangle_T.
\end{align}
$\langle\cdot\rangle_T$ denotes a time average over 1 ms at each location. 

We again foliate the RMNS and its envelope in terms of the rest-mass density and estimate volume-averaged convergence metrics defined by Eq.~(\ref{eq:v-ave}). 
Comparison of the global simulations with local box simulations for Newtonian accretion-disk systems 
suggests that $\langle Q_z \rangle \gtrsim 15$, $\langle Q_\varphi \rangle \gtrsim 20$, 
$\langle {\cal R} \rangle \gtrsim 0.2$, and 
$\langle \alpha_{\rm mag} \rangle \gtrsim 0.45$ are necessary to sustain 
the MRI-driven turbulence~\cite{Hawley:2011tq,Hawley:2013lga}. 
$Q_z$ and $Q_\varphi$ measure how many grid points are assigned for resolving the MRI wavelength of the fastest growing mode. 
$\alpha_{\rm mag}$ is similar to the $\alpha$ parameter, but it is defined only by the magnetic-field component in the stress tensor. 
This quantity measures the correlation between the radial and azimuthal components of the magnetic field. 
$\cal R$ measures the capability to generate the poloidal field due to the nonlinear development of the MRI-driven turbulence. 

Figure~\ref{fig6} shows $\langle Q_z \rangle_a$, $\langle {\cal R} \rangle_a \equiv \langle b_R^2 / b_\varphi^2 \rangle_a$, 
and $\langle \alpha_{\rm mag} \rangle_a$ as functions of time for $a=10$--$14$. 
Because $\langle Q_\varphi \rangle_a$ is always larger than $\langle Q_z \rangle_a$, 
we do not show the evolution of $\langle Q_{\varphi} \rangle_a$. For $a=14$, $\langle Q_z \rangle$ satisfies the criterion for the $12.5$ m run and 
does not for either $70$ m nor $110$ m runs. $\langle {\cal R} \rangle$ decreases with time irrespective of the grid resolution and 
reaches a value below the criterion. Note the RMNS is in a highly dynamical state for $t-t_{\rm merger}\lesssim 13$--$15$ ms (see Fig.~\ref{fig2}). 
$\langle \alpha_{\rm mag} \rangle$ is always below the criterion irrespective of the grid resolution. 
However, this is a natural consequence because most part of this density region (in our simulation results) is not subject to the MRI as discussed 
in the previous subsection.
We note that as already mentioned in Sec.~\ref{subsec:dyn},
for a high-density range with $\rho \gtrsim 10^{14}{\rm~g~cm^{-3}}$, we have not yet obtained a convergent result. In the assumption that 
the results in our current best-resolution runs are not far from the convergence, we may conclude not the MRI but the winding due to the strong poloidal 
field is likely to play a 
dominant role in the angular momentum transport. However, we need to keep in mind that 
a more resolved study is required for clarifying the turbulent state of the high-density region. 

For $a=13$, $\langle Q_z \rangle$ is larger than the criterion for the $12.5$ m run and far below the criterion 
for the $70$ m and $110$ m runs. 
$\langle {\cal R} \rangle$ decreases with time and reaches below the criterion. Even for the $12.5$ m run, this convergence metric 
approaches $\approx 0.1$ at $t-t_{\rm merger} \approx 30$ ms. This asymptotic value is increased with improving the grid resolution. 
$\langle \alpha_{\rm mag} \rangle$ is smaller than the criterion for all the runs. 
Thus, convergence is not achieved even for the 12.5 m run. However, the time-averaged value of this convergence metric 
is increased with improving the grid resolution (see Table~\ref{tab1}). This suggests that the sustainability of the MRI-driven turbulence is improved with the grid resolution and the magneto-turbulent state may be partially sustained for the $12.5$ m run. 

For $a=12$, $\langle Q_z \rangle$ is larger than the criterion for the $12.5$ m and $70$ m runs. It is smaller than the criterion for the $110$ m run. 
A mean value of $\langle {\cal R} \rangle $ with respect to the time is $\approx 0.2$ for the $12.5$ m and $70$ m runs
(thus, the criterion to sustain the turbulence is marginally satisfied),
but it is much smaller than $0.2$ for the $110$ m run for $t-t_{\rm merger}\gtrsim 15$ ms. 
$\langle \alpha_{\rm mag} \rangle$ fluctuates around $0.52$ for the $12.5$ m and $70$ m runs. Therefore, the magneto-turbulence is sustained in these runs. 
For the $110$ m run, $\langle \alpha_{\rm mag} \rangle$ starts decreasing at $t-t_{\rm merger}\approx 17$--$18$ ms and 
reaches a value below the criterion. This might be ascribed to numerical resistivity because the low value of 
$\langle Q_z \rangle$ suggests that the MRI is not developed for the $110$ m run. 
On the other hand, the $12.5$ m and $70$ m runs likely have the capability to sustain the MRI-driven turbulence in this density range. 

For $a=11$, $\langle Q_z \rangle$ satisfies the criterion at $t-t_{\rm merger} \gtrsim 10$ ms irrespective of the grid resolution. 
The asymptotic value of $\langle {\cal R} \rangle$ with respect to the time is $\approx 0.17$ for the $12.5$ m run, $\approx 0.15$ 
for the $70$ m run, and $\approx 0.12$ for the $110$ m run. 
Therefore, the criterion is approximately satisfied because the value is close to $0.2$. 
Irrespective of the grid resolution, $\langle \alpha_{\rm mag}\rangle$ is larger than 
the criterion. We find a similar trend in the convergence metrics for $a=10$ although $\langle \alpha_{\rm mag}\rangle$ is slightly smaller than the criterion. 
This indicates that the MRI-driven turbulence is marginally sustained in these low-density regions irrespective of the grid resolution.

We take the following time-average of the volume-averaged convergence metrics for $15~{\rm ms}\le t - t_{\rm merger} \le 30$ ms 
in each density range and summarize in Table~\ref{tab1};
\begin{align}
\langle\langle q \rangle\rangle_a = \frac{1}{T}\int^{30{\rm ms}}_{15{\rm ms}} \langle q \rangle_a {\rm d}t~,
\end{align}
with $T=15~{\rm ms}$. We choose this time window because the RMNS settles to a quasi-stationary state for $t-t_{\rm merger}\gtrsim 15$ ms as shown in 
Fig.~\ref{fig2}. 
Table~\ref{tab1} shows that the convergence metrics in a region with $\rho < 10^{13}~{\rm g~cm^{-3}}$ 
is likely to satisfy the criterion~\cite{Hawley:2011tq,Hawley:2013lga} for the $12.5$ m and $70$ m runs. 
For the $110$ m run, the MRI-driven turbulence is decayed by the large numerical diffusion due to the insufficient grid resolution. 
In the high-density range with $\rho \ge 10^{13}~{\rm g~cm^{-3}}$, all the convergence metrics increase with
improving the grid resolution. This indicates that the MRI cannot be fully resolved and 
the MRI-driven turbulence still suffers from the numerical diffusion even 
for the highest resolution run. 

\subsection{Effective $\alpha$ parameter and angular momentum transport timescale}\label{sec:ang_transport}
We evaluate an effective $\alpha$-viscosity parameter defined by
\begin{align}
\alpha = \frac{1}{\langle P \rangle}_T \left\langle \rho\delta v^{R} \delta v^{\varphi} - \frac{b_R b_\varphi}{4\pi} \right\rangle_T,
\end{align}
where $\delta v^i = v^i - \langle v^i \rangle_T$ is the velocity fluctuation in time~\cite{Hawley:2011tq}. 
Again, $\langle \cdot \rangle_T$ denotes a time average over $1$ ms.
Figure~\ref{fig7} shows the time evolution of the volume-averaged values, $\langle\alpha\rangle_a$, with $a=10$--$14$. 
Table~\ref{tab1} summarizes the values of $\langle\langle \alpha \rangle\rangle$. 

For $a=14$, the $\alpha$ parameter is $O(10^{-4})$ for $t-t_{\rm merger}\gtrsim 15$ ms.
A possible reason for this small value is that this high-density region (i.e., central region of the RMNS) might not be subject to the MRI. 
However for the high-density region, our simulation cannot fully resolve the Kelvin-Helmholtz instability and we cannot draw a definite conclusion. 

For $a=13$ (i.e., for the outer region of the RMNS) for which the MRI should play a key role for the angular momentum transport,
the mean value of the $\alpha$ parameter with respect to the time is $\approx 5\times10^{-3}$ 
for the $12.5$ m run, $\approx 3 \times 10^{-3}$ for the $70$ m run, and $\approx 2 \times 10^{-3}$ for the $110~{\rm m}$ run.
Thus, the value increases with improving the grid resolution (see also Table~\ref{tab1}) 
and hence $\alpha$ is likely to be larger than this value in reality. 

For $a=12$, the time evolution curves of the $\alpha$ parameter approximately overlap for the $12.5$ m and $70$ m runs for $t-t_{\rm merger}\gtrsim 15$ ms 
and the mean values with respect to the time is $\approx 0.01-0.02$ for these runs. The $\alpha$ parameter for the $110$ m run 
is always smaller than those for the higher resolution runs. Note that dependence of the convergence metrics $\mathcal{R}$ and $\alpha_{\rm mag}$ 
on the grid resolution exhibits a similar trend as discussed in the previous subsection. 
For $a=11$ and $10$, the evolution feature and the dependence on the grid resolution of the $\alpha$ parameters are similar to 
those for $a=12$~\cite{footnote1}. 

Because the $\alpha$ parameter as well as the convergence metrics in the RMNS tends to increase with improving the grid resolution, 
we conclude that the MRI-driven turbulence is not fully sustained even for the highest resolution run 
resulting in underestimate of the effective viscosity. Therefore, the $\alpha$ parameter of the RMNS derived in 
this work should be regarded as a lower limit. 

For the envelope, the detailed analysis of the convergence metrics indicates that the grid resolution with $\Delta x \lesssim 70$ m 
has a capability to sustain the MRI-driven turbulence and the resultant effective viscosity parameter is $\approx 0.01-0.02$. 
These values of the $\alpha$ parameter and the convergence metrics discussed above 
are consistent with those in the local shearing box simulations~\cite{Hawley:2011tq}. 

We estimate the angular momentum transport timescale by the shear viscous effect by $j/(\alpha c_s^2)$~\cite{Balbus:1999fk}. 
Table~\ref{tab2} shows the estimated viscous timescale in each density range. 
Note that the viscous timescale of the RMNS $(a=13)$ is longer than that of the envelope $(a=12,11)$ 
because the $\alpha$ parameter would be underestimated inside the RMNS due to the limitation of the grid resolution as discussed above. 
On the other hand, the $\alpha$ parameter of the envelope is not likely to depend significantly on the grid resolution. 
For $a=14$, the viscous timescale is shorter than that for $a=13$ even though the $\alpha$ parameter is much smaller than that for $a=13$ 
(see Table~\ref{tab1}). 
This is because the specific angular momentum is small and the sound speed is high. 
As we discuss in Sec.~\ref{sec:dis}, the magnetic braking associated with the magnetic winding could play a role for the angular momentum redistribution 
in this high-density region.

\subsection{Power spectrum of magnetic field}

Figure~\ref{fig8} plots the power spectrum of the poloidal magnetic-field energy. 
To calculate the power spectrum, we first define the Fourier component of the poloidal magnetic-field strength by 
\begin{align}
\tilde{b}_p(\vec{k})\equiv \iiint b_p(\vec{x}) e^{-i\vec{k}\cdot\vec{x}}{\rm d}^3x,
\end{align}
where $\vec{k}=(k_x,k_y,k_z)$, $\vec{x}=(x,y,z)$, and $b_p^2=b_R^2 + b_z^2$. 
Then, we define the power spectrum of the poloidal magnetic-field energy by
\begin{align}
P_B(k) \equiv \frac{1}{(2\pi)^3}\int \frac{1}{8\pi}\tilde{b}_p(\vec{k})\tilde{b}_p^{*}(\vec{k})k^2 d\Omega_k,
\end{align}
where $\tilde{b}_p^*$ is a complex conjugate of $\tilde{b}_p$, $k=|\vec{k}|$, and $d\Omega_k$ is a solid angle in the 
$k$-space. Integration of the power spectrum with respect to $k$ gives the poloidal magnetic-field energy. 
Figure~\ref{fig8} shows the power spectrum, $kP_B(k)$, at $t-t_{\rm merger}=1$ ms, $15$ ms, and $30$ ms. 
Because of the Kelvin-Helmholtz instability, the power spectrum amplitude at $t-t_{\rm merger}=1$ ms for the $12.5$ m run is 
much larger than those for the $70$ and $110$ m runs. 
This feature is remarkable at high wavenumber, i.e., at small scale. 
The amplitude for the $12.5$ m run is larger than those for the $70$ m and $110$ m run because the kinetic energy of the 
turbulence is converted to the magnetic-field energy more efficiently in the higher resolution runs~\cite{Kiuchi:2015sga}. 
However even for the $12.5$ m run we do not obtain the convergence for the magnetic-field amplification due to the Kelvin-Helmholtz instability because the saturated magnetic-field energy is much smaller than the rotational kinetic and internal energy as discussed in Sec.~\ref{subsec:dyn}.

The spectrum amplitude at low wavenumber $k \lesssim 10^{-6}~{\rm cm^{-1}}$ increases from $t-t_{\rm merger}=1$ ms to $15$ ms, 
which may indicate inverse cascade due to the MRI~\cite{Balbus:1999fk}. 
The spectrum is flat in the inertial range of the turbulent cascade, $10^{-6} \lesssim k~[{\rm cm^{-1}}] \lesssim 10^{-4}$ for the highest resolution run, which is likely to be consistent with a feature found
in the local simulations of a large-scale dynamo during the kinematical amplification phase~\cite{Brandenburg}.

\begin{figure*}[t]
\hspace{-50mm}
\begin{minipage}{0.27\hsize}
\begin{center}
\includegraphics[width=9.0cm,angle=0]{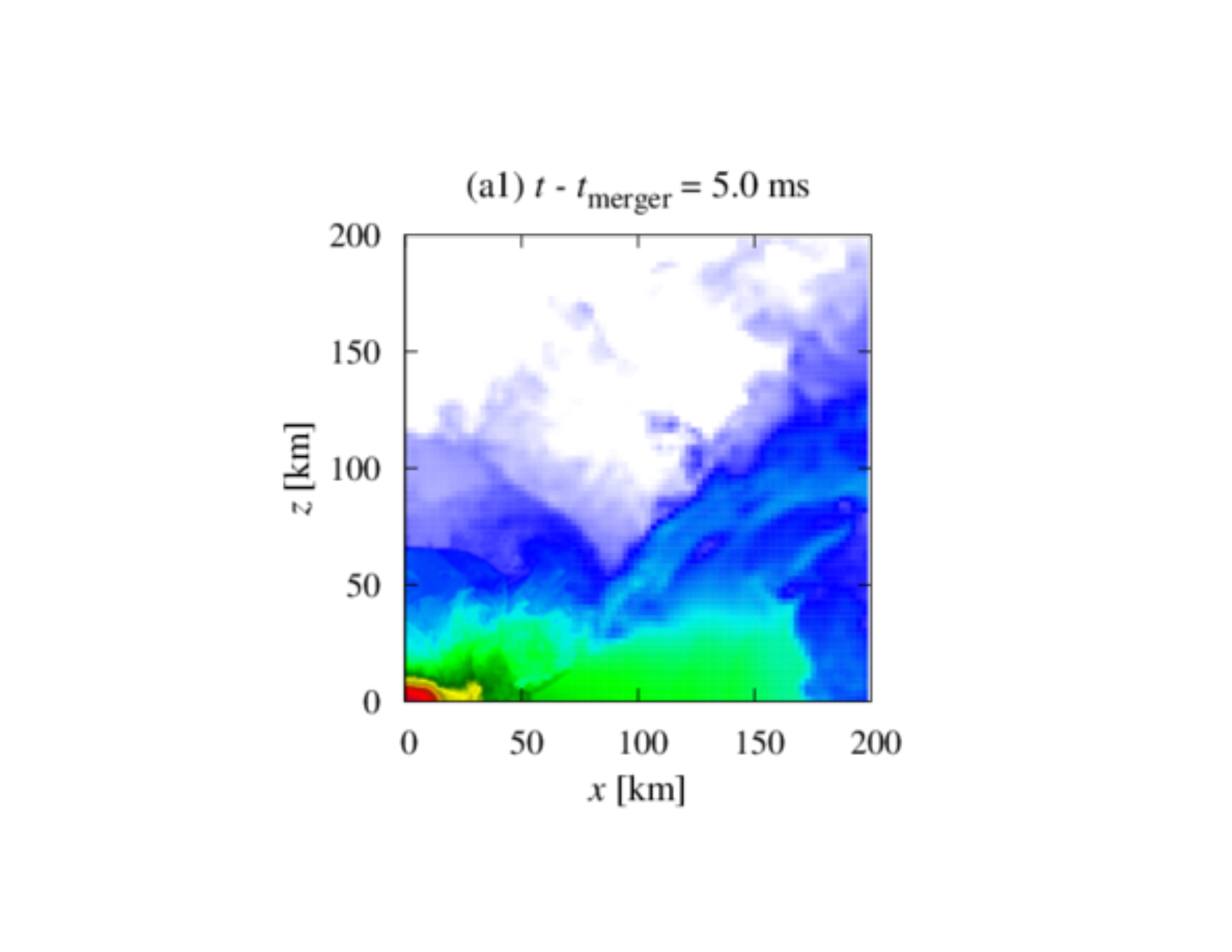}
\end{center}
\end{minipage}
\hspace{-12mm}
\begin{minipage}{0.27\hsize}
\begin{center}
\includegraphics[width=9.0cm,angle=0]{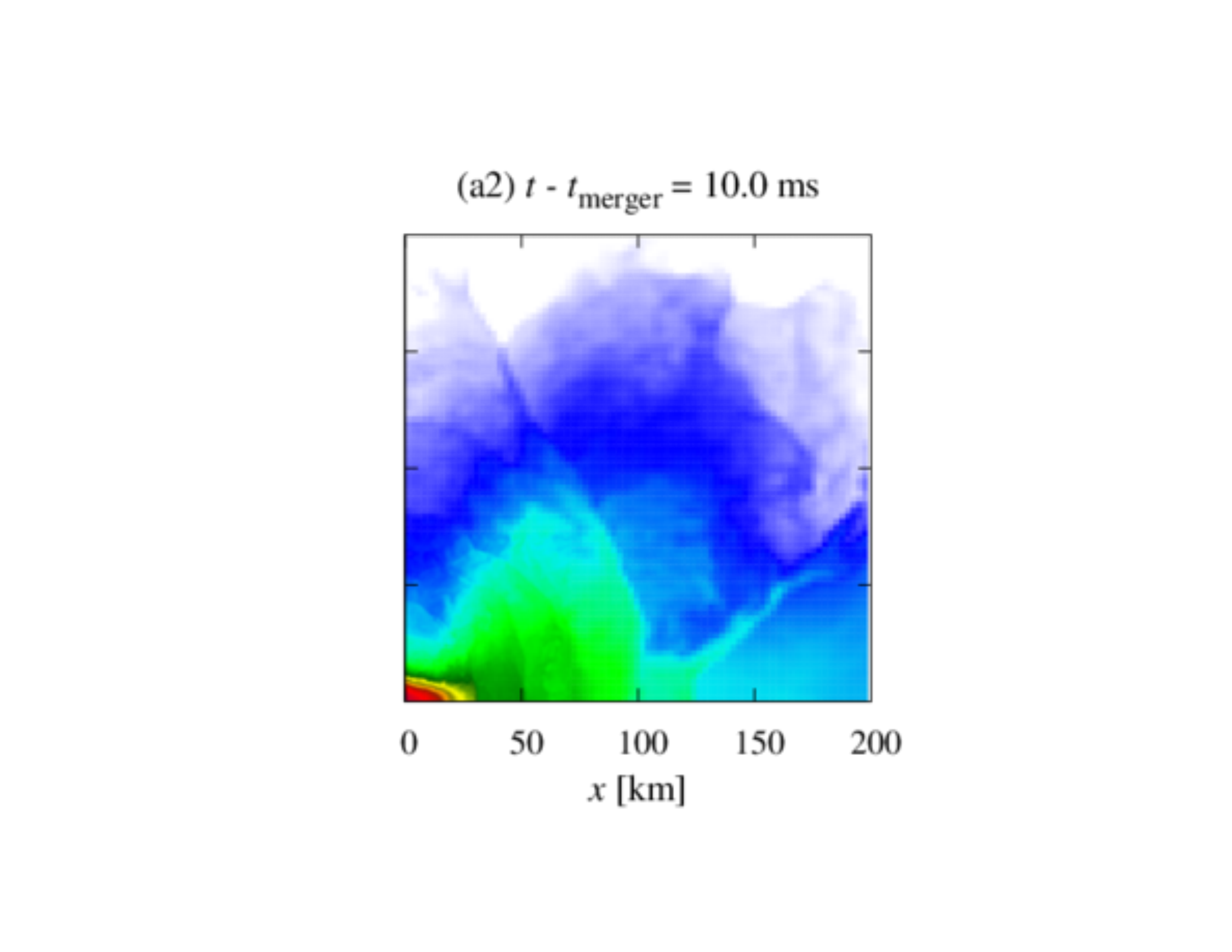}
\end{center}
\end{minipage}
\hspace{-12mm}
\begin{minipage}{0.27\hsize}
\begin{center}
\includegraphics[width=9.0cm,angle=0]{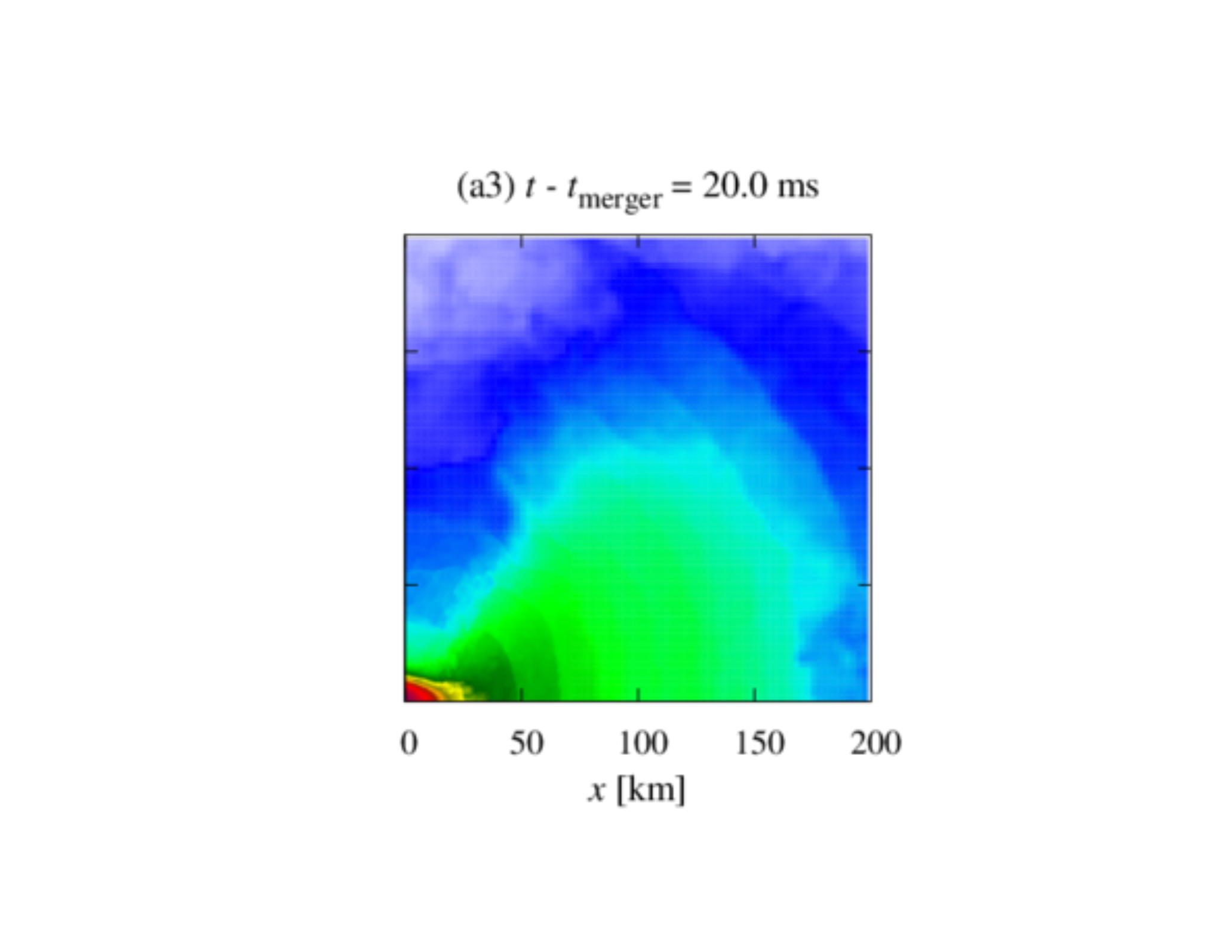}
\end{center}
\end{minipage}
\hspace{-12mm}
\begin{minipage}{0.27\hsize}
\begin{center}
\includegraphics[width=9.0cm,angle=0]{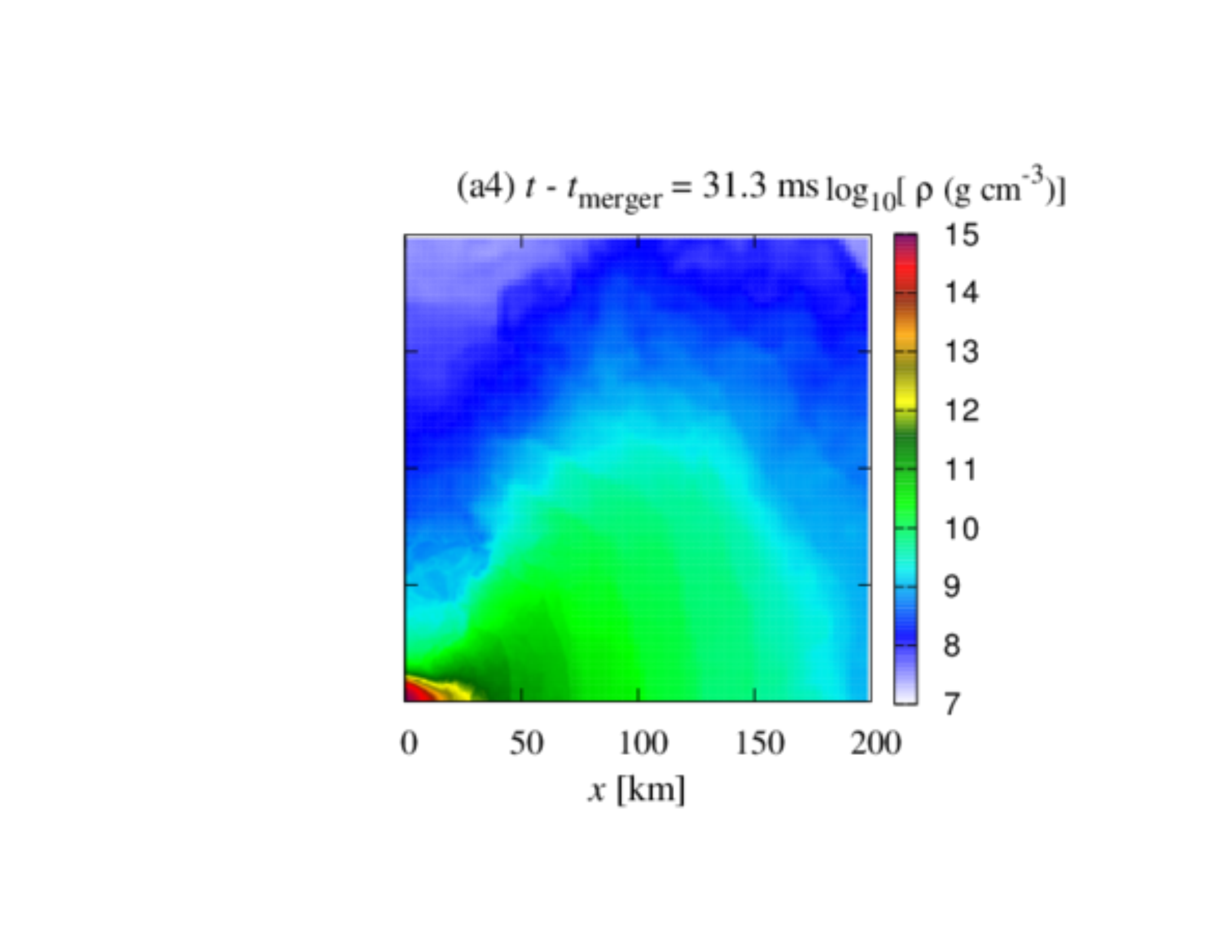}
\end{center}
\end{minipage}\\
\vspace{-18mm}
\hspace{-50mm}
\begin{minipage}{0.27\hsize}
\begin{center}
\includegraphics[width=9.0cm,angle=0]{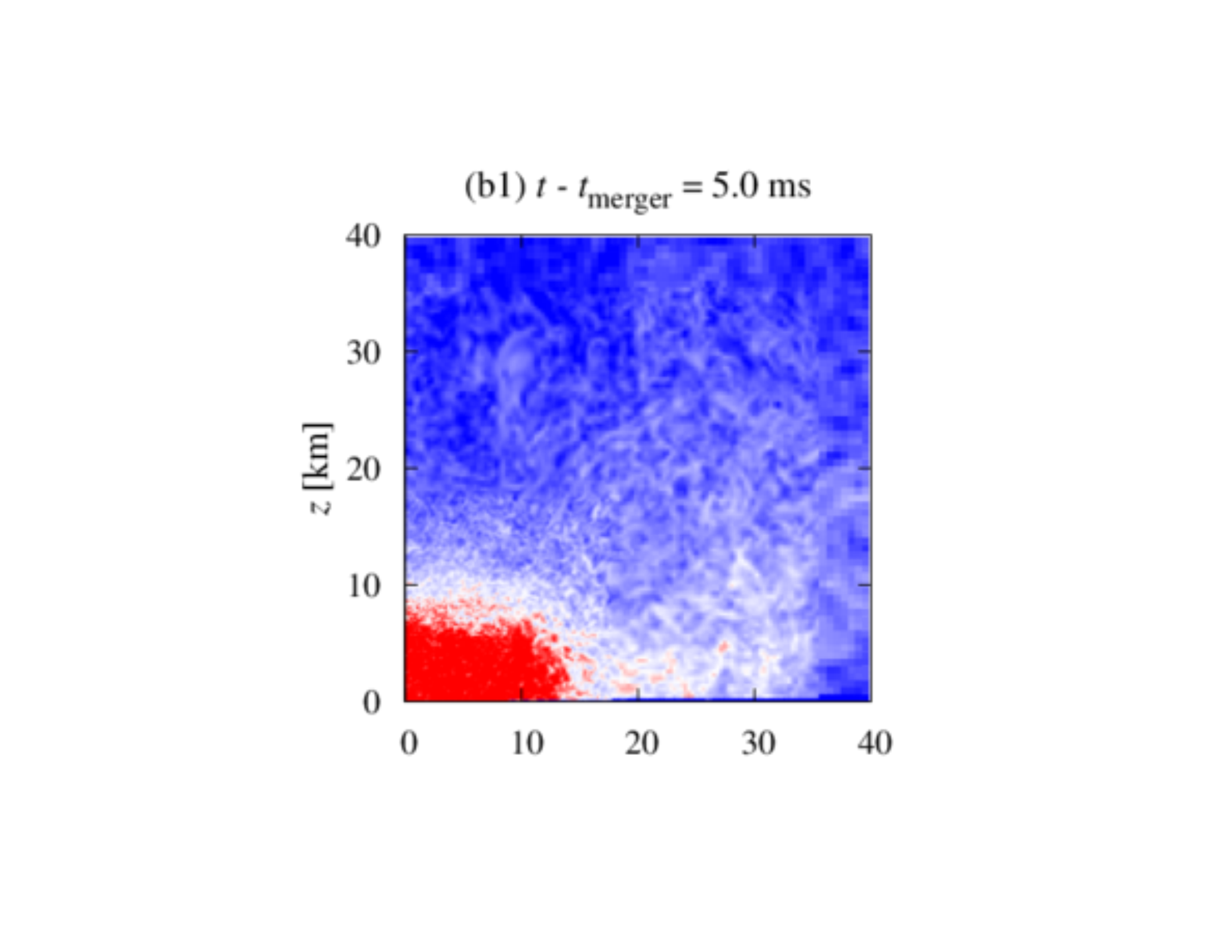}
\end{center}
\end{minipage}
\hspace{-12mm}
\begin{minipage}{0.27\hsize}
\begin{center}
\includegraphics[width=9.0cm,angle=0]{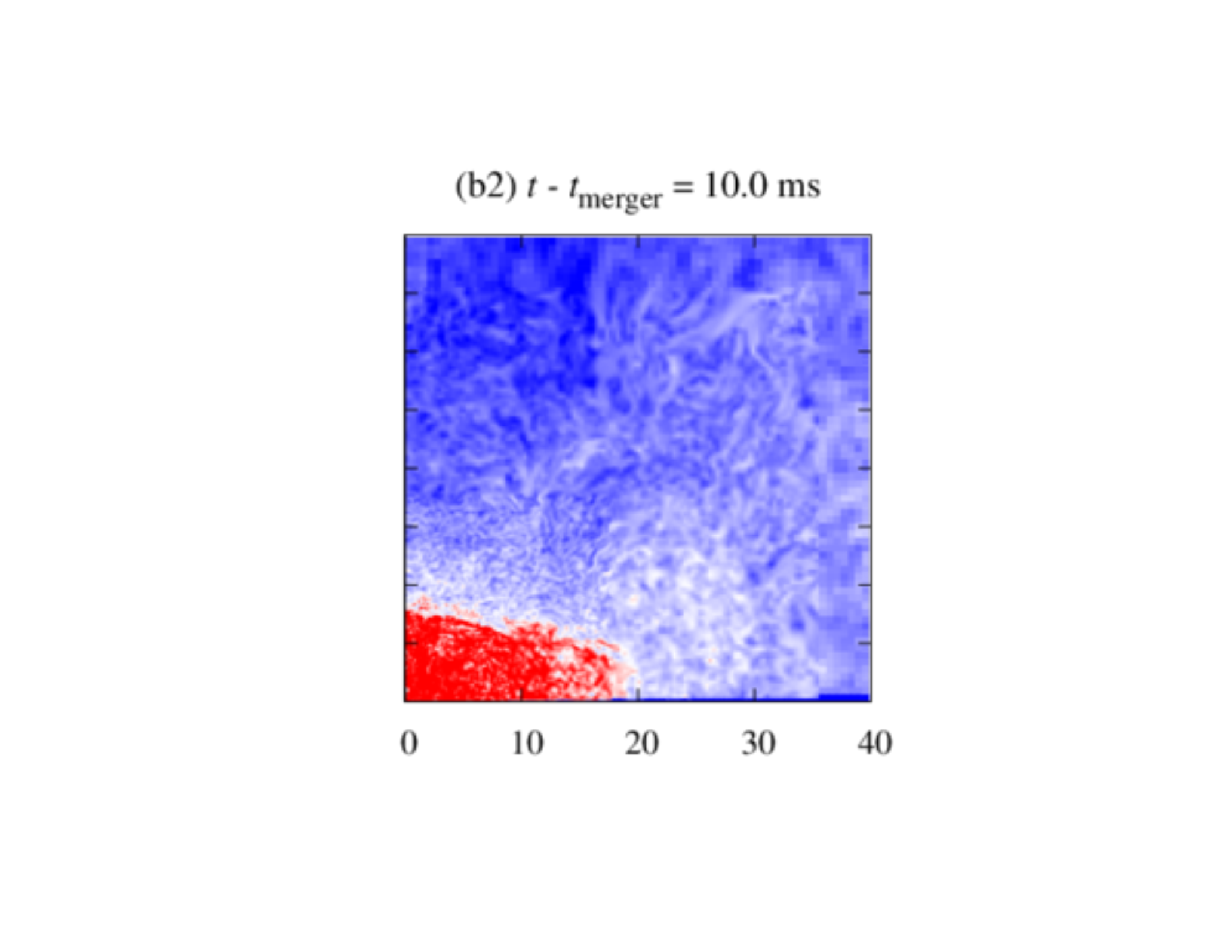}
\end{center}
\end{minipage}
\hspace{-12mm}
\begin{minipage}{0.27\hsize}
\begin{center}
\includegraphics[width=9.0cm,angle=0]{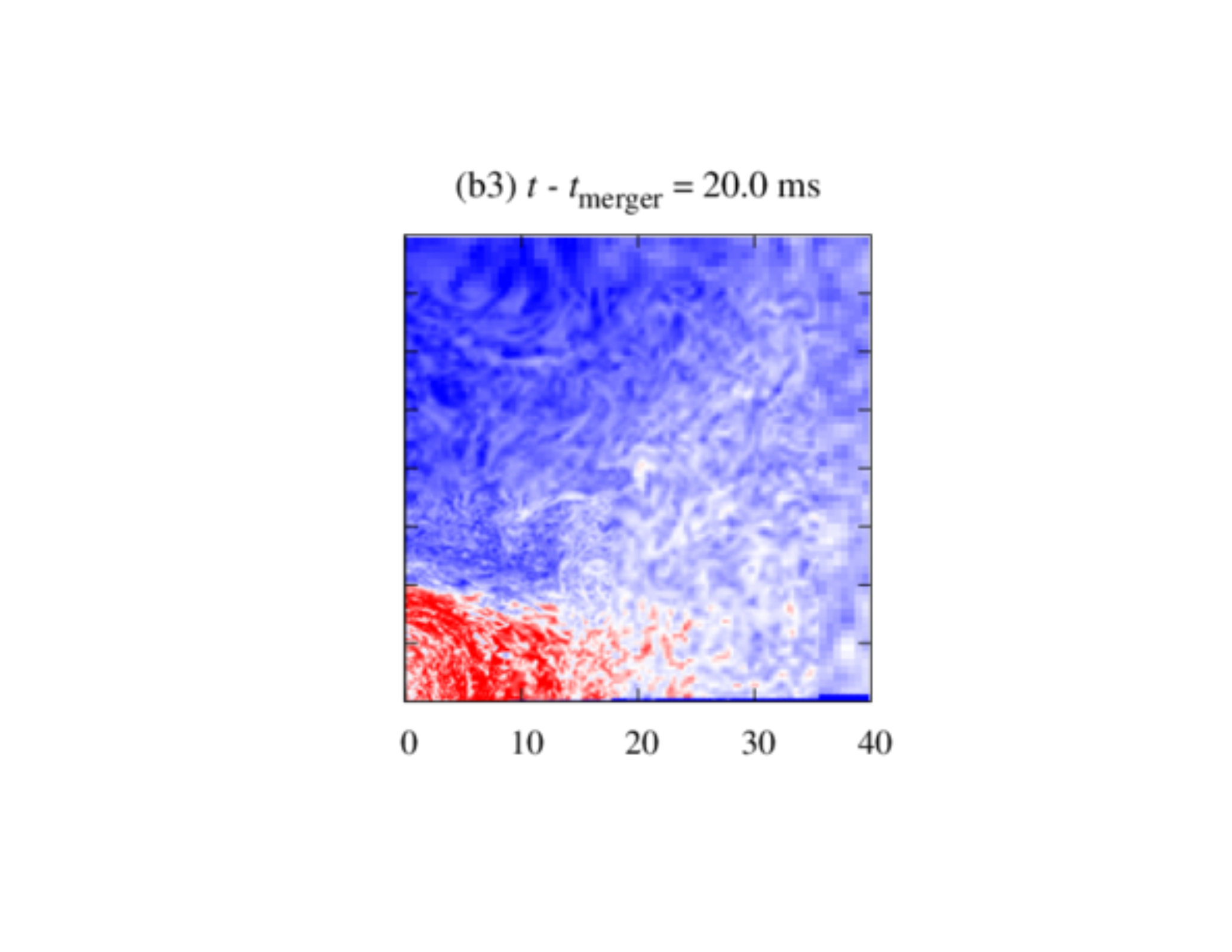}
\end{center}
\end{minipage}
\hspace{-12mm}
\begin{minipage}{0.27\hsize}
\begin{center}
\includegraphics[width=9.0cm,angle=0]{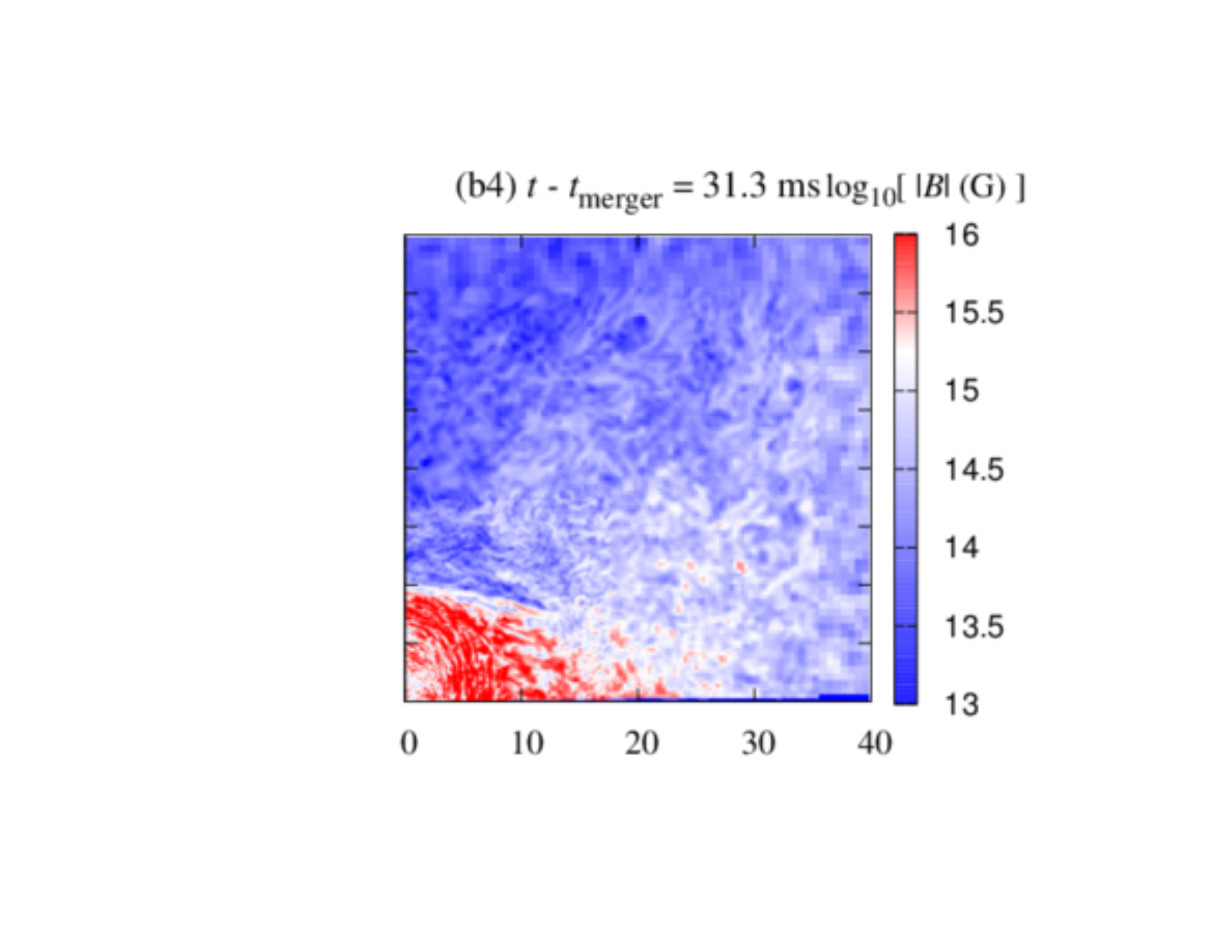}
\end{center}
\end{minipage}\\
\vspace{-18mm}
\hspace{-50mm}
\begin{minipage}{0.27\hsize}
\begin{center}
\includegraphics[width=9.0cm,angle=0]{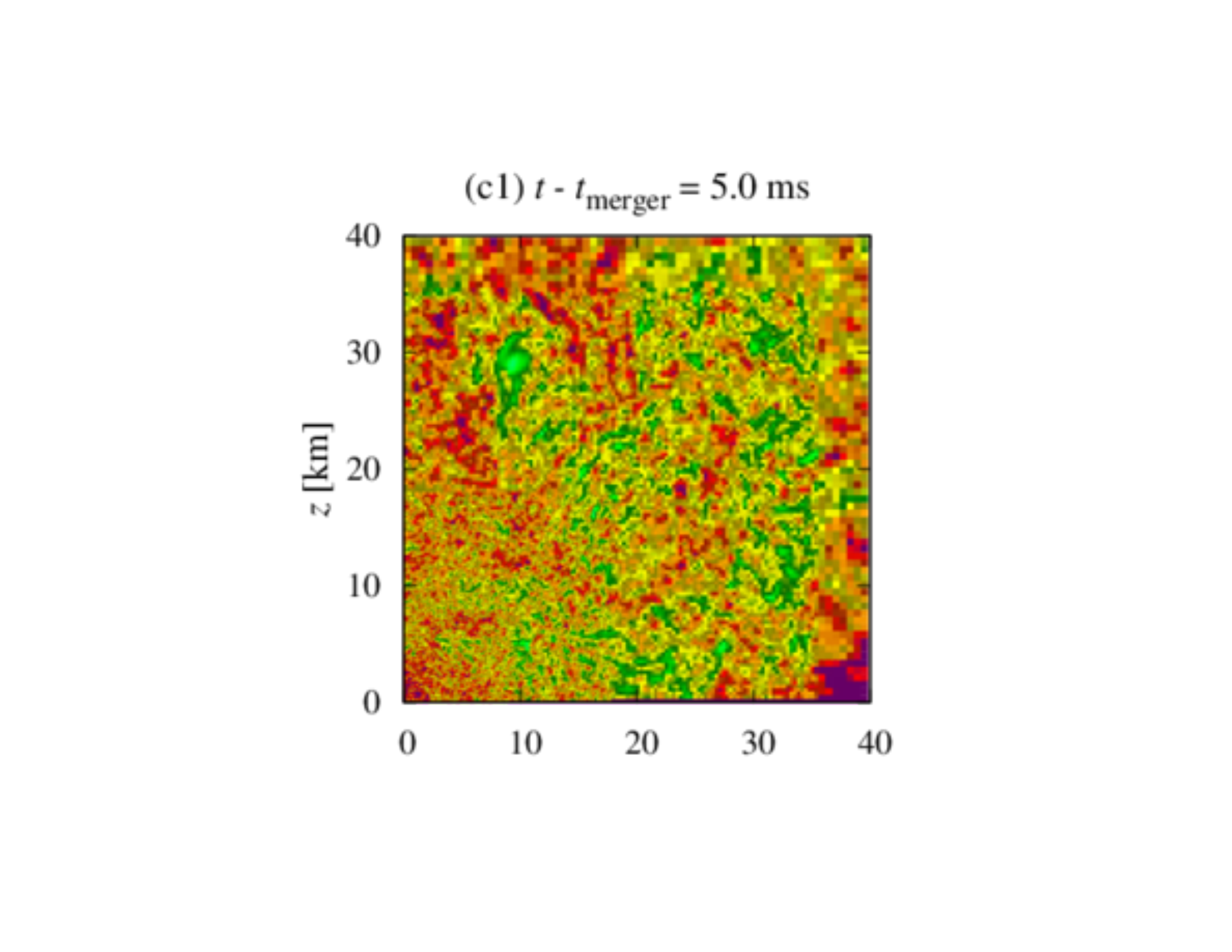}
\end{center}
\end{minipage}
\hspace{-12mm}
\begin{minipage}{0.27\hsize}
\begin{center}
\includegraphics[width=9.0cm,angle=0]{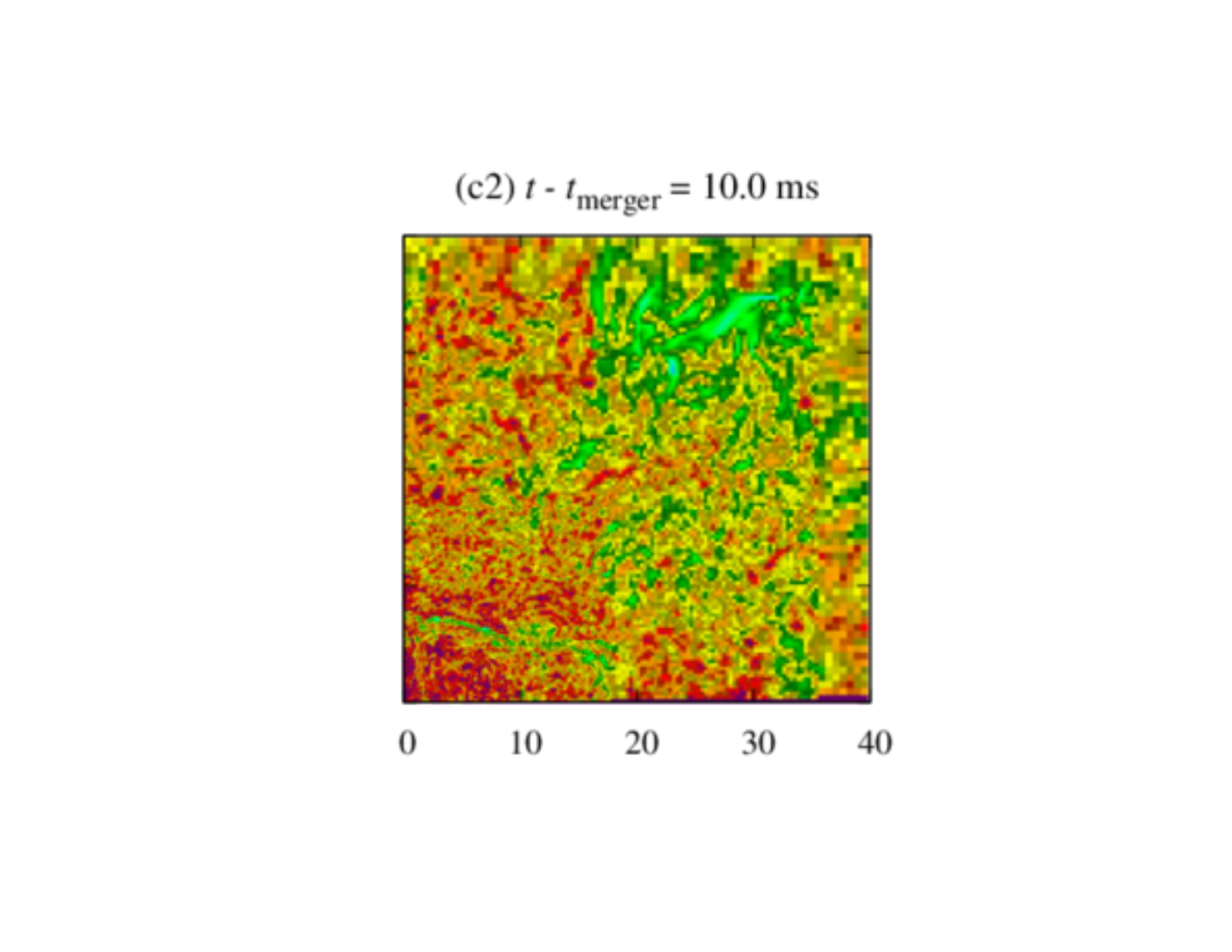}
\end{center}
\end{minipage}
\hspace{-12mm}
\begin{minipage}{0.27\hsize}
\begin{center}
\includegraphics[width=9.0cm,angle=0]{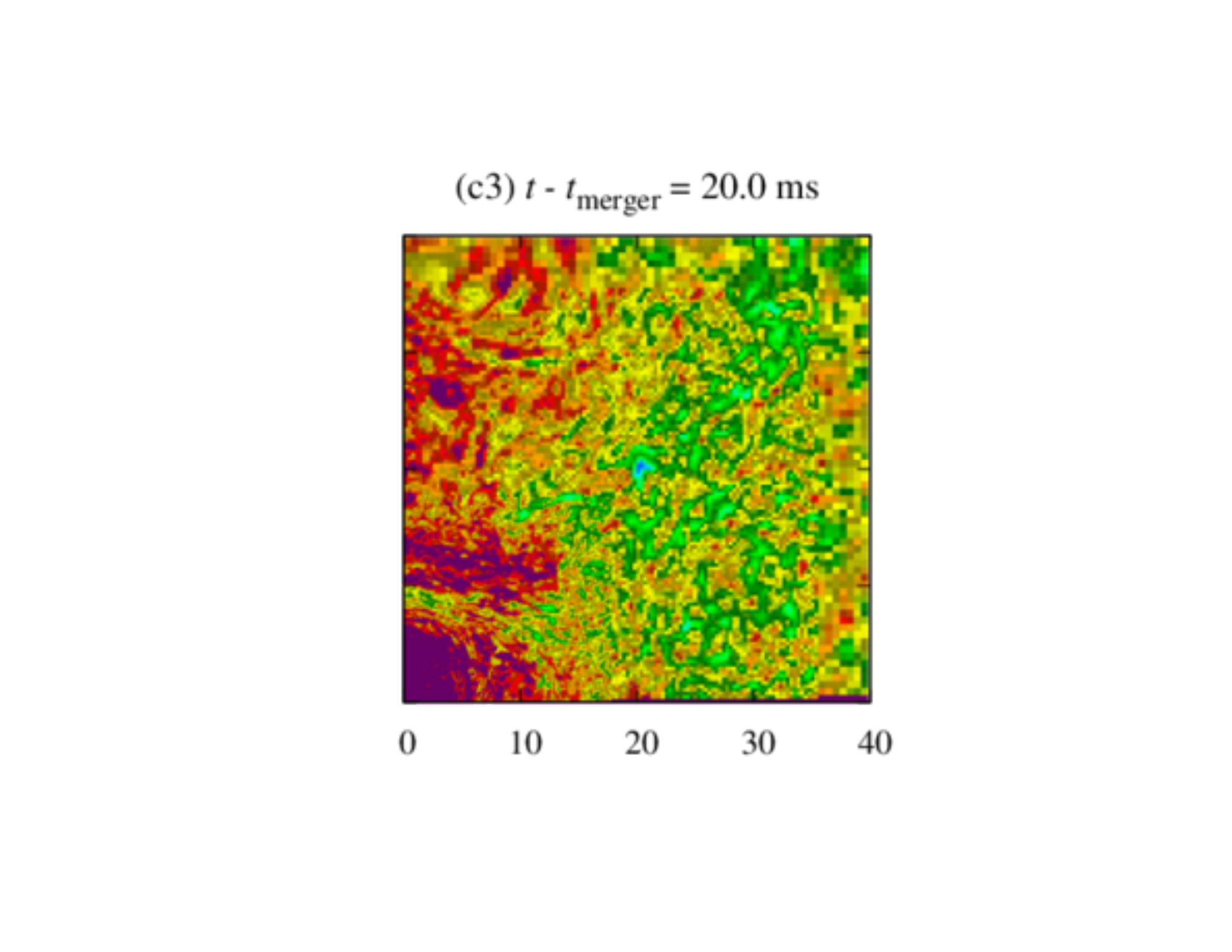}
\end{center}
\end{minipage}
\hspace{-12mm}
\begin{minipage}{0.27\hsize}
\begin{center}
\includegraphics[width=9.0cm,angle=0]{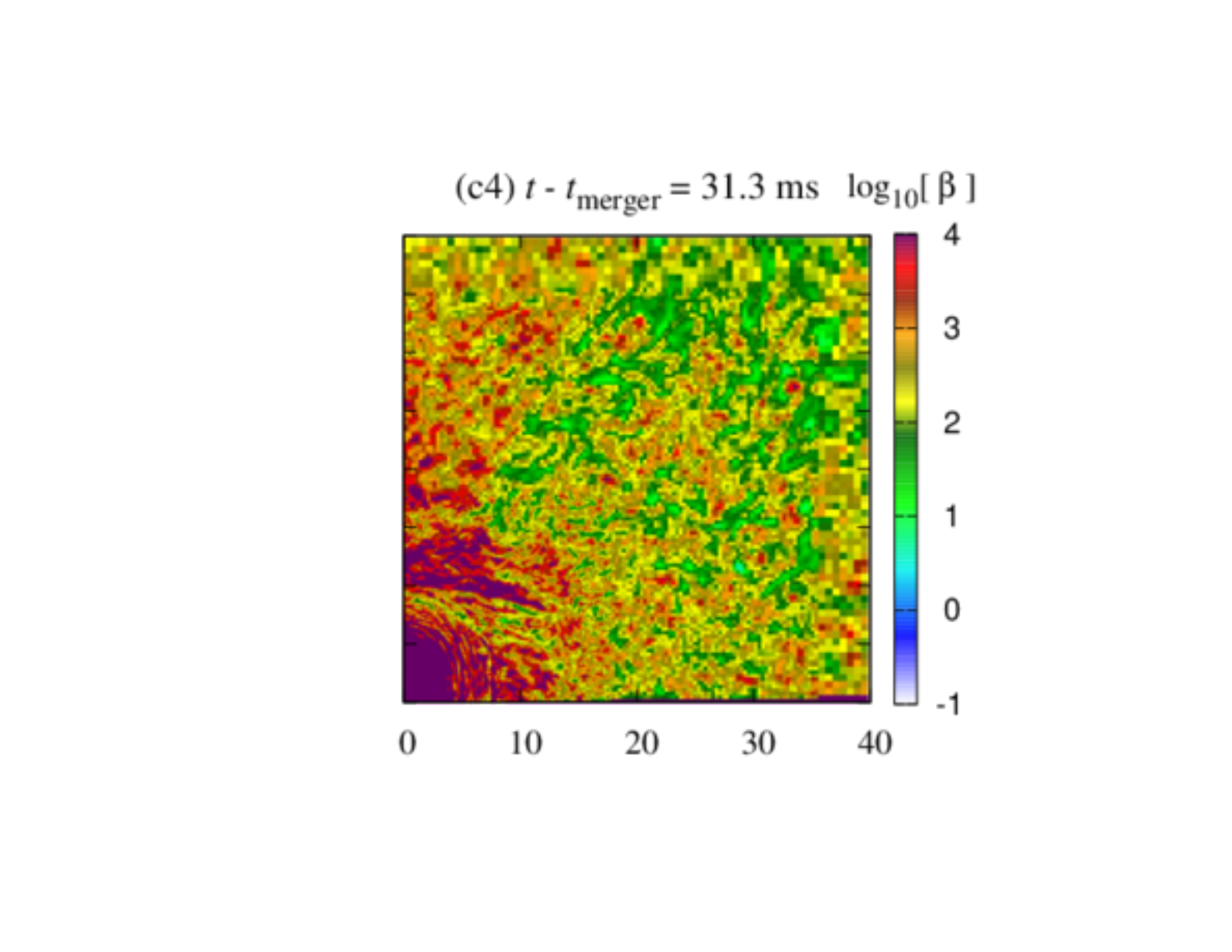}
\end{center}
\end{minipage}\\
\vspace{-18mm}
\hspace{-50mm}
\begin{minipage}{0.27\hsize}
\begin{center}
\includegraphics[width=9.0cm,angle=0]{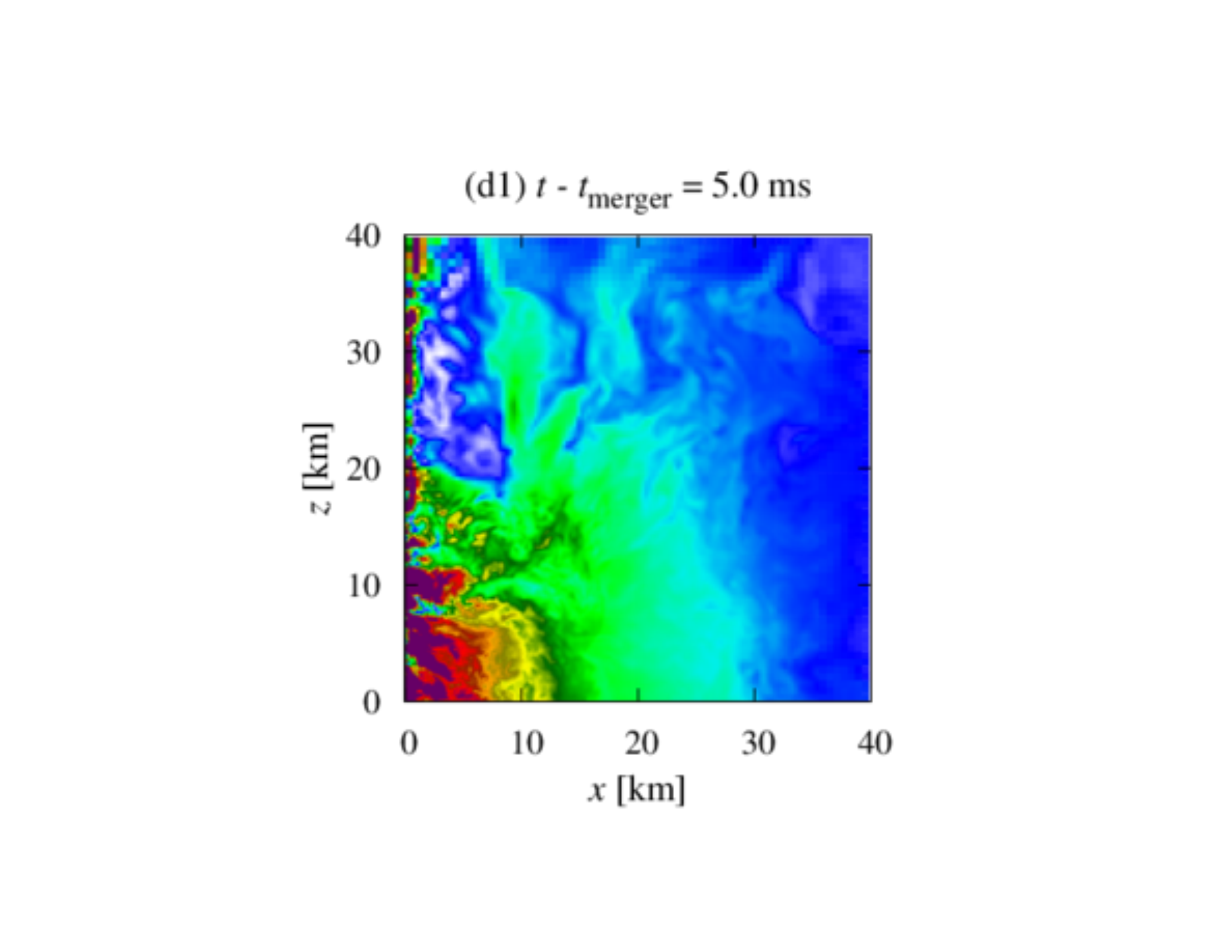}
\end{center}
\end{minipage}
\hspace{-12mm}
\begin{minipage}{0.27\hsize}
\begin{center}
\includegraphics[width=9.0cm,angle=0]{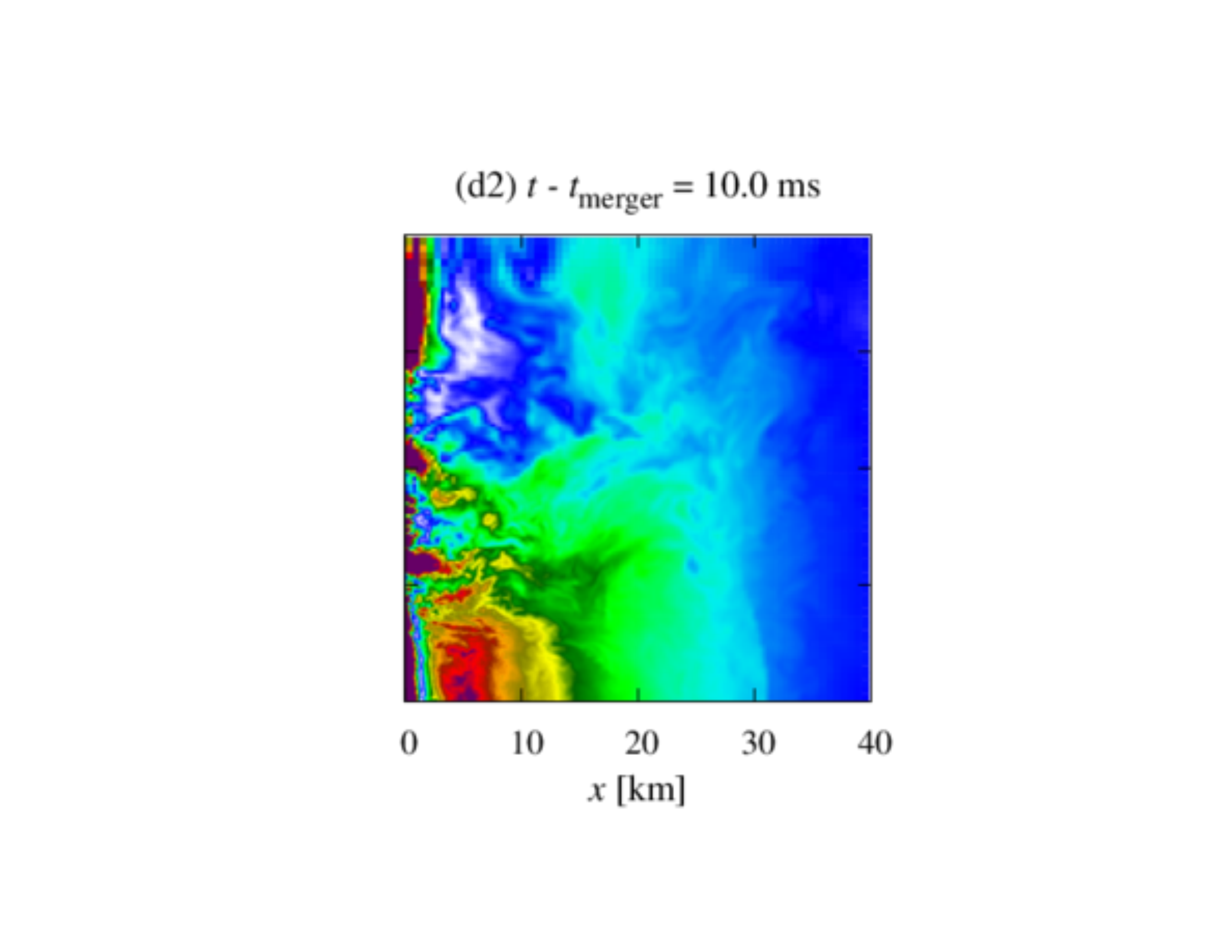}
\end{center}
\end{minipage}
\hspace{-12mm}
\begin{minipage}{0.27\hsize}
\begin{center}
\includegraphics[width=9.0cm,angle=0]{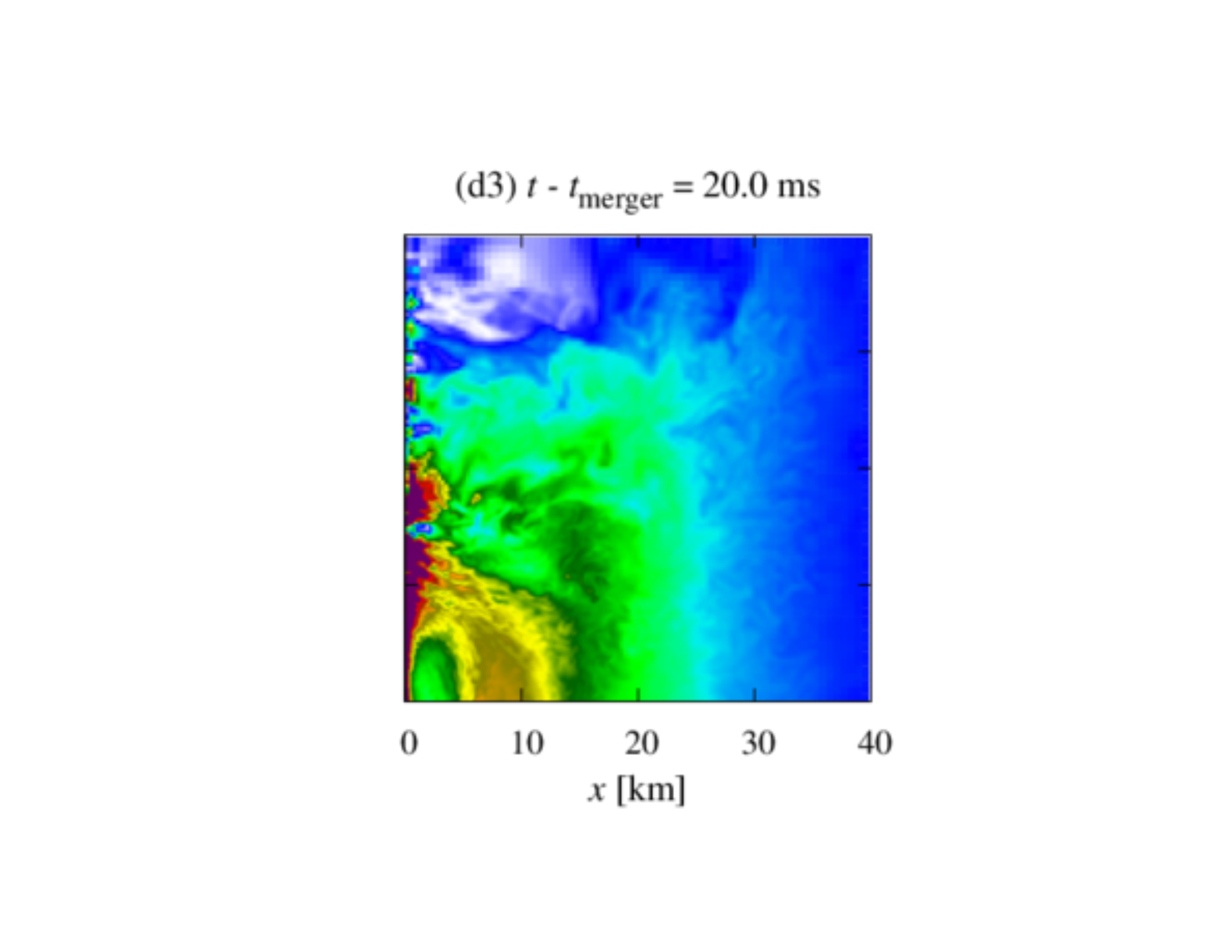}
\end{center}
\end{minipage}
\hspace{-12mm}
\begin{minipage}{0.27\hsize}
\begin{center}
\includegraphics[width=9.0cm,angle=0]{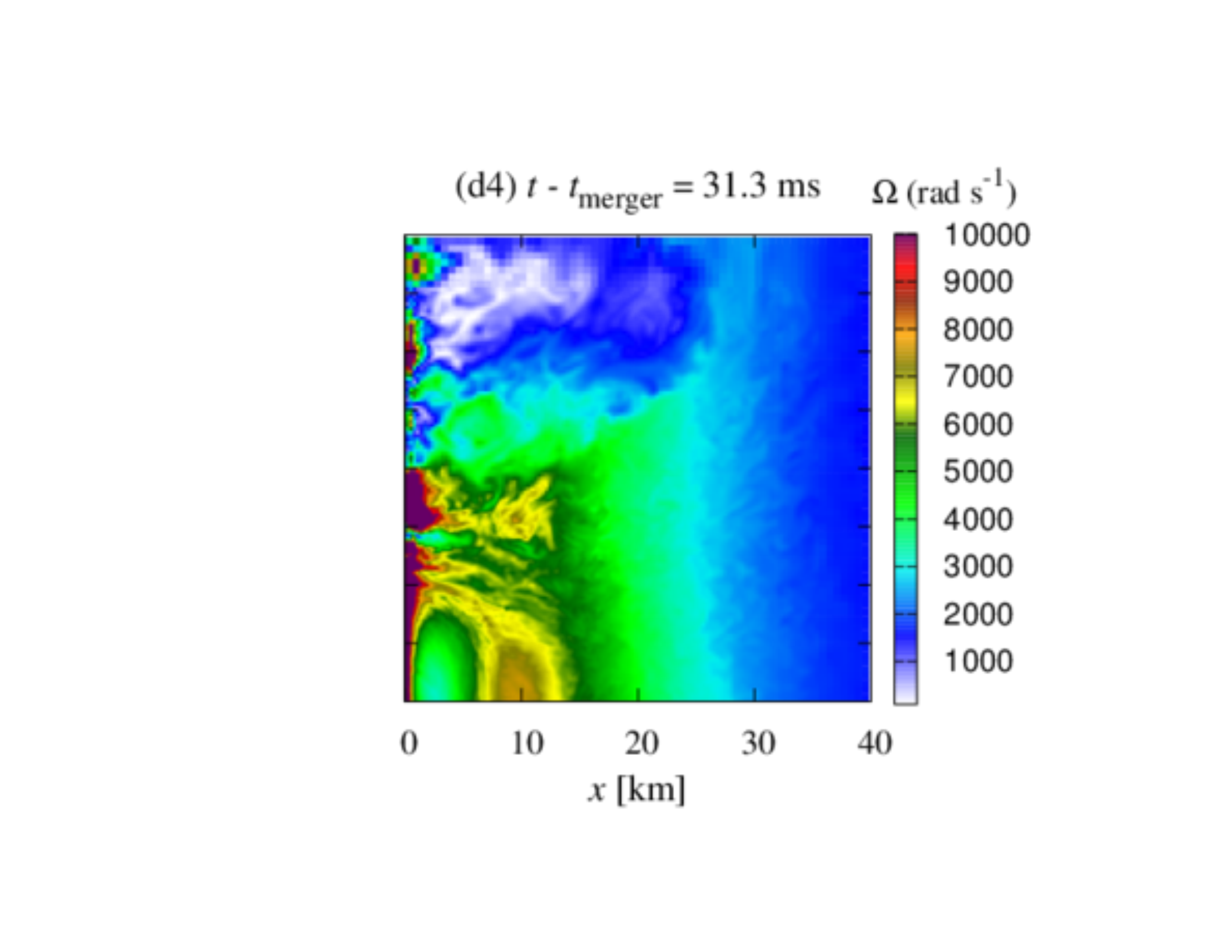}
\end{center}
\end{minipage}

\caption{\label{fig1}
Profiles of the rest-mass density (panels a1--a4), the magnetic-field strength (panels b1--b4), 
the plasma beta (panels c1--c4), and the angular velocity (panels d1--d4) on a meridional plane 
for the $12.5$ m run. $t_{\rm merger}$ is the merger time (see text in detail). 
Note that the panels (a1--a4) show a wider region than the other panels. 
}
\end{figure*}

Note that a coherent large-scale magnetic field such as a dipole field is not developed during the simulation time although 
small-scale magnetic fields are amplified by the Kelvin-Helmholtz instability and the MRI. 
One possibility is that the grid resolution in this work is still insufficient to simulate a large-scale dynamo 
(see Ref.~\cite{Mosta:2015ucs} for the large-scale dynamo of the toroidal magnetic field).
The other possibility is that a part of the dynamical ejecta falls back onto the RMNS and the matter inertia still dominates the electromagnetic force around the RMNS as shown in Fig.~\ref{fig1} (c1--c4).
Therefore, the magnetic pressure alone cannot drive an outflow and this indicates the formation of the coherent poloidal magnetic field in the early time of $\lesssim 30$ ms is unlikely.

\begin{figure*}[t]
\hspace{-50mm}
\begin{minipage}{0.27\hsize}
\begin{center}
\includegraphics[width=9.5cm,angle=0]{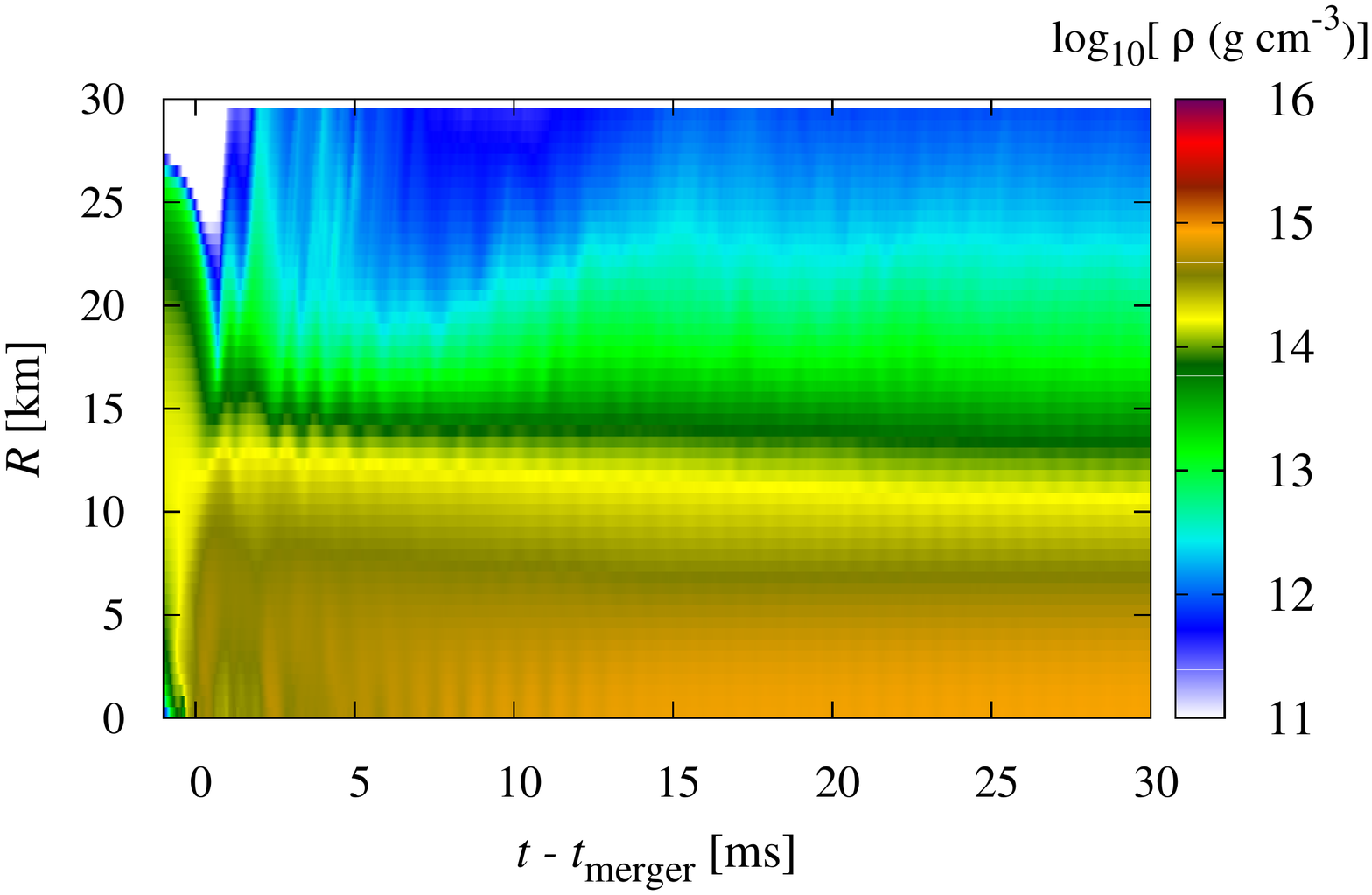}
\end{center}
\end{minipage}
\hspace{30mm}
\begin{minipage}{0.27\hsize}
\begin{center}
\includegraphics[width=9.5cm,angle=0]{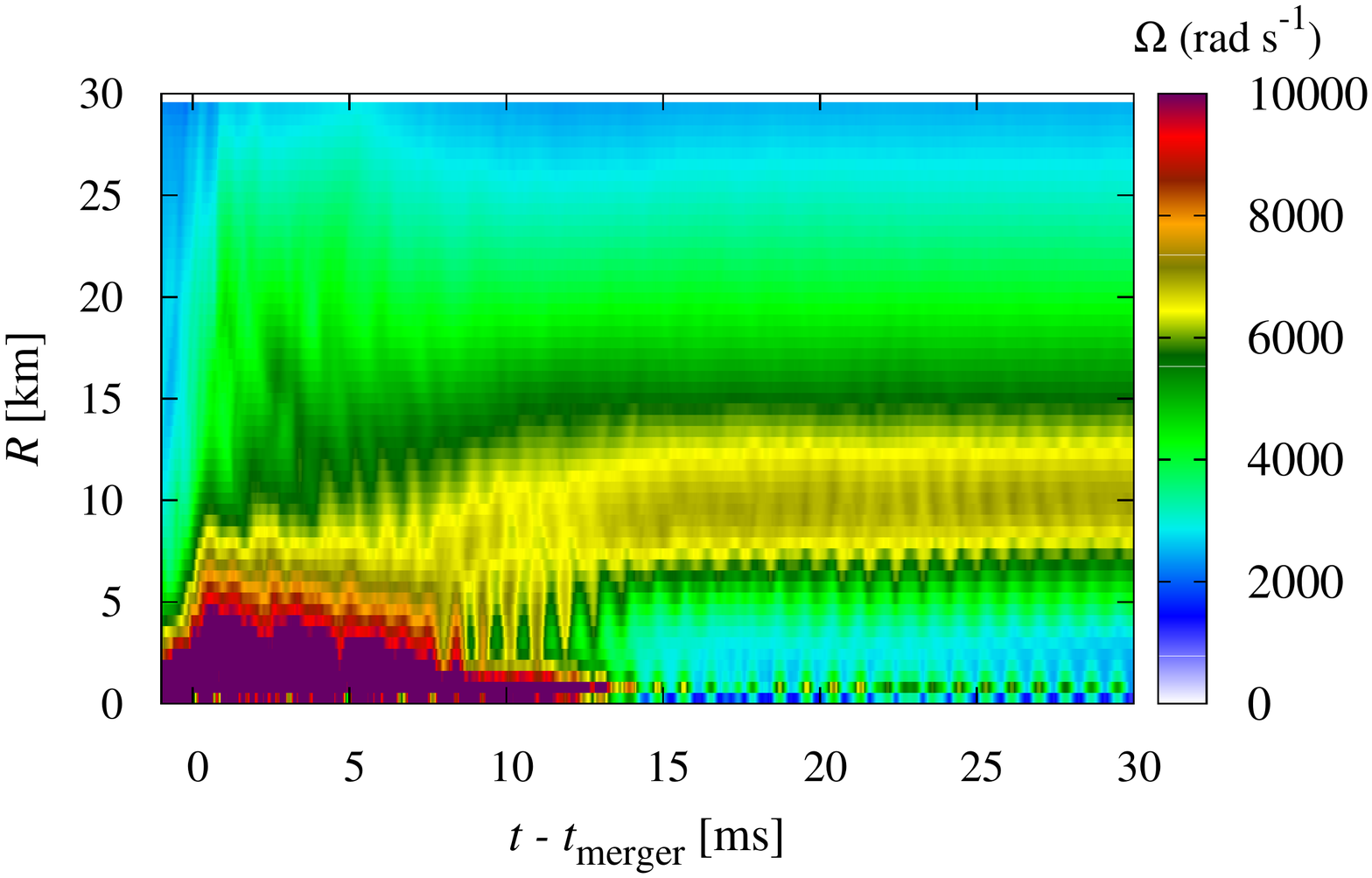}
\end{center}
\end{minipage}\\
\vspace{-20mm}
\hspace{-50mm}
\begin{minipage}{0.27\hsize}
\begin{center}
\includegraphics[width=9.5cm,angle=0]{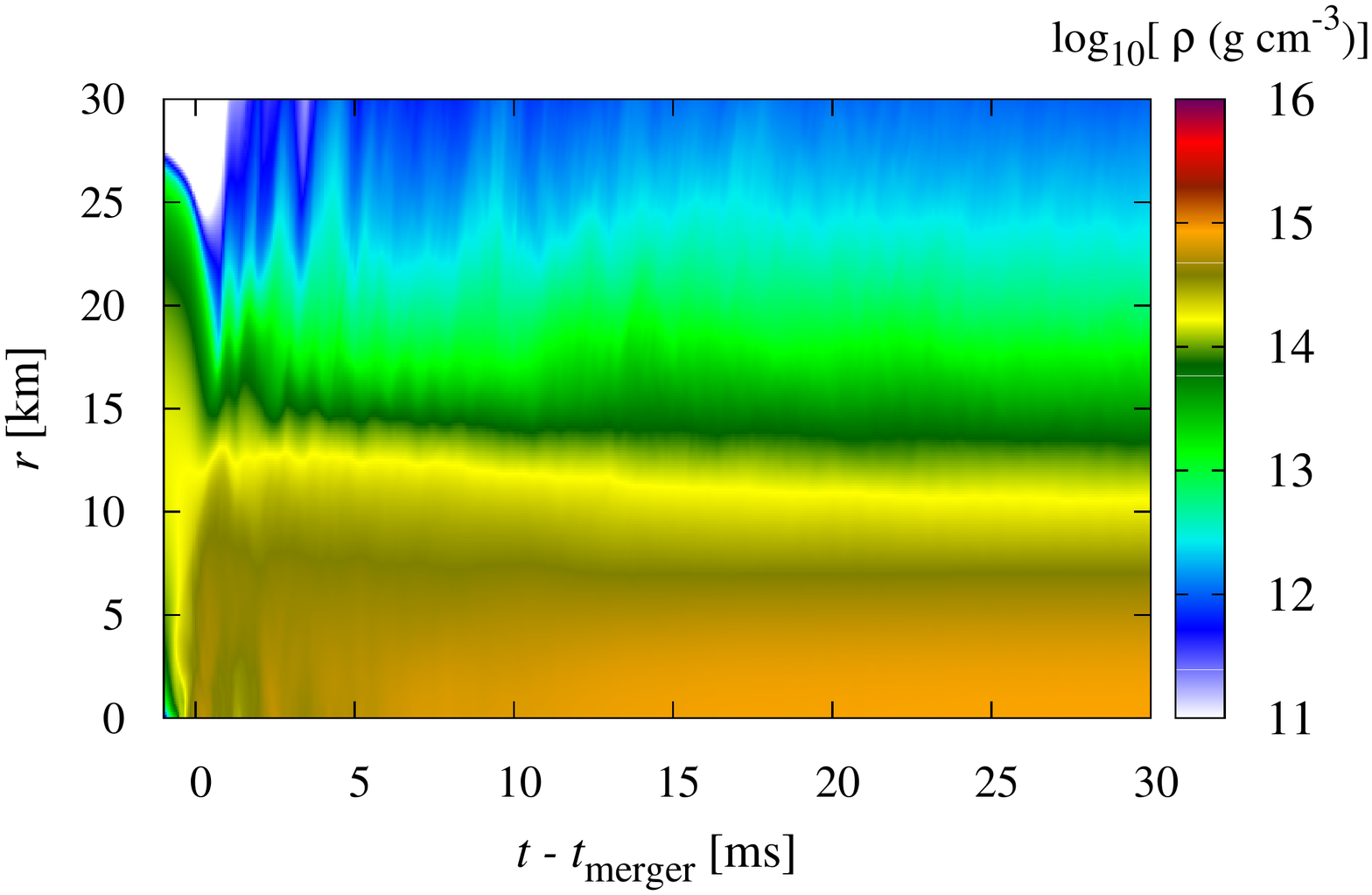}
\end{center}
\end{minipage}
\hspace{30mm}
\begin{minipage}{0.27\hsize}
\begin{center}
\includegraphics[width=9.5cm,angle=0]{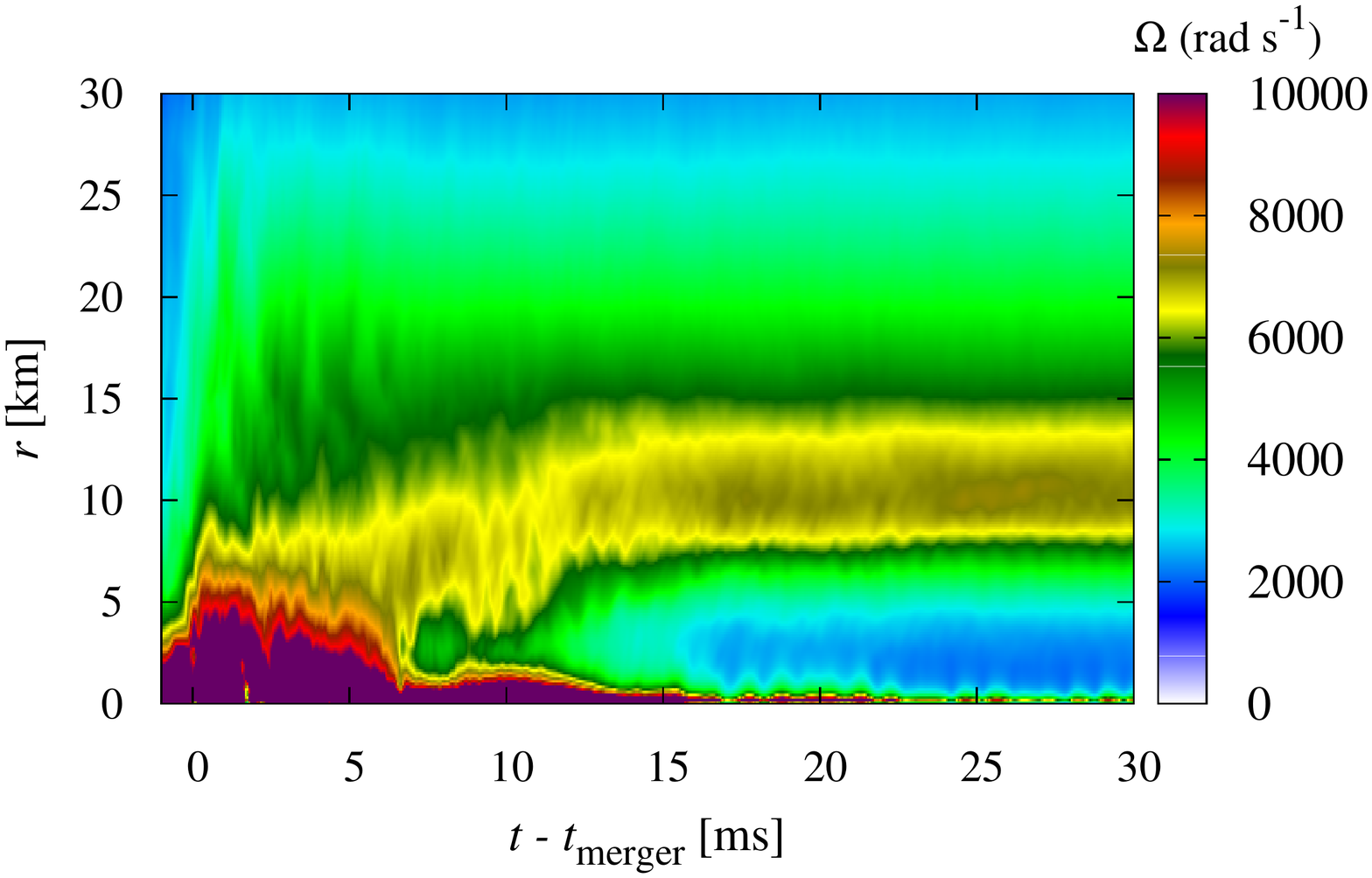}
\end{center}
\end{minipage}\\
\vspace{-20mm}
\hspace{-50mm}
\begin{minipage}{0.27\hsize}
\begin{center}
\includegraphics[width=9.5cm,angle=0]{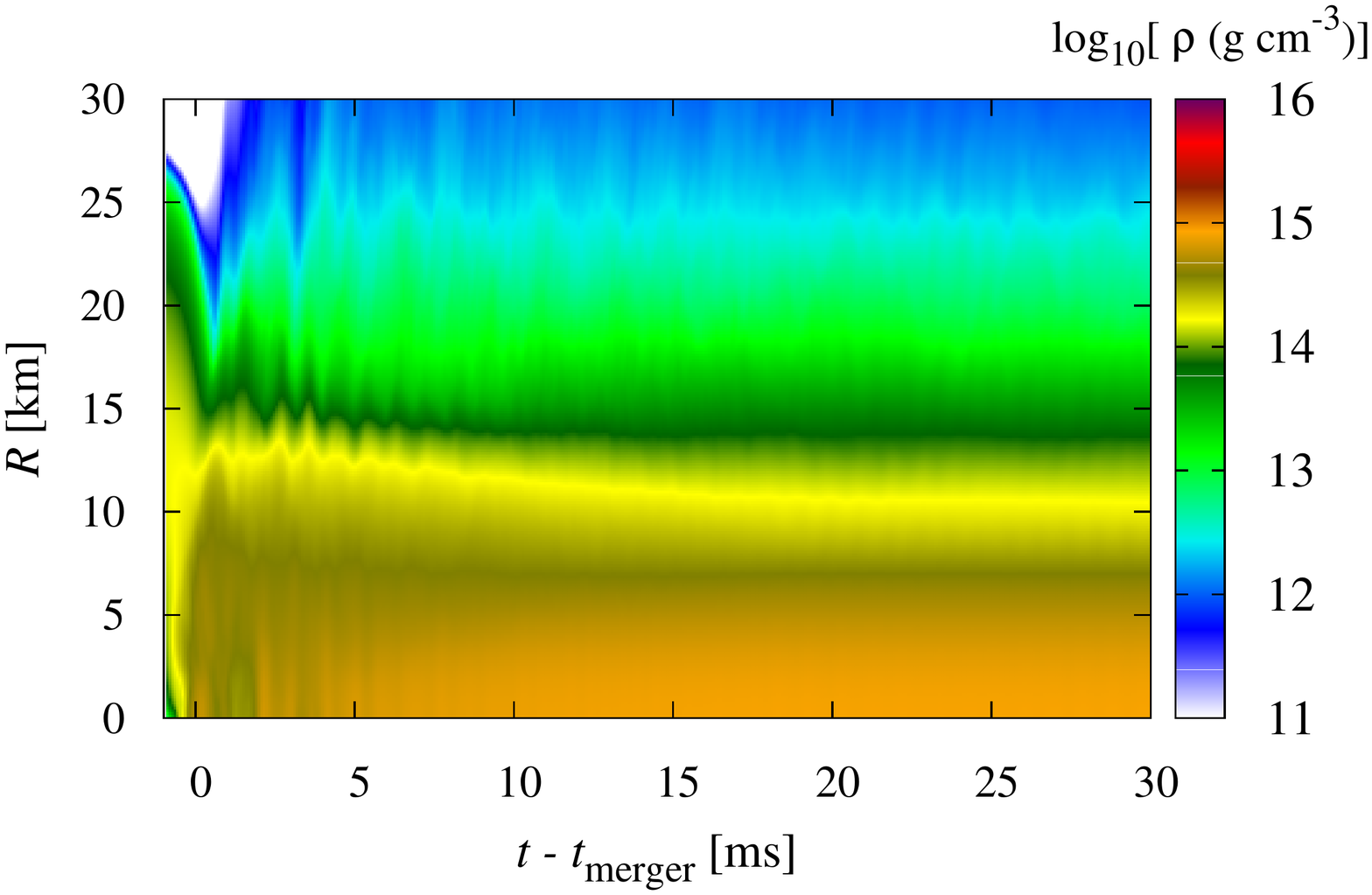}
\end{center}
\end{minipage}
\hspace{30mm}
\begin{minipage}{0.27\hsize}
\begin{center}
\includegraphics[width=9.5cm,angle=0]{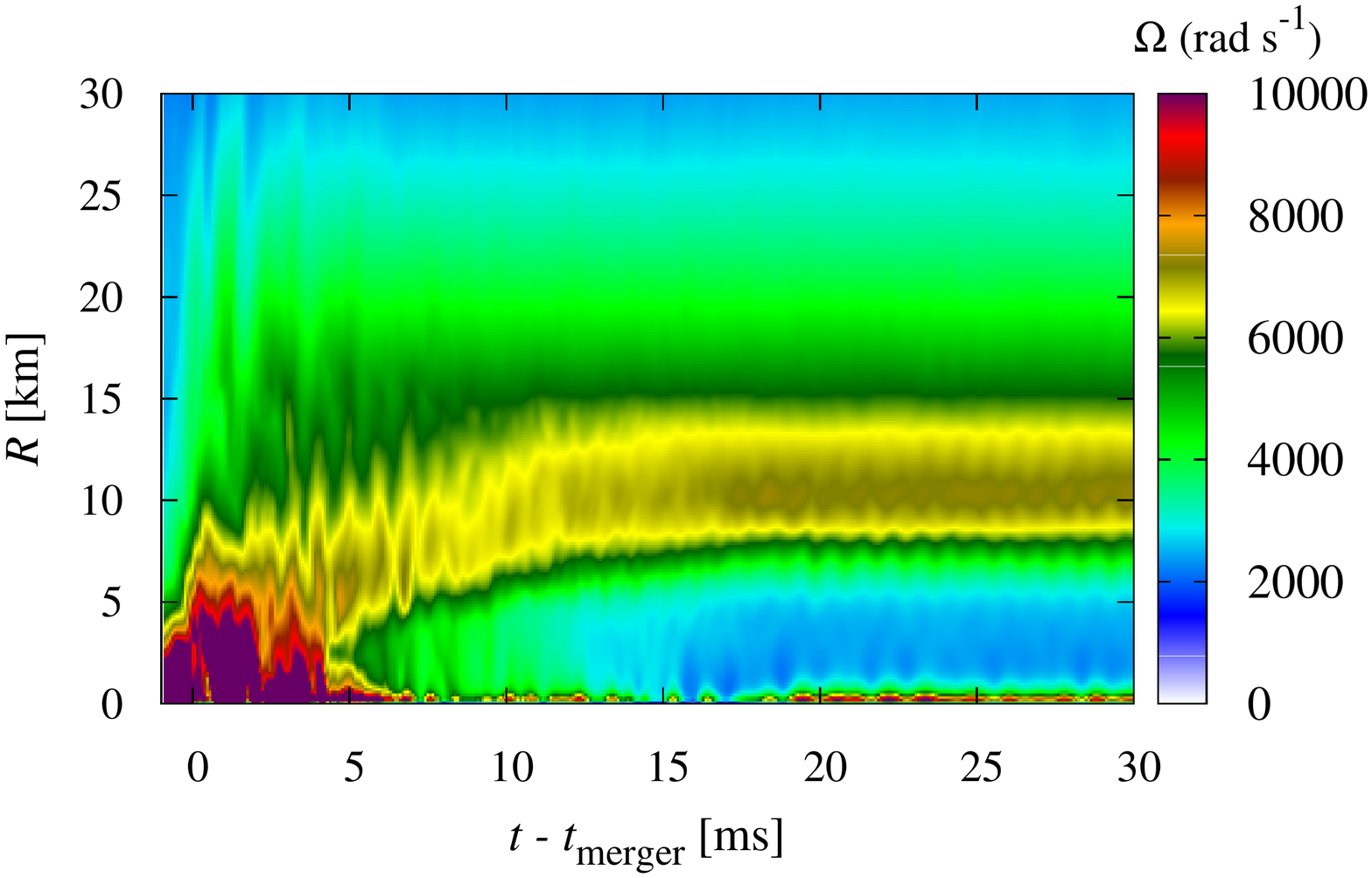}
\end{center}
\end{minipage}
\caption{\label{fig2}
Spacetime diagrams for the rest-mass density (left) and the angular velocity (right) on the orbital plane 
for the $12.5$ m run (top), the $70$ m run (middle), and the $110$ m run (bottom). 
Both profiles are generated by averaging the corresponding quantities along the azimuthal direction. 
}
\end{figure*}

\begin{figure*}[t]
\hspace{-50mm}
\begin{minipage}{0.27\hsize}
\begin{center}
\includegraphics[width=9.5cm,angle=0]{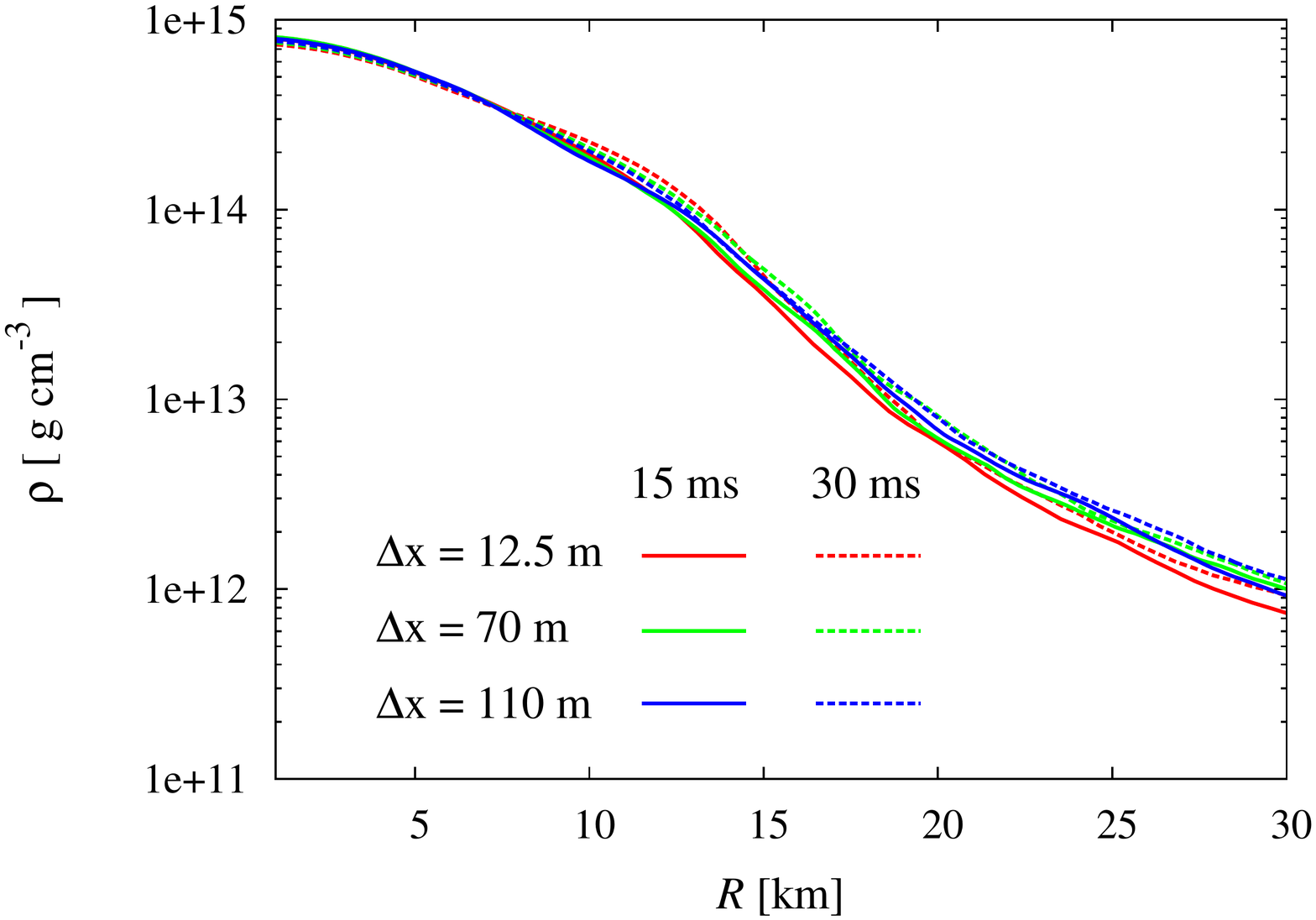}
\end{center}
\end{minipage}
\hspace{35mm}
\begin{minipage}{0.27\hsize}
\begin{center}
\includegraphics[width=9.5cm,angle=0]{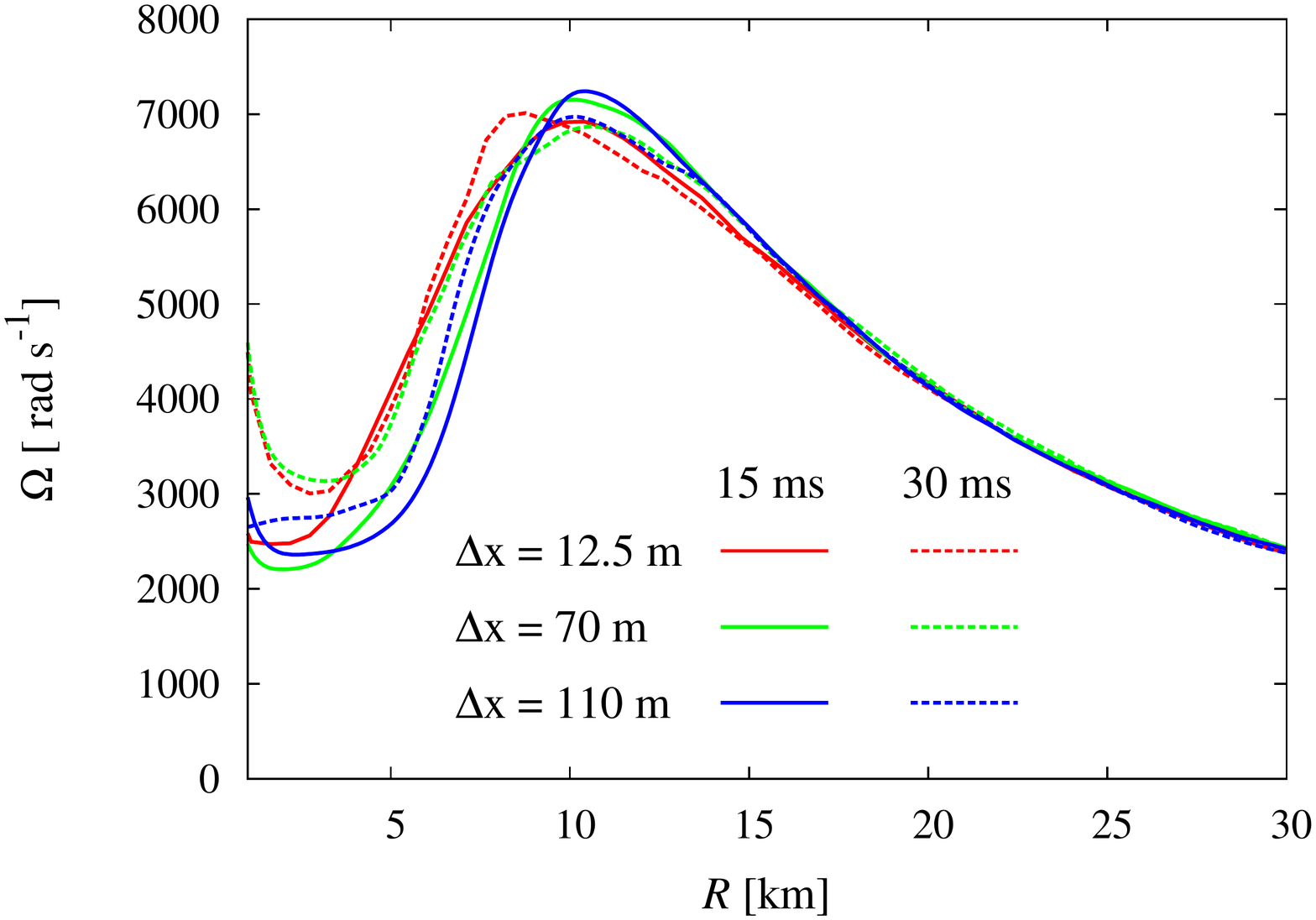}
\end{center}
\end{minipage}
\caption{\label{fig3}
Radial profiles of the rest-mass density (left) and the angular velocity (right) 
averaged along the azimuthal direction on the orbital plane at $t-t_{\rm merger}=15$ ms and $30$ ms. 
}
\end{figure*}

\begin{figure*}[t]
\hspace{-30mm}
\begin{minipage}{0.27\hsize}
\begin{center}
\includegraphics[width=7.5cm,angle=0]{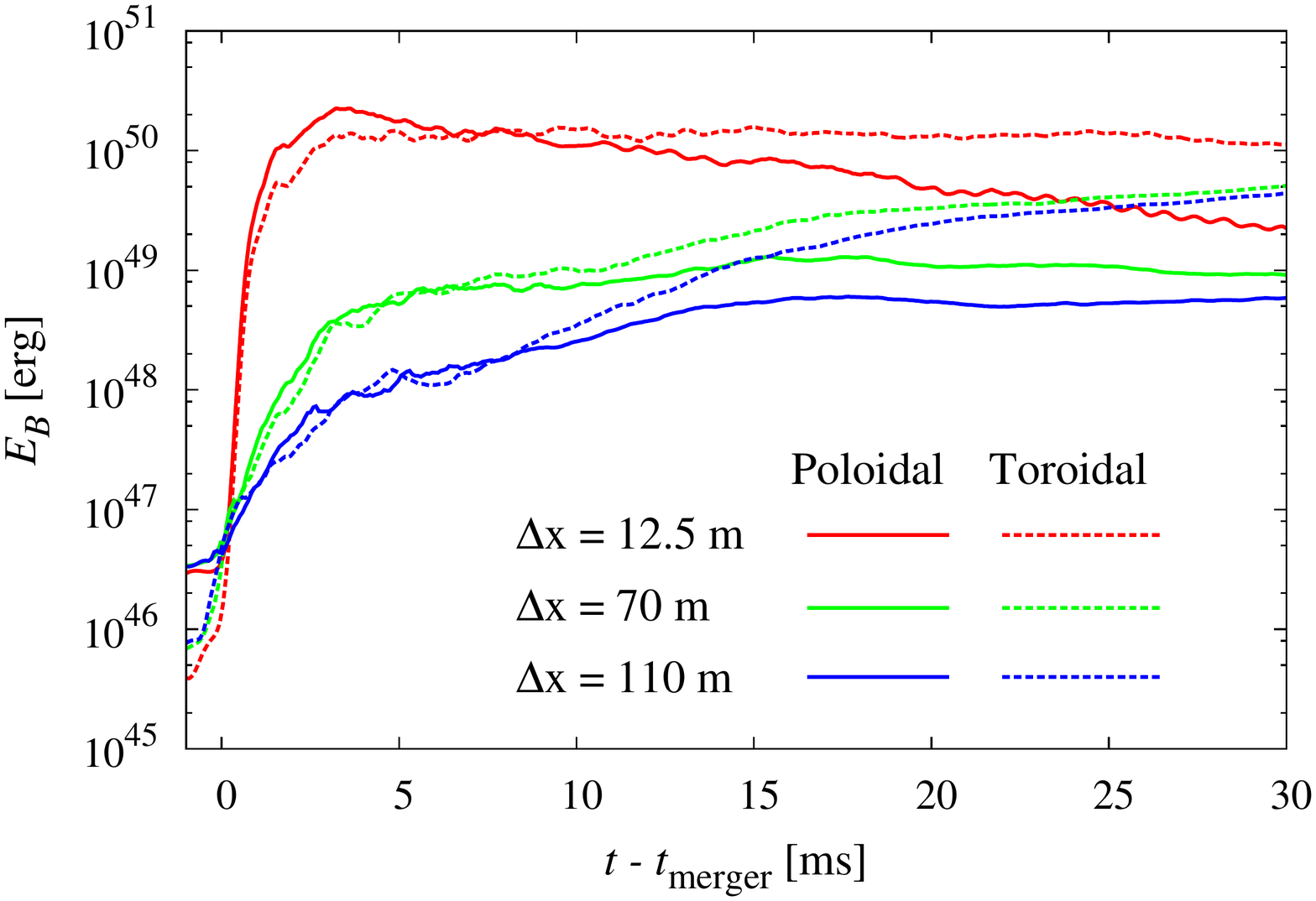}
\end{center}
\end{minipage}
\hspace{15mm}
\begin{minipage}{0.27\hsize}
\begin{center}
\includegraphics[width=7.5cm,angle=0]{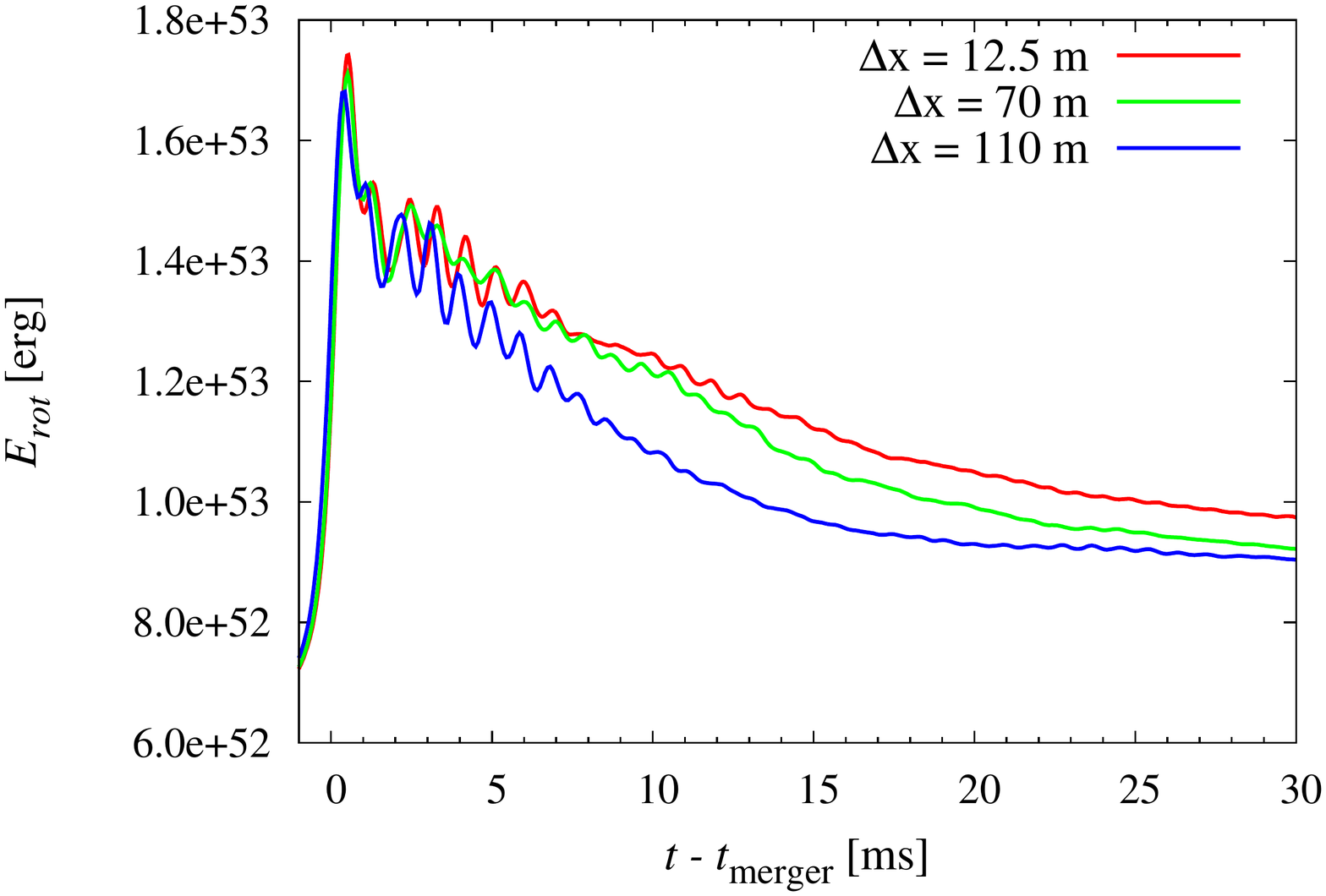}
\end{center}
\end{minipage}
\hspace{15mm}
\begin{minipage}{0.27\hsize}
\begin{center}
\includegraphics[width=7.5cm,angle=0]{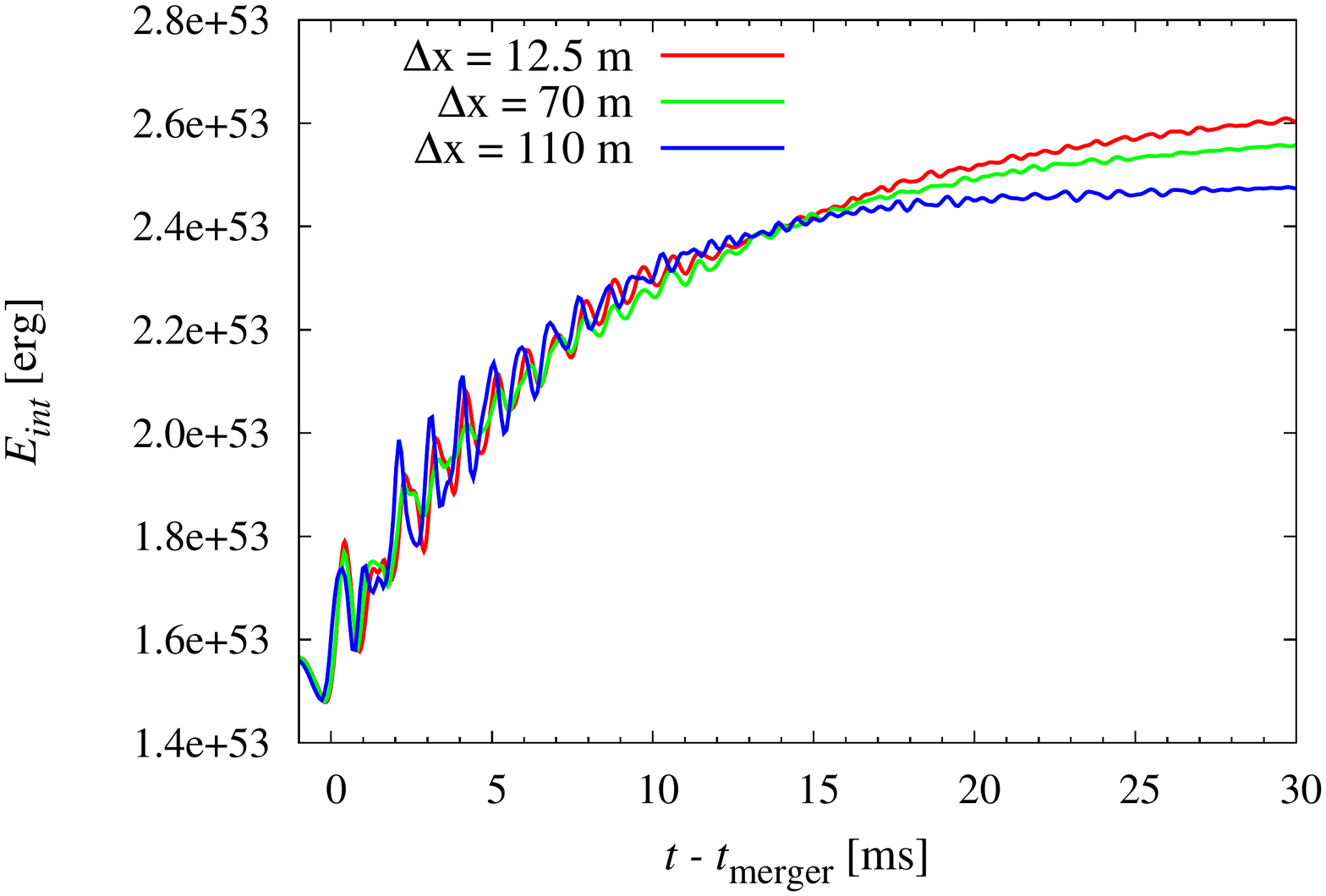}
\end{center}
\end{minipage}
\caption{\label{fig4}
Magnetic-field energy (left), rotational kinetic energy (center), and internal energy (right) as functions of time for all the runs. 
The solid and dashed curves in the left panel show the poloidal and toroidal components, respectively.
}
\end{figure*}

\begin{figure*}[t]
\hspace{-50mm}
\begin{minipage}{0.27\hsize}
\begin{center}
\includegraphics[width=7.0cm,angle=0]{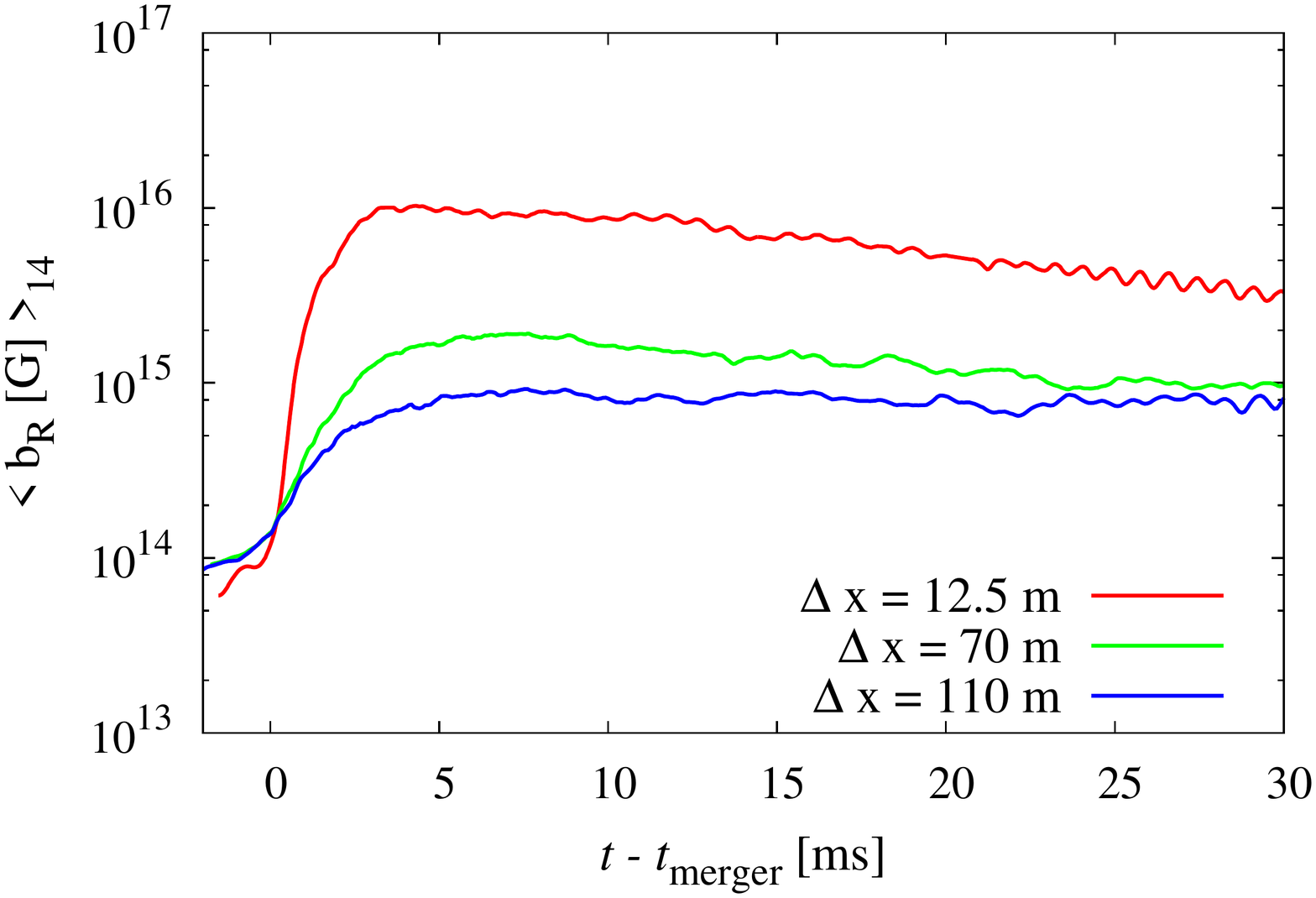}
\end{center}
\end{minipage}
\hspace{20mm}
\begin{minipage}{0.27\hsize}
\begin{center}
\includegraphics[width=7.0cm,angle=0]{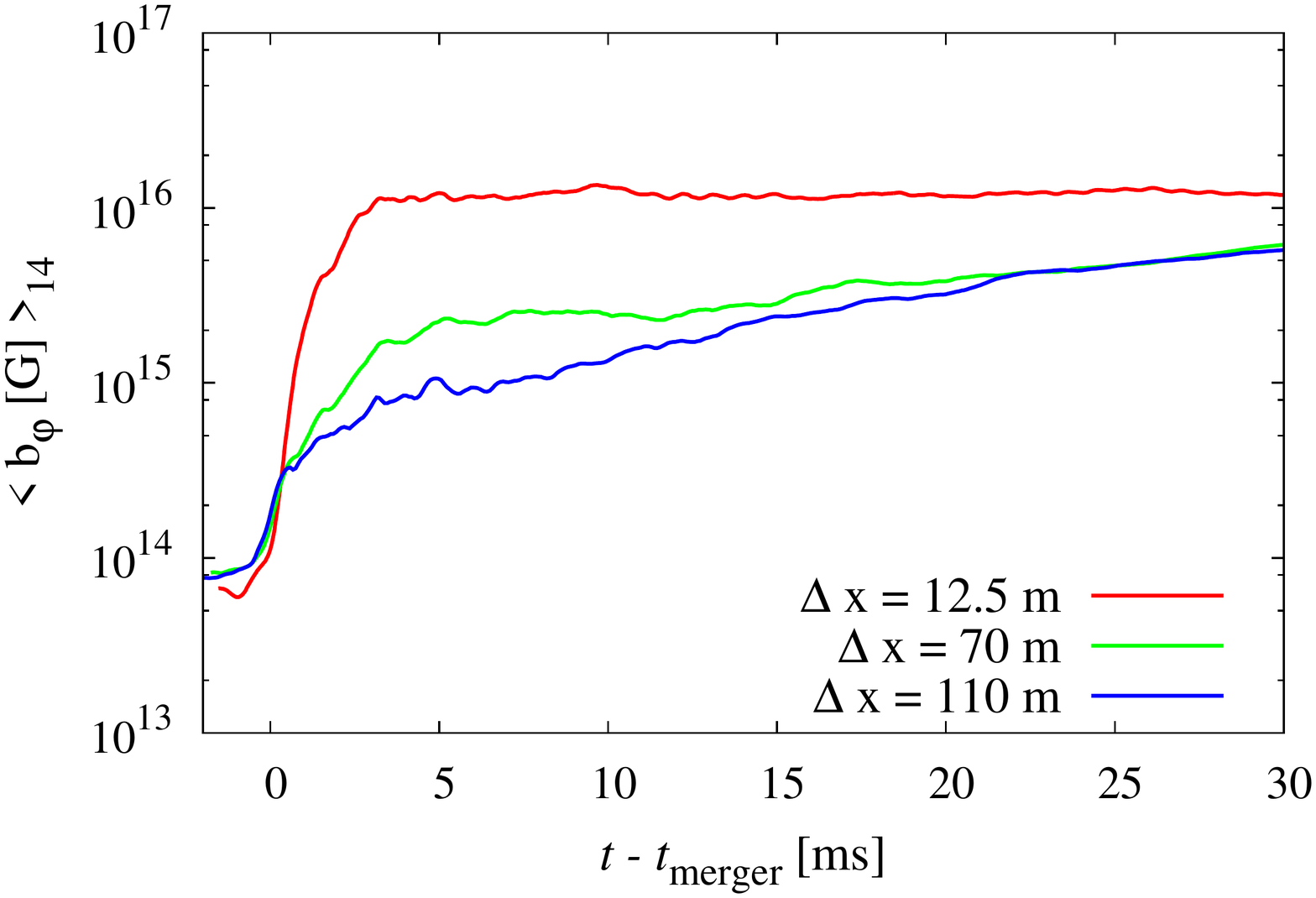}
\end{center}
\end{minipage}\\
\vspace{-10mm}
\hspace{-50mm}
\begin{minipage}{0.27\hsize}
\begin{center}
\includegraphics[width=7.0cm,angle=0]{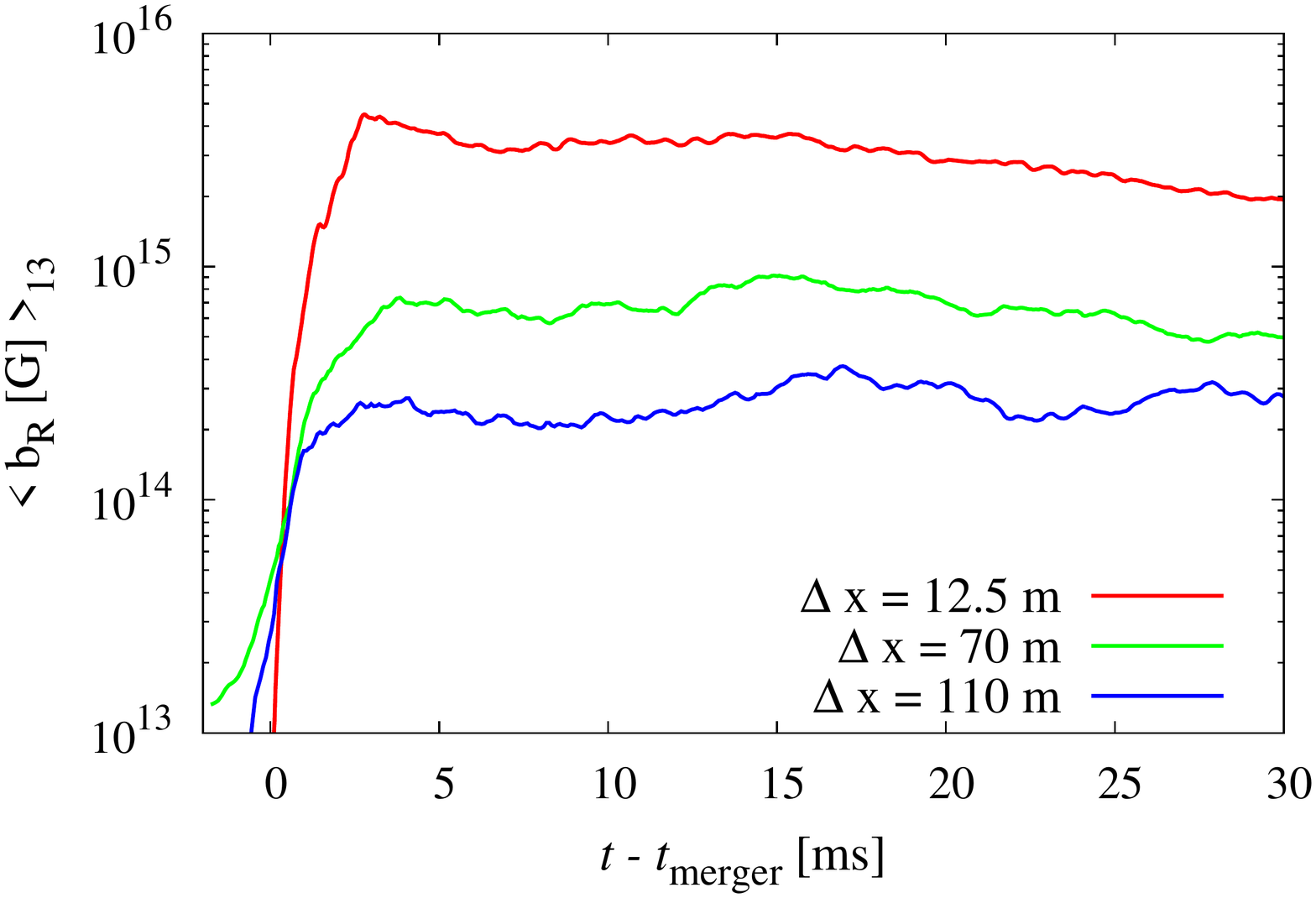}
\end{center}
\end{minipage}
\hspace{20mm}
\begin{minipage}{0.27\hsize}
\begin{center}
\includegraphics[width=7.0cm,angle=0]{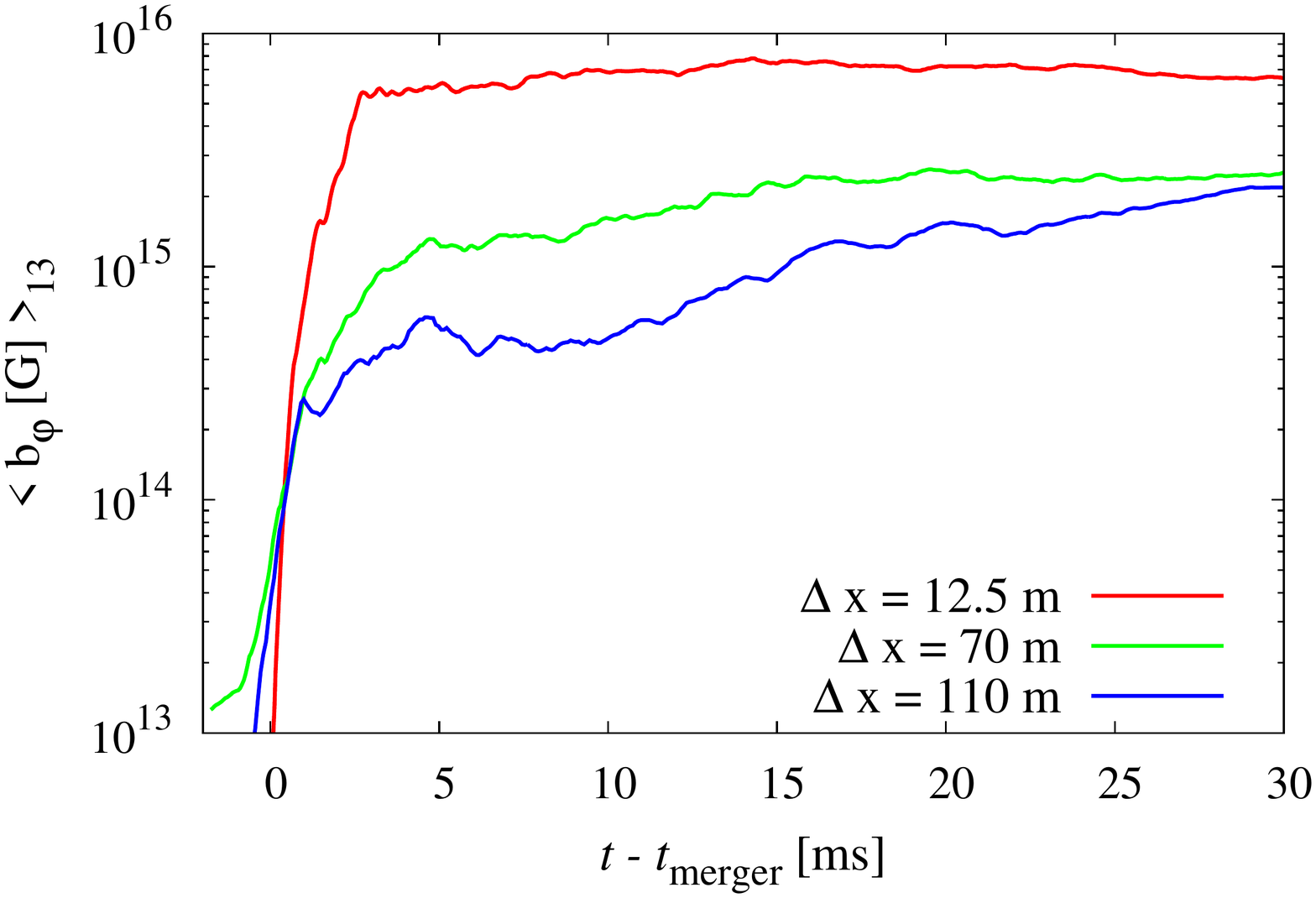}
\end{center}
\end{minipage}\\
\vspace{-10mm}
\hspace{-50mm}
\begin{minipage}{0.27\hsize}
\begin{center}
\includegraphics[width=7.0cm,angle=0]{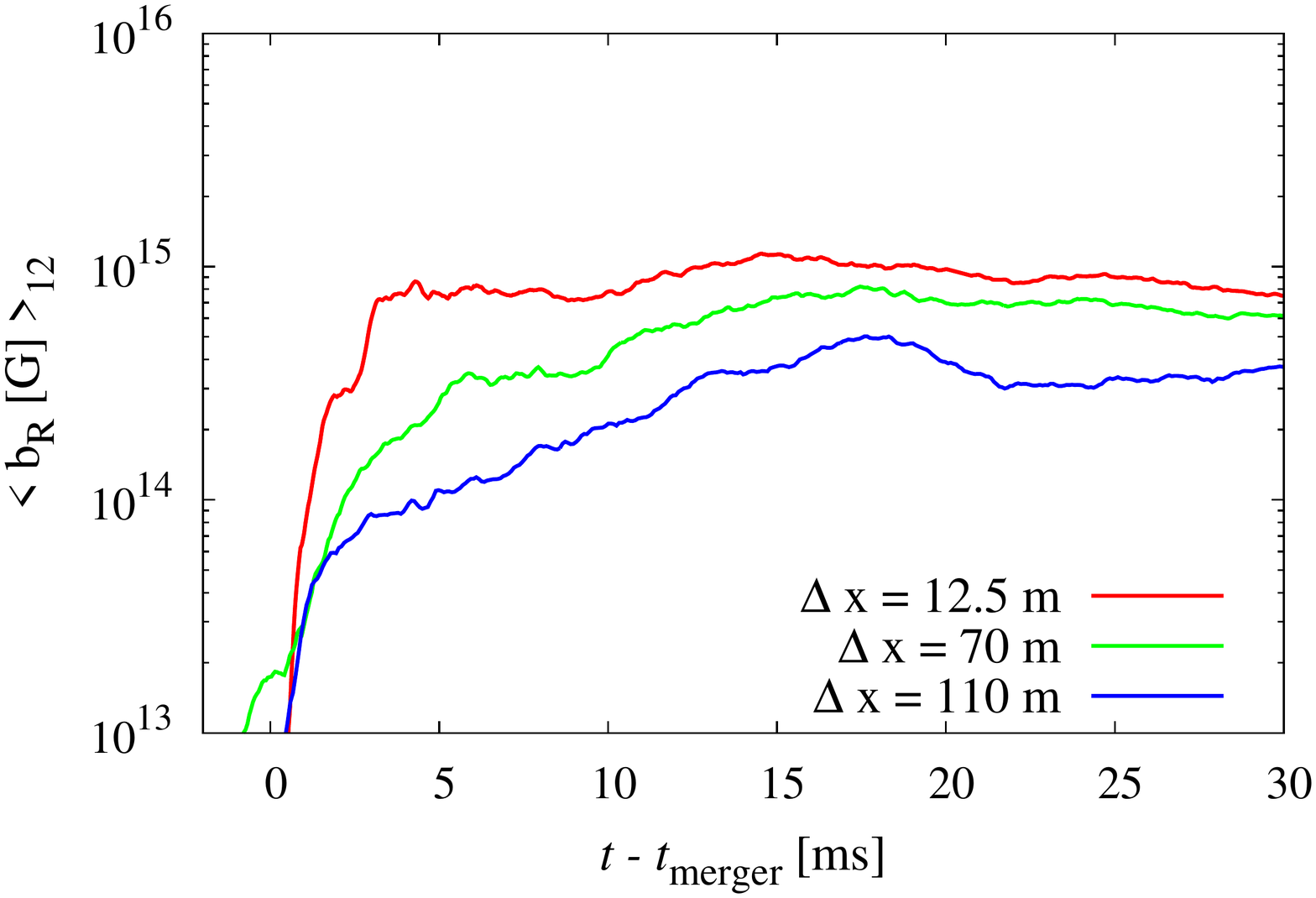}
\end{center}
\end{minipage}
\hspace{20mm}
\begin{minipage}{0.27\hsize}
\begin{center}
\includegraphics[width=7.0cm,angle=0]{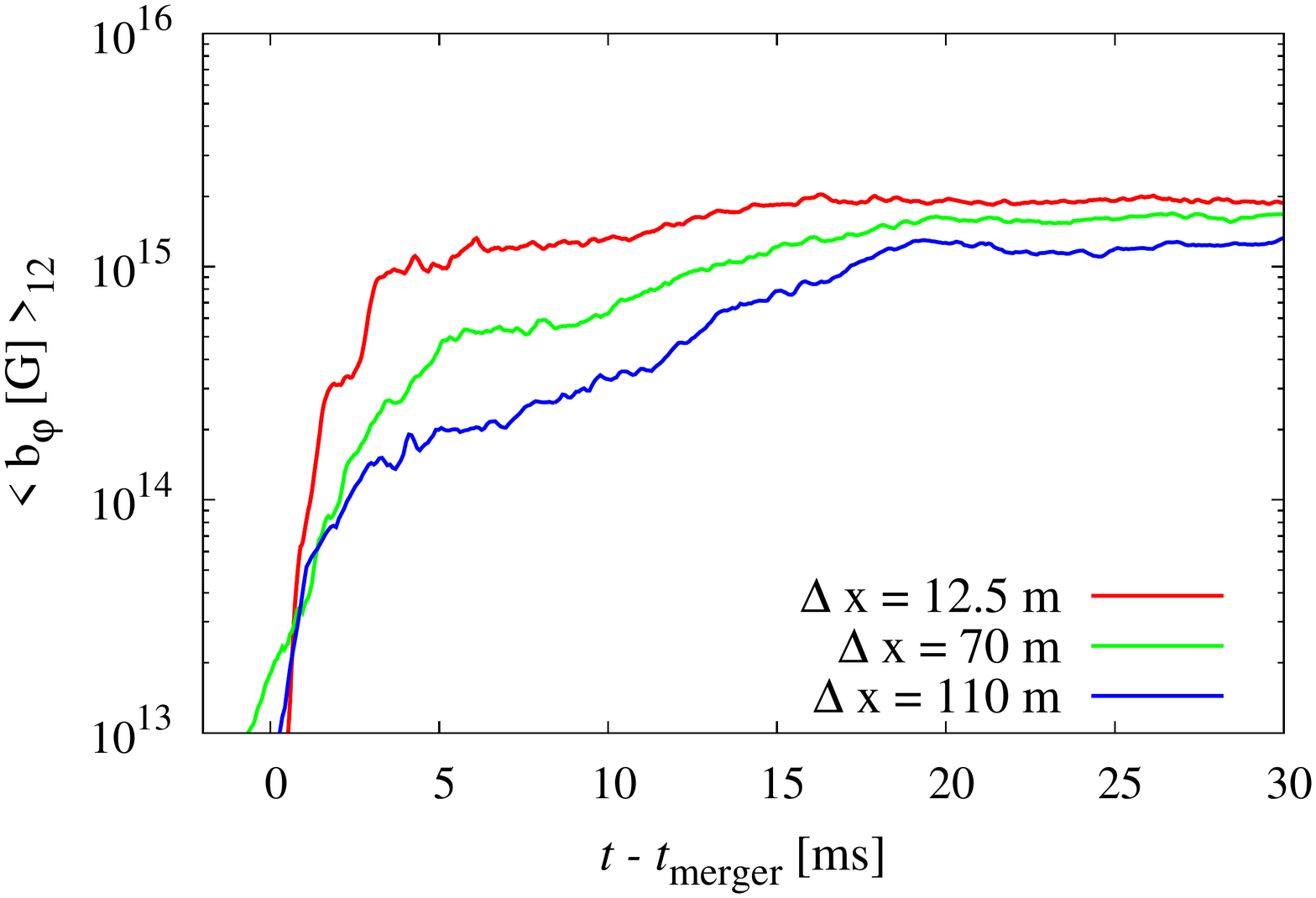}
\end{center}
\end{minipage}\\
\vspace{-10mm}
\hspace{-50mm}
\begin{minipage}{0.27\hsize}
\begin{center}
\includegraphics[width=7.0cm,angle=0]{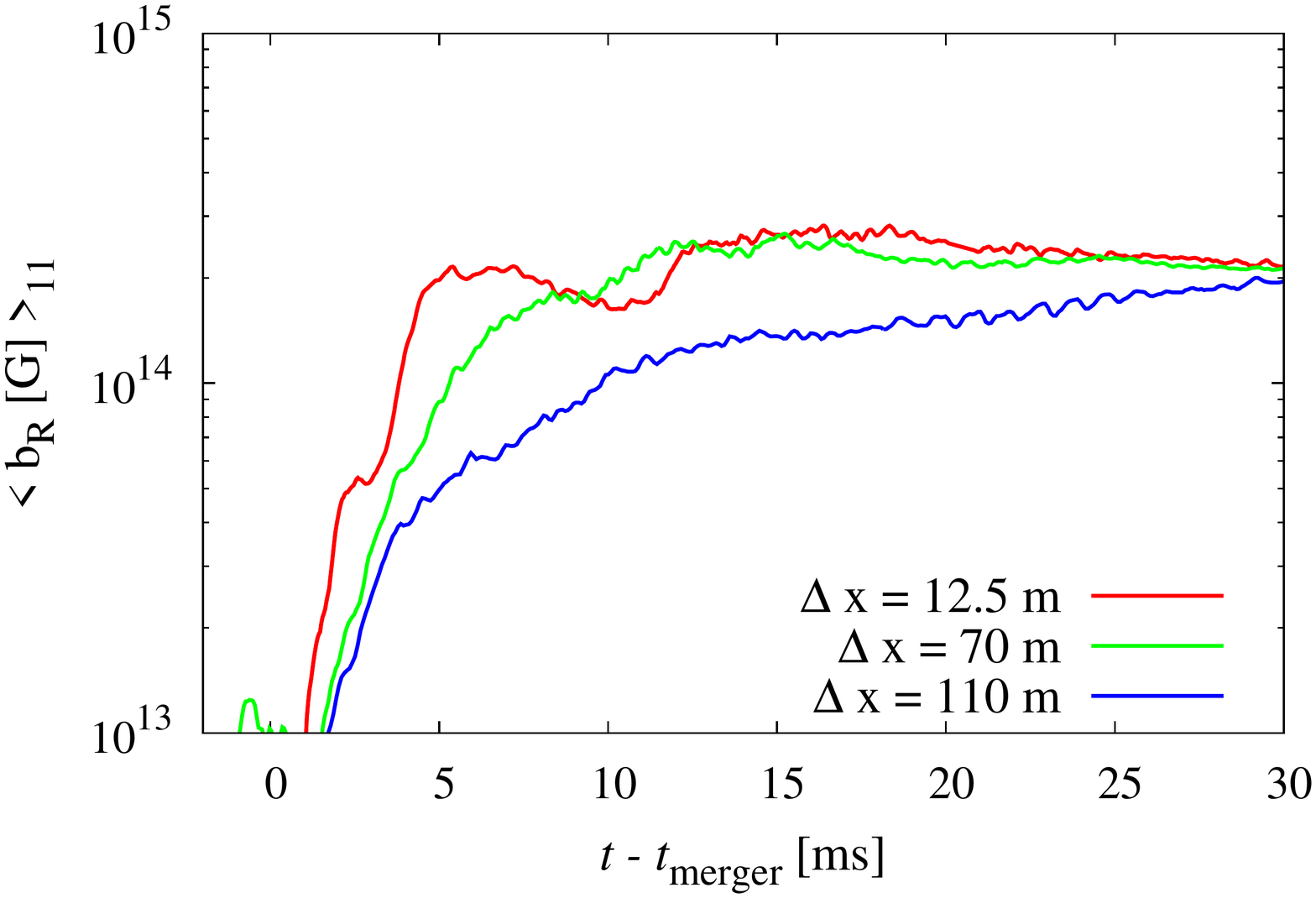}
\end{center}
\end{minipage}
\hspace{20mm}
\begin{minipage}{0.27\hsize}
\begin{center}
\includegraphics[width=7.0cm,angle=0]{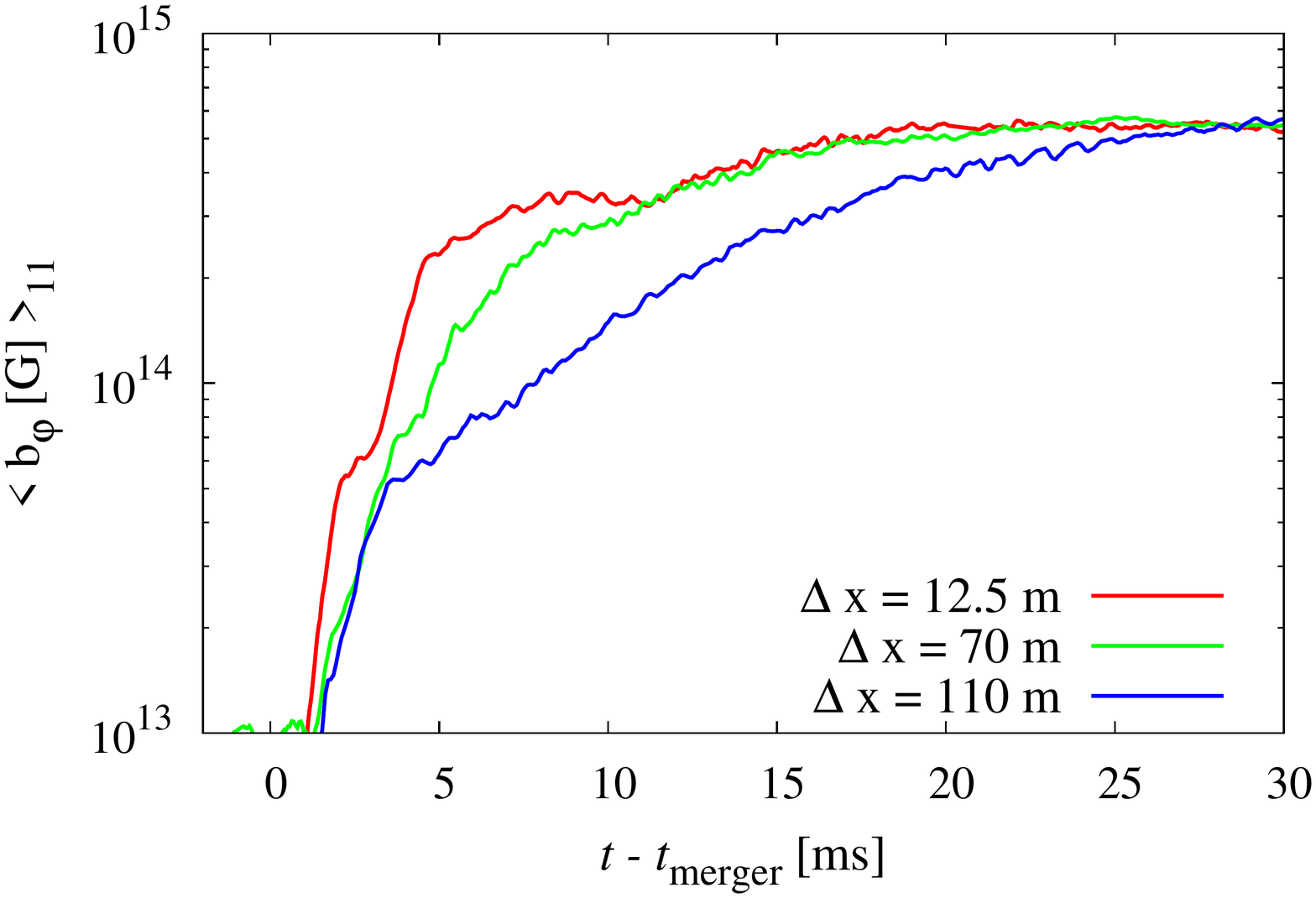}
\end{center}
\end{minipage}\\
\vspace{-10mm}
\hspace{-50mm}
\begin{minipage}{0.27\hsize}
\begin{center}
\includegraphics[width=7.0cm,angle=0]{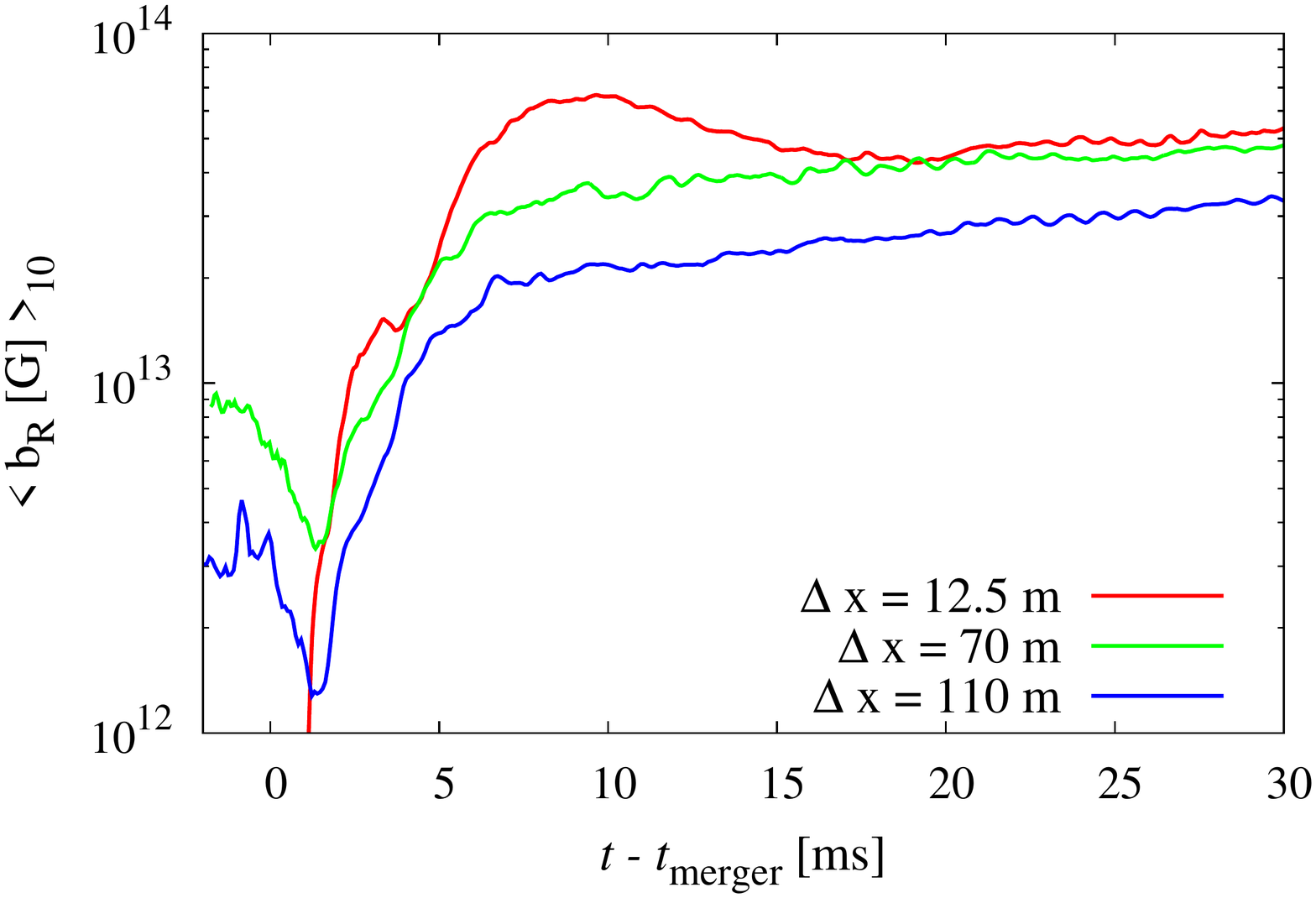}
\end{center}
\end{minipage}
\hspace{20mm}
\begin{minipage}{0.27\hsize}
\begin{center}
\includegraphics[width=7.0cm,angle=0]{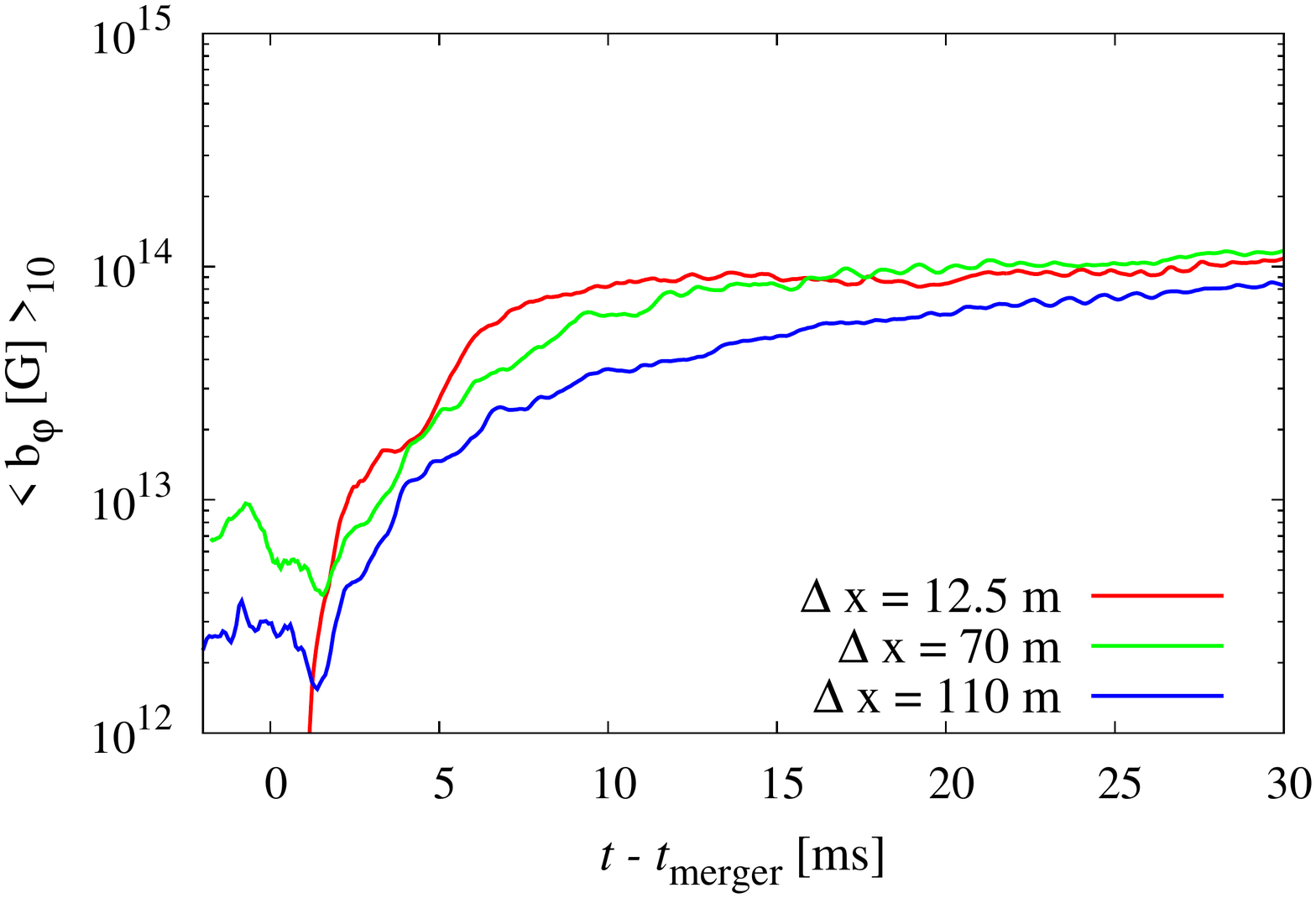}
\end{center}
\end{minipage}\\
\caption{\label{fig5}
Volume-averaged radial and azimuthal components of the magnetic field as functions of time. 
$\langle\cdot\rangle_a$ indicates a volume-average in a density range for $a \le\log_{10}[\rho~({\rm g~cm^{-3}})] < a+1$ with $a=10$--$14$. 
}
\end{figure*}

\begin{figure*}[t]
\hspace{-10mm}
\begin{minipage}{0.27\hsize}
\begin{center}
\includegraphics[width=6.3cm,angle=0]{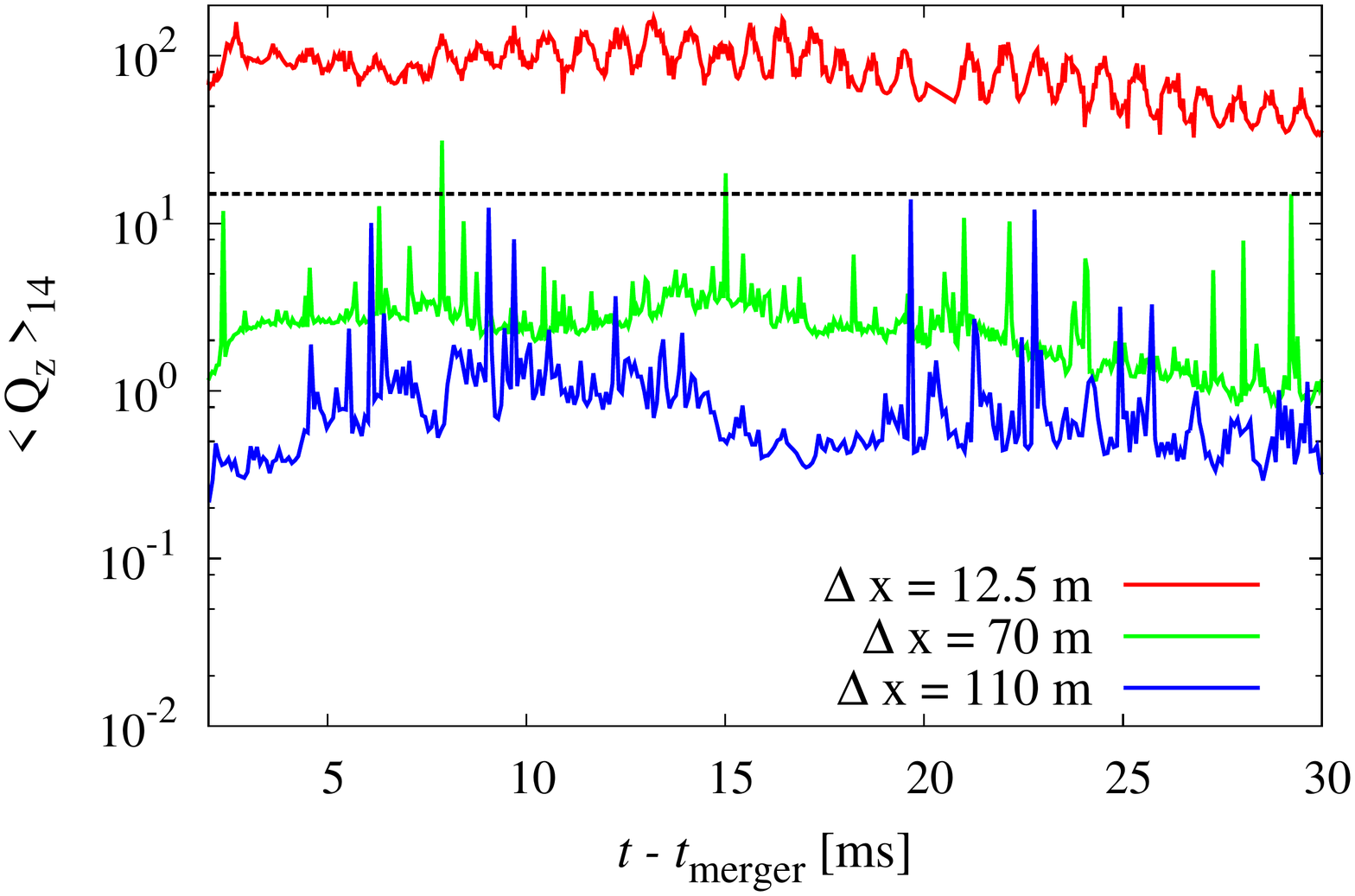}
\end{center}
\end{minipage}
\hspace{9mm}
\begin{minipage}{0.27\hsize}
\begin{center}
\includegraphics[width=6.3cm,angle=0]{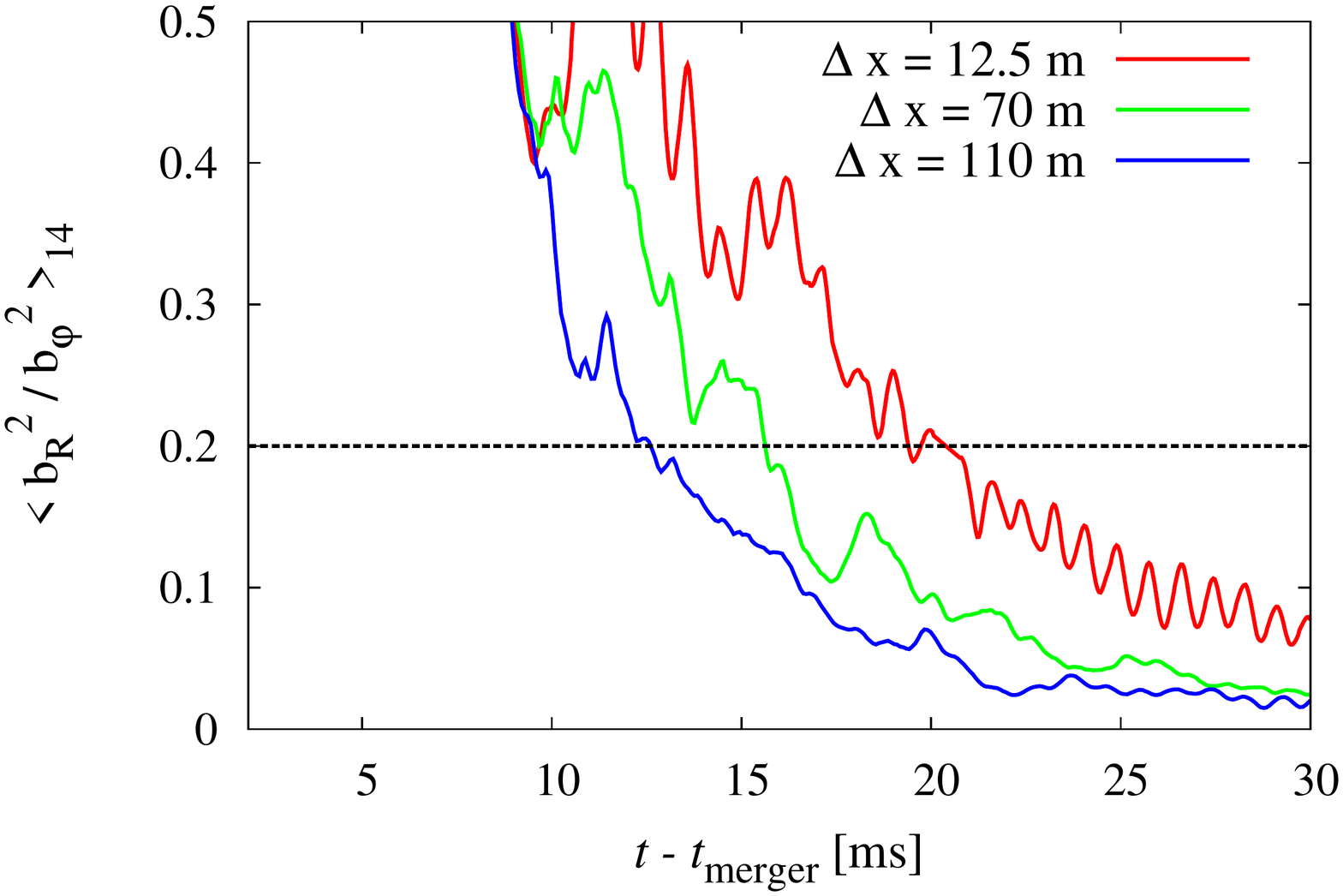}
\end{center}
\end{minipage}
\hspace{9mm}
\begin{minipage}{0.27\hsize}
\begin{center}
\includegraphics[width=6.3cm,angle=0]{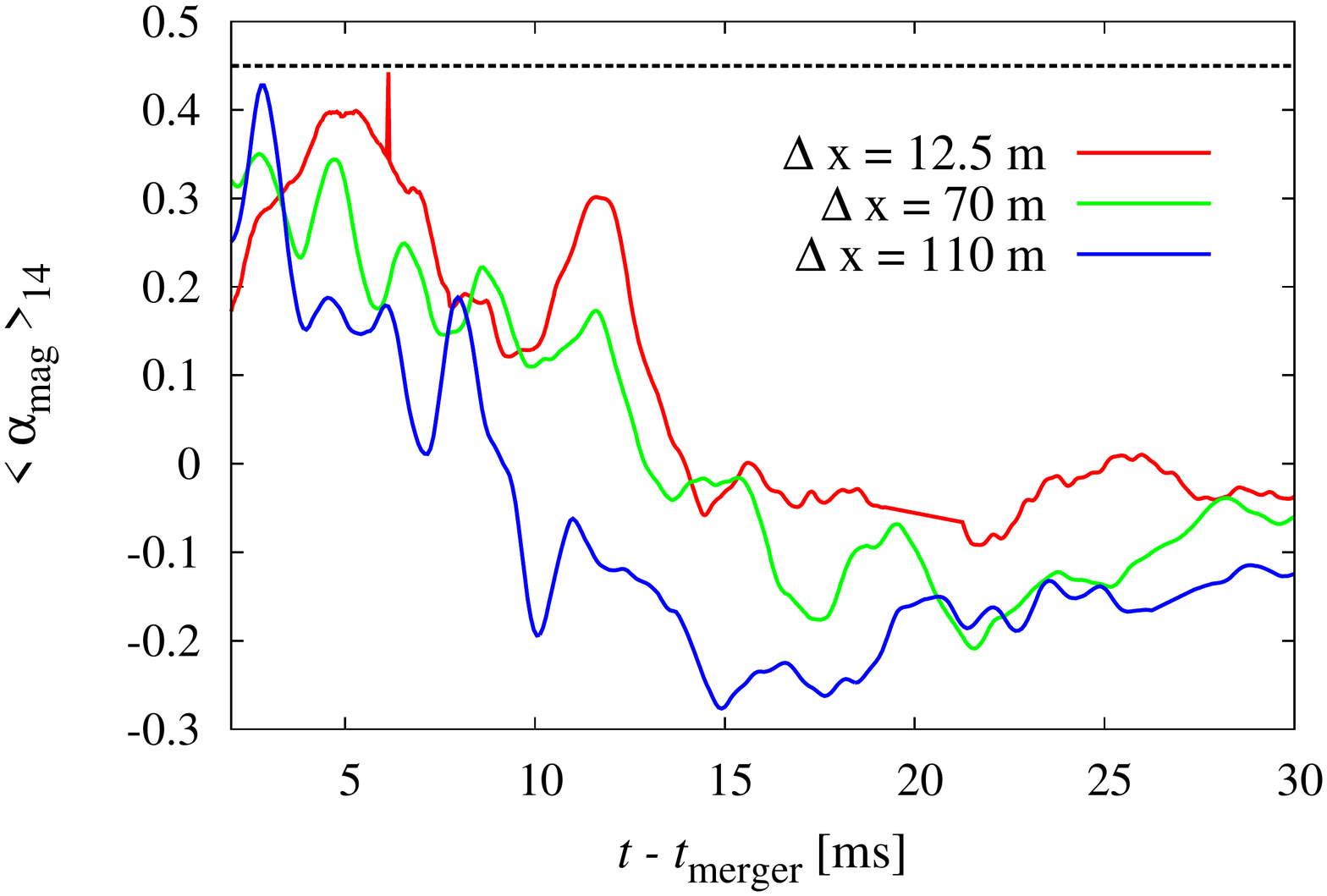}
\end{center}
\end{minipage}\\
\vspace{-10mm}
\hspace{-10mm}
\begin{minipage}{0.27\hsize}
\begin{center}
\includegraphics[width=6.3cm,angle=0]{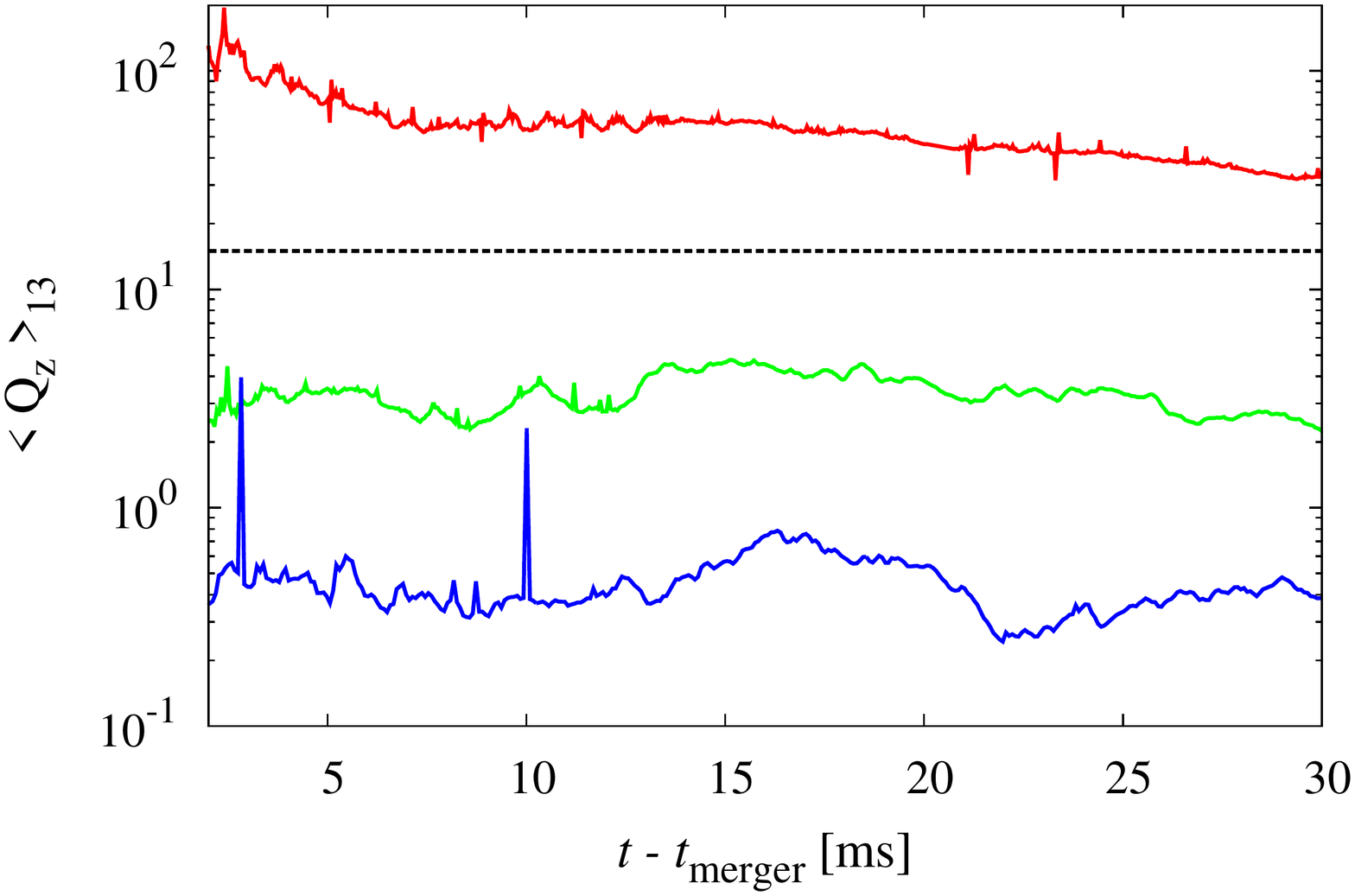}
\end{center}
\end{minipage}
\hspace{9mm}
\begin{minipage}{0.27\hsize}
\begin{center}
\includegraphics[width=6.3cm,angle=0]{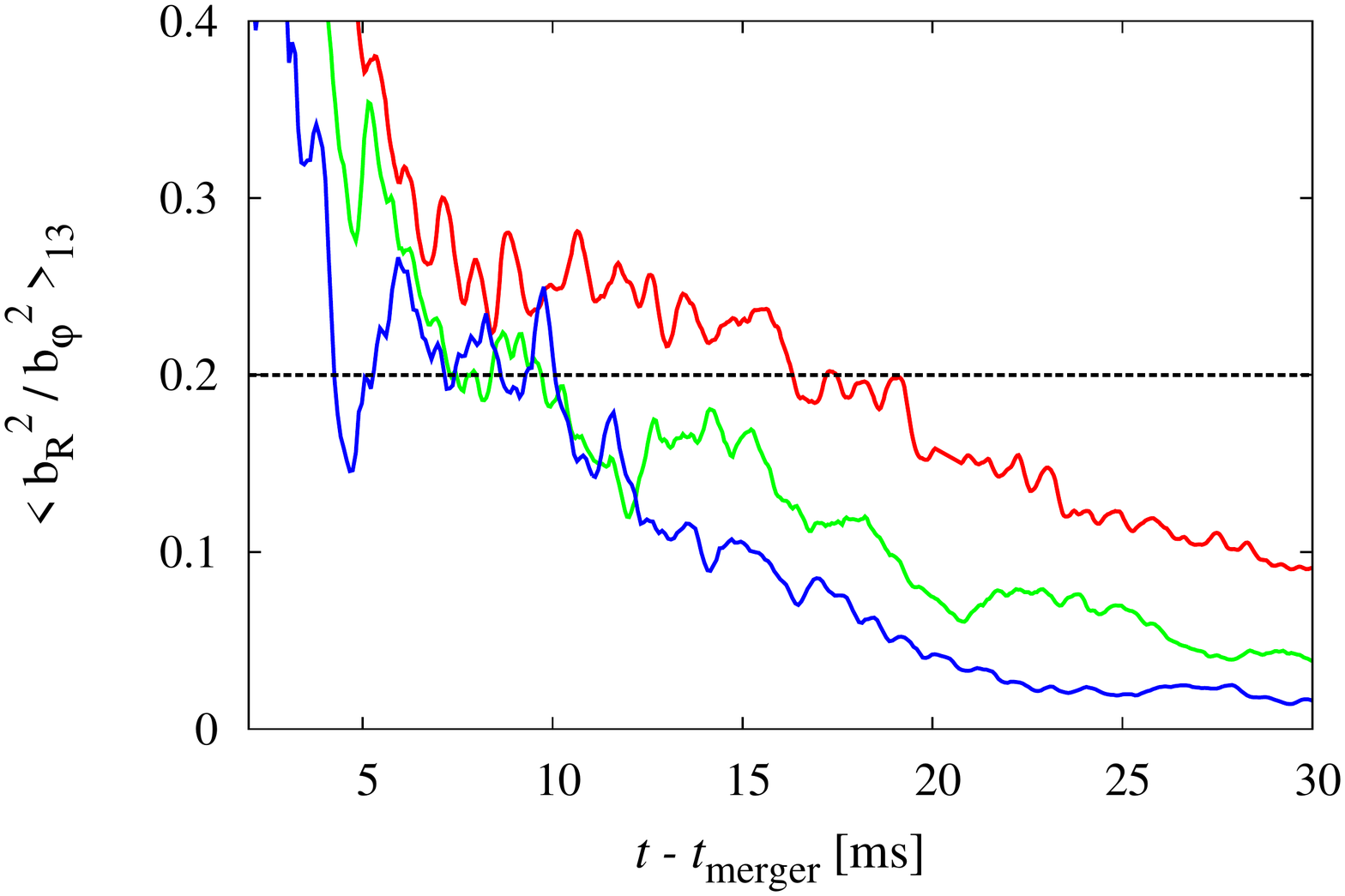}
\end{center}
\end{minipage}
\hspace{9mm}
\begin{minipage}{0.27\hsize}
\begin{center}
\includegraphics[width=6.3cm,angle=0]{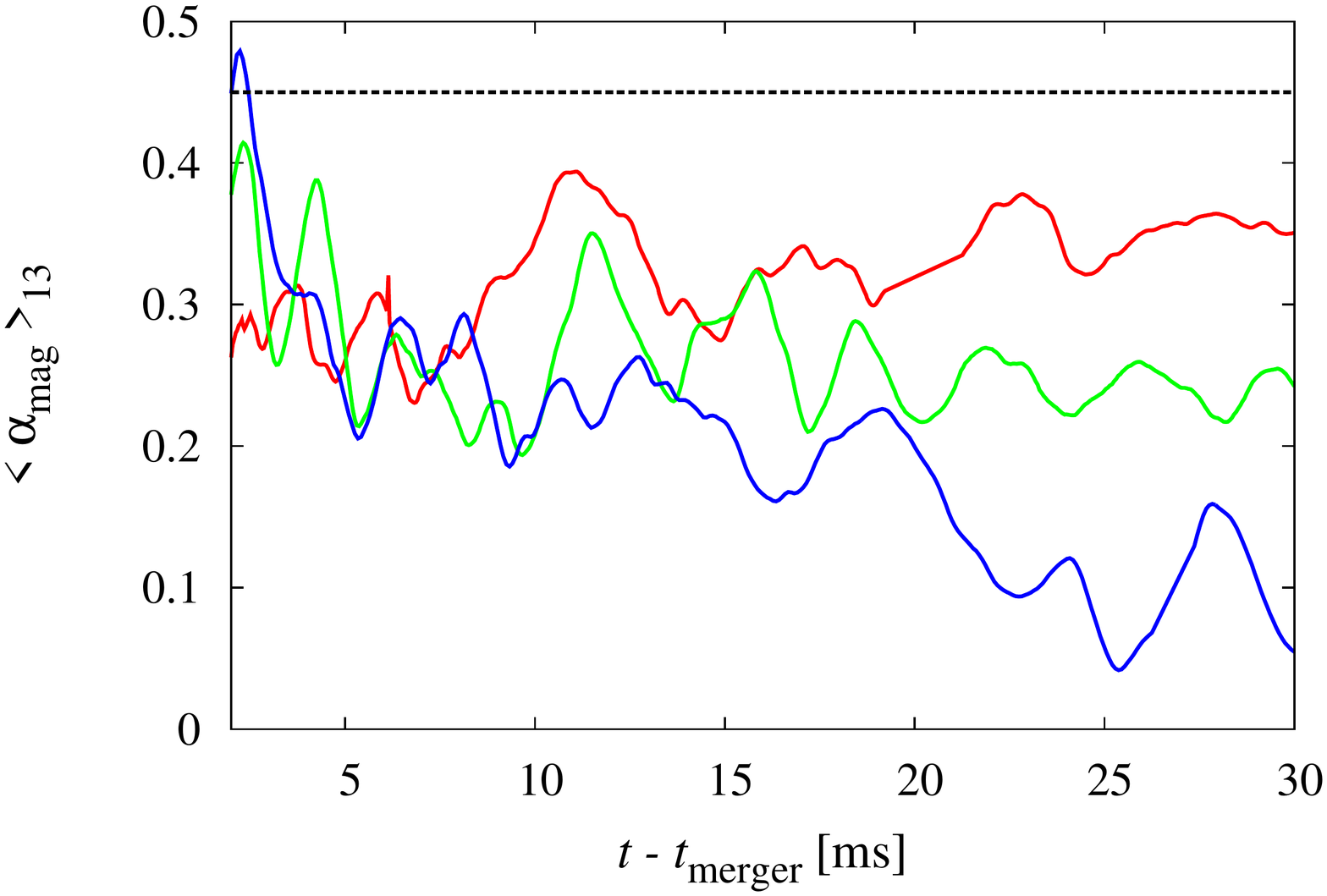}
\end{center}
\end{minipage}\\
\vspace{-10mm}
\hspace{-10mm}
\begin{minipage}{0.27\hsize}
\begin{center}
\includegraphics[width=6.3cm,angle=0]{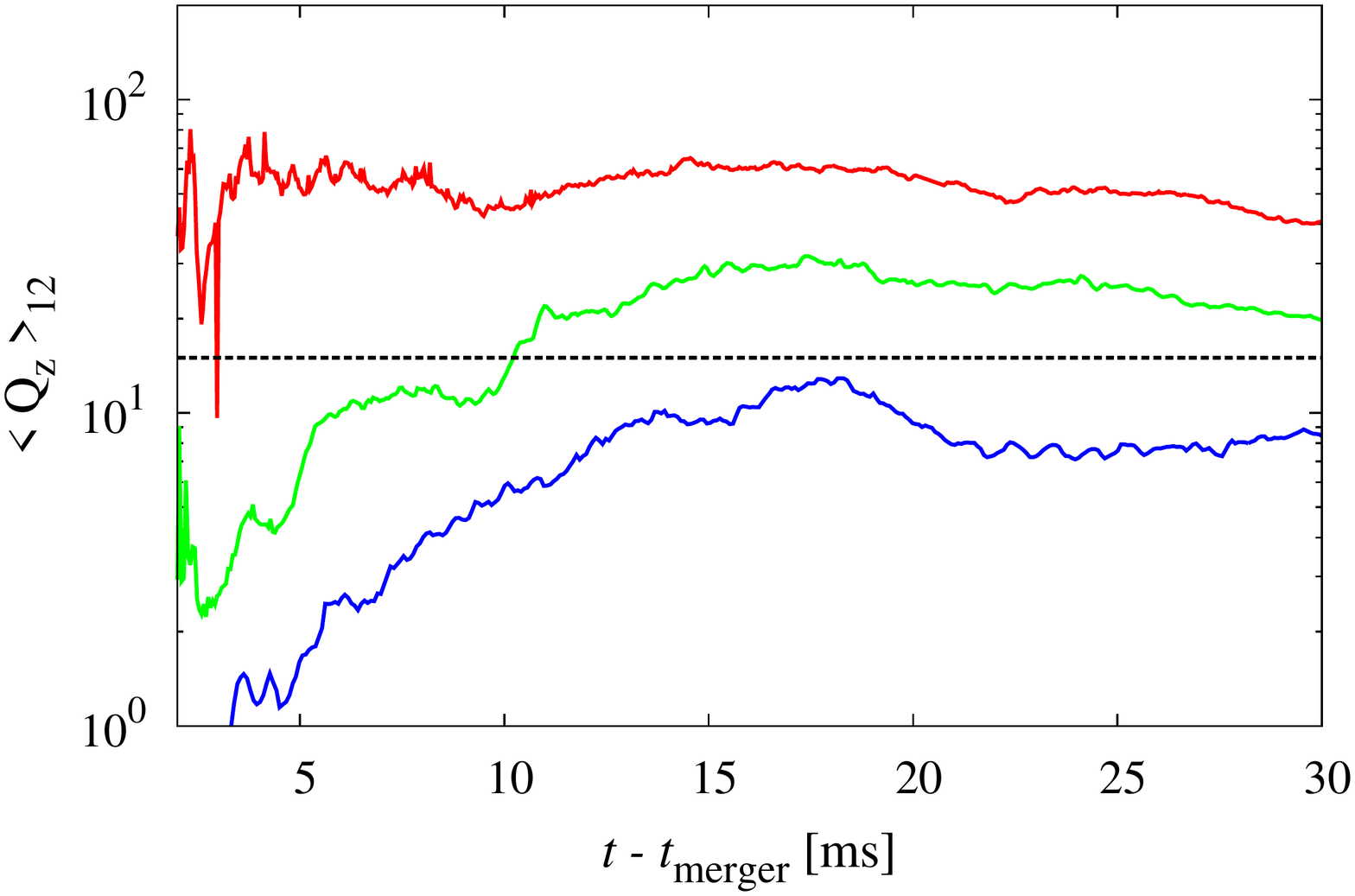}
\end{center}
\end{minipage}
\hspace{9mm}
\begin{minipage}{0.27\hsize}
\begin{center}
\includegraphics[width=6.3cm,angle=0]{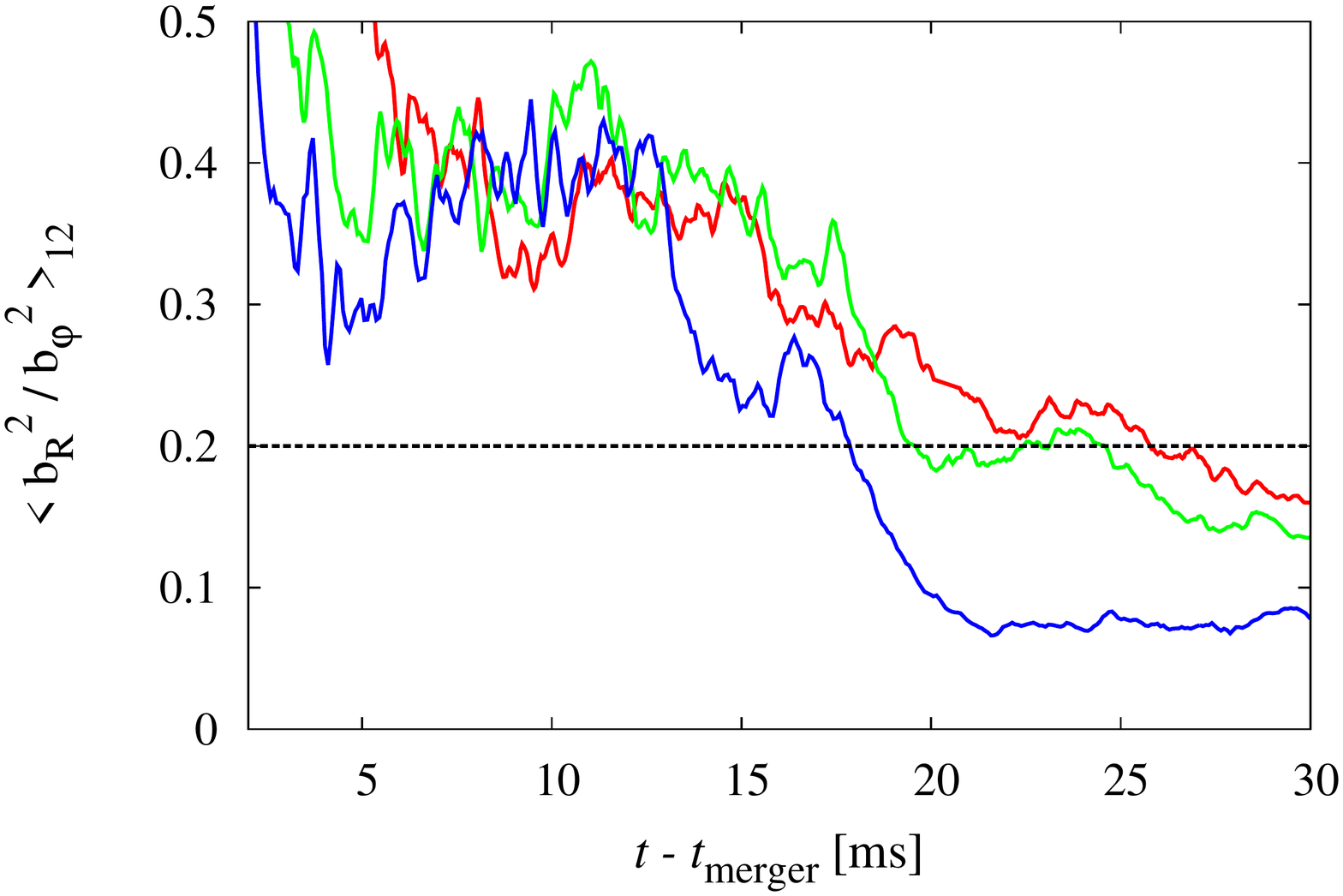}
\end{center}
\end{minipage}
\hspace{9mm}
\begin{minipage}{0.27\hsize}
\begin{center}
\includegraphics[width=6.3cm,angle=0]{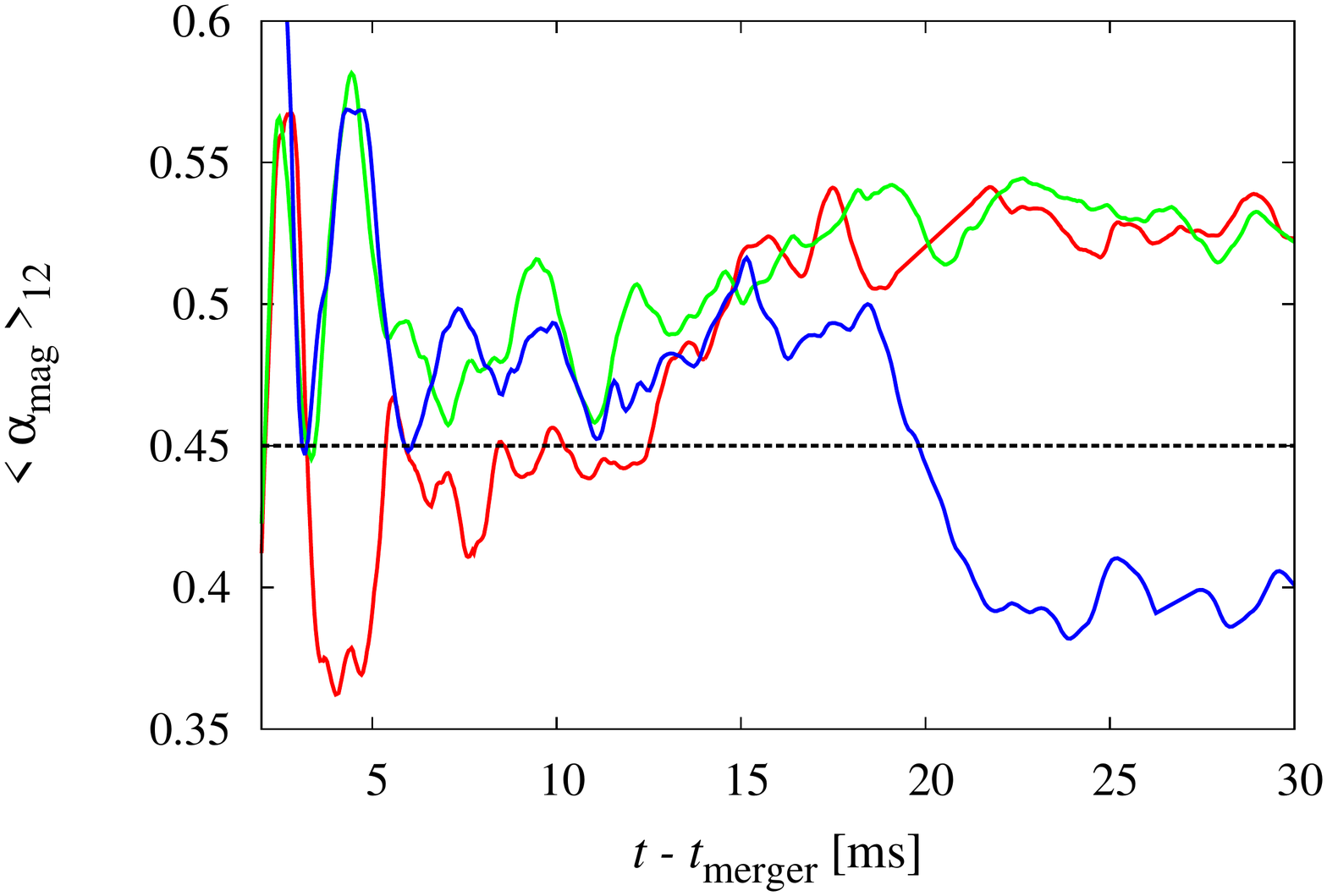}
\end{center}
\end{minipage}\\
\vspace{-10mm}
\hspace{-10mm}
\begin{minipage}{0.27\hsize}
\begin{center}
\includegraphics[width=6.3cm,angle=0]{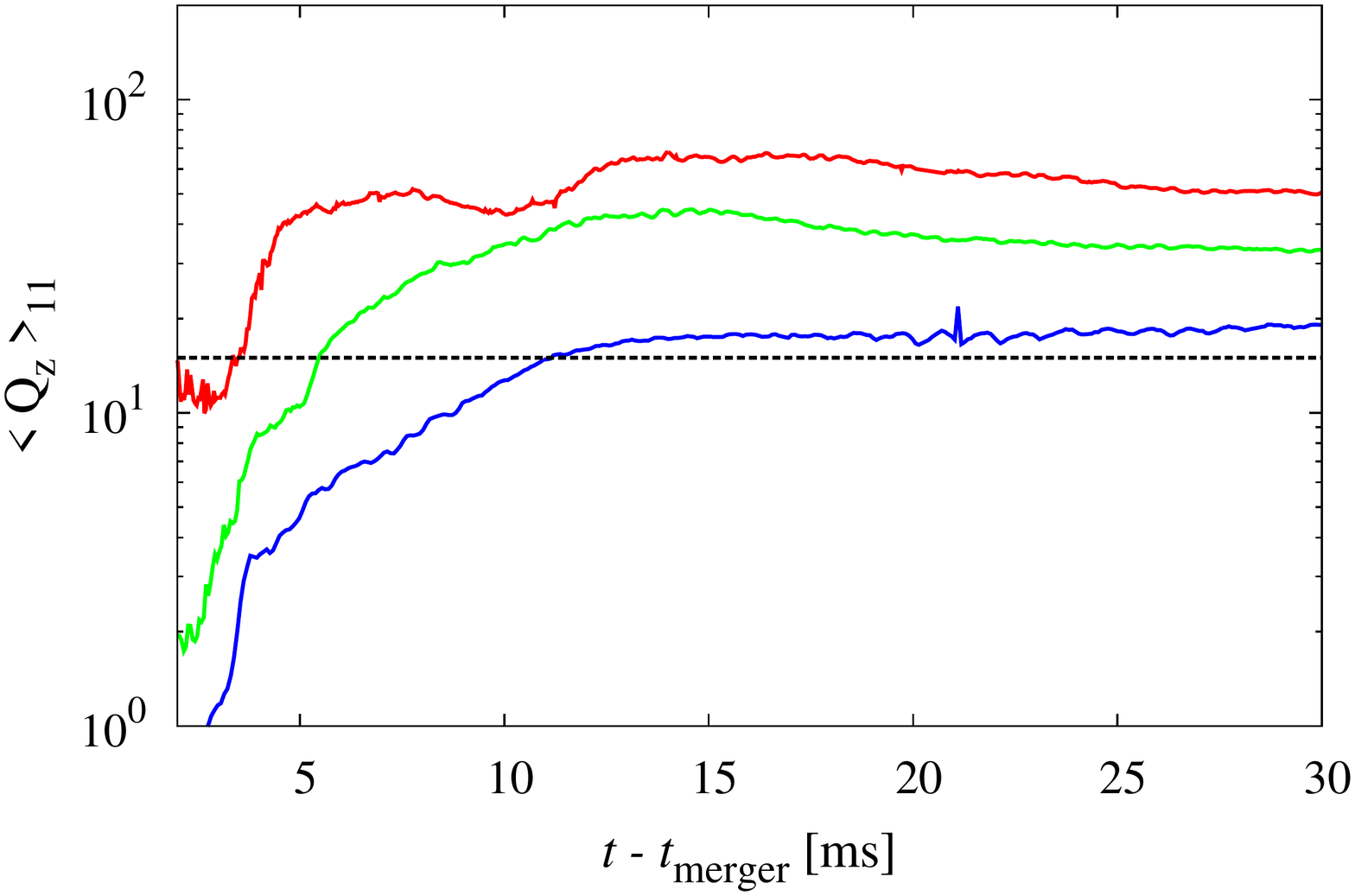}
\end{center}
\end{minipage}
\hspace{9mm}
\begin{minipage}{0.27\hsize}
\begin{center}
\includegraphics[width=6.3cm,angle=0]{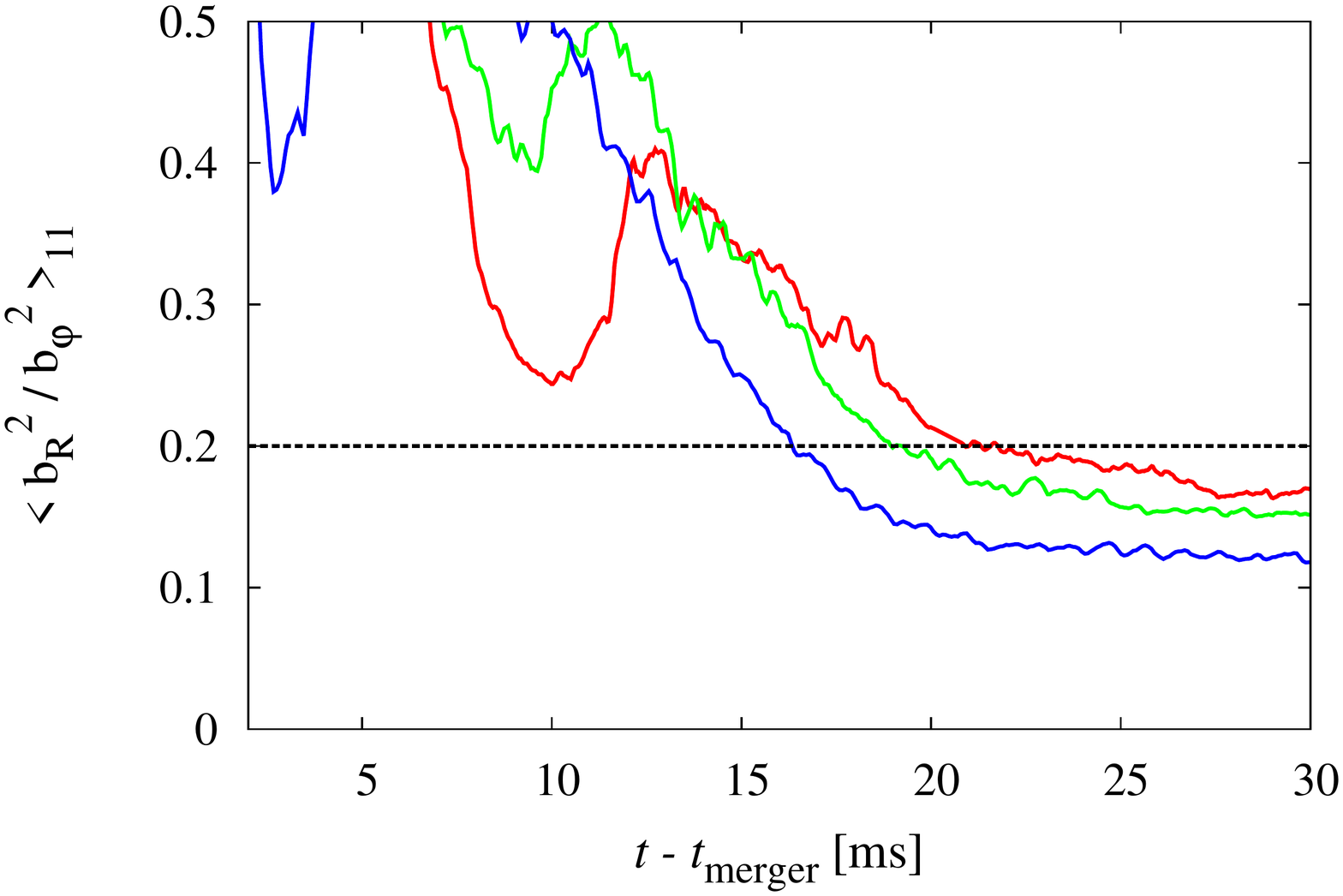}
\end{center}
\end{minipage}
\hspace{9mm}
\begin{minipage}{0.27\hsize}
\begin{center}
\includegraphics[width=6.3cm,angle=0]{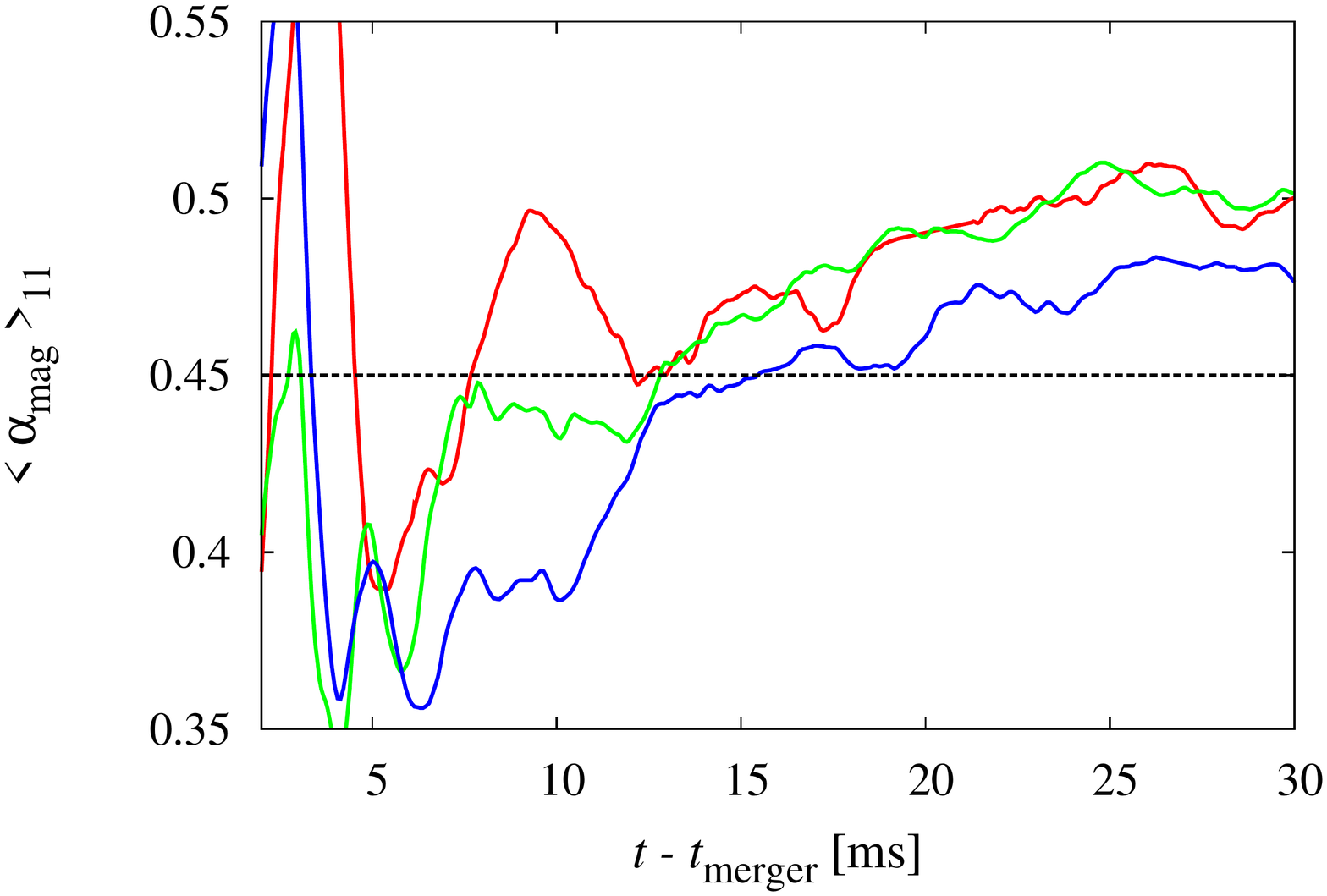}
\end{center}
\end{minipage}\\
\vspace{-10mm}
\hspace{-10mm}
\begin{minipage}{0.27\hsize}
\begin{center}
\includegraphics[width=6.3cm,angle=0]{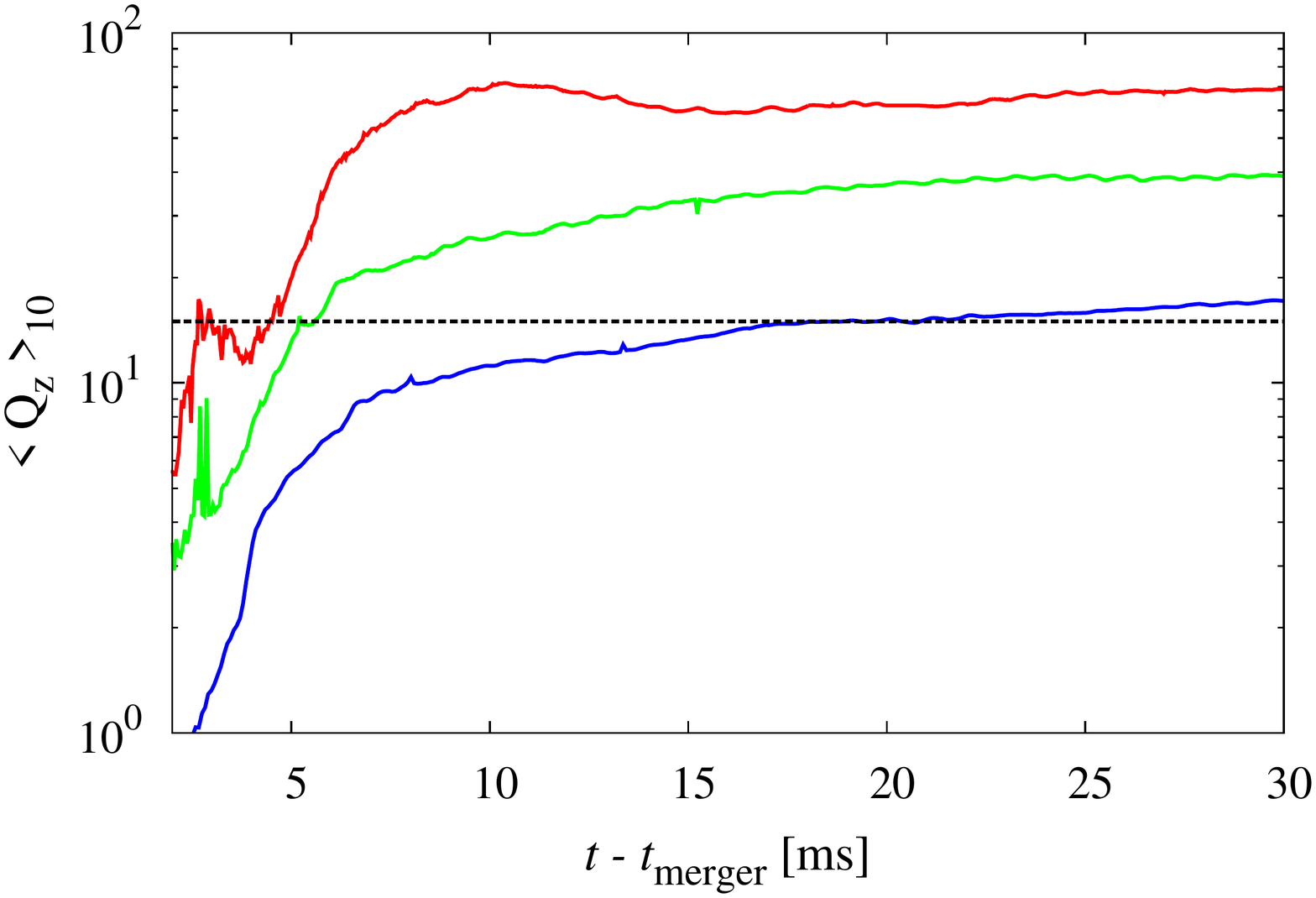}
\end{center}
\end{minipage}
\hspace{9mm}
\begin{minipage}{0.27\hsize}
\begin{center}
\includegraphics[width=6.3cm,angle=0]{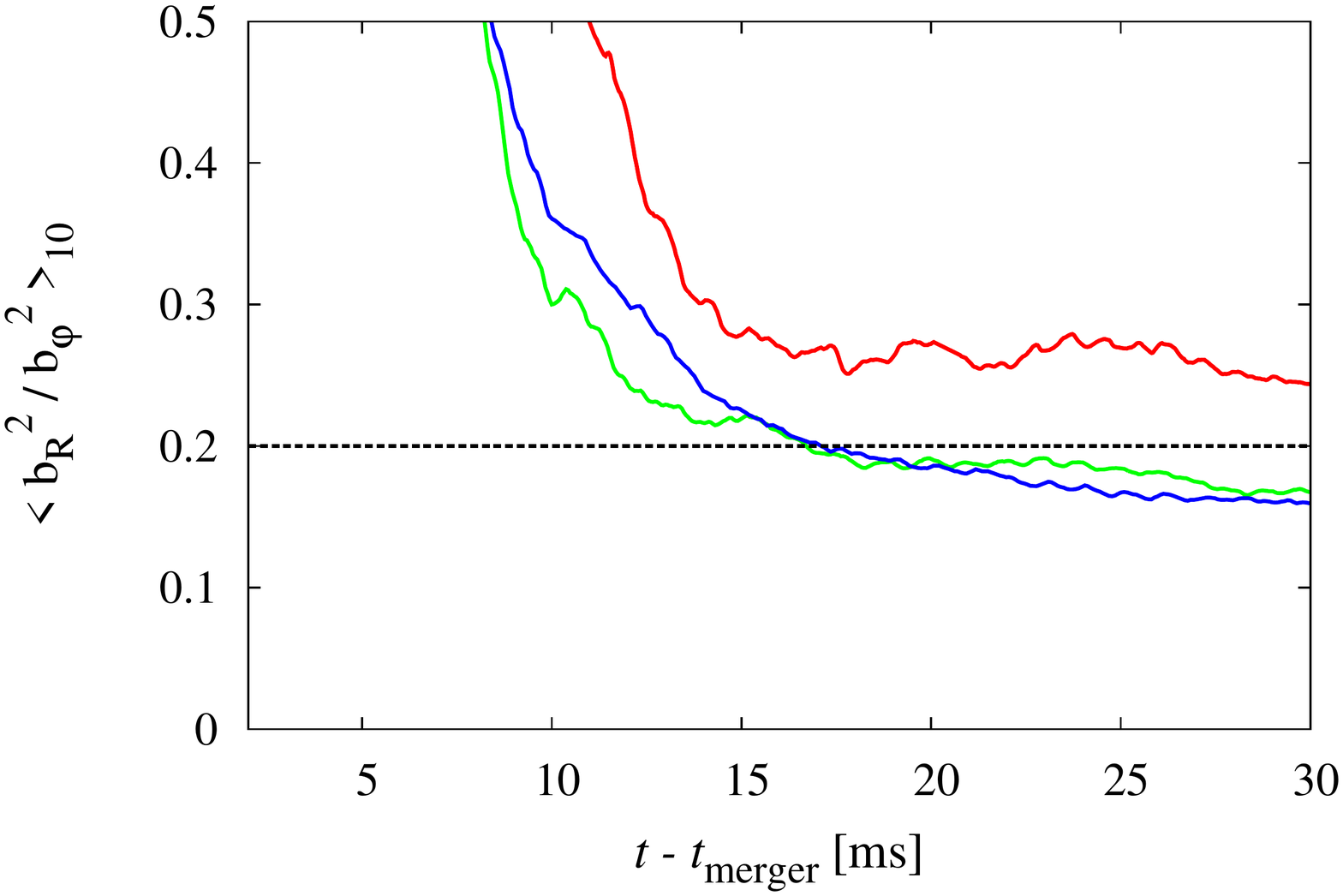}
\end{center}
\end{minipage}
\hspace{9mm}
\begin{minipage}{0.27\hsize}
\begin{center}
\includegraphics[width=6.3cm,angle=0]{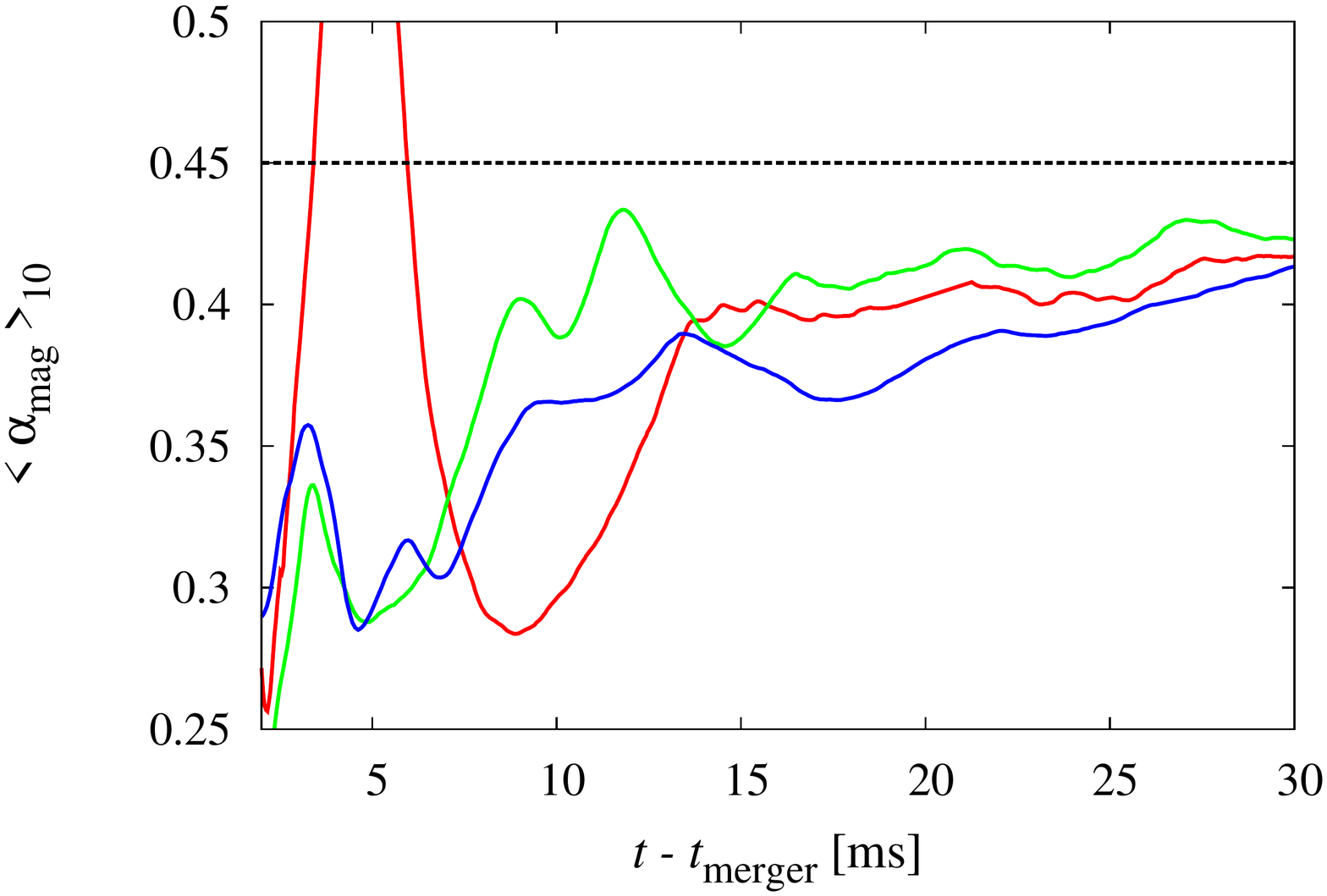}
\end{center}
\end{minipage}\\
\caption{\label{fig6}
Volume-averaged convergence metrics $Q_z$, ${\cal R}=b_R^2/b_\varphi^2$, and $\alpha_{\rm mag}$ as functions of time. 
$\langle\cdot\rangle_a$ indicates a volume-average in a 
density range with $a \le\log_{10}[\rho~({\rm g~cm^{-3}})] < a+1$ with $a=10$--$14$. 
The black-dashed horizontal lines are the criterion above which the MRI-driven turbulence 
is sustained~\cite{Hawley:2011tq,Hawley:2013lga}. 
}
\end{figure*}

\begin{table*}
\centering
\caption{\label{tab1}
Time- and volume-averaged convergence metrics $Q_z$, $Q_\varphi$, ${\cal R}$, and $\alpha_{\rm mag}$. 
The time average is carried out for $15~{\rm ms} \le t - t_{\rm merger} \le 30~{\rm ms}$. We show the result for 
three different grid resolutions. The criterion for the sustainability of the MRI-driven turbulence is given by
$\langle Q_z \rangle \gtrsim 15$, $\langle Q_\varphi \rangle \gtrsim 20$, $\langle {\cal R} \rangle \gtrsim 0.2$, and $\langle \alpha_{\rm mag} \rangle \gtrsim 0.45$, respectively.
Note that in the low-density ranges with $a=11$ and 10 the grid resolution for the highest resolution run is improved by only a factor of $1.4$ and $2.2$
compared to the middle and low resolution runs, respectively because of our choice of the grid structure in the FMR algorithm. This results in a moderate improvement of $\langle Q_z \rangle$ in these density ranges compared
to the high-density ranges with $a=12-$14. 
}
\begin{tabular}{ccccccc}
\hline\hline
$\Delta x_{l_{\rm max}}$ [m] & $\langle\langle Q_z \rangle\rangle_{14}$ & $\langle\langle Q_z \rangle\rangle_{13}$ & $\langle\langle Q_z \rangle\rangle_{12}$ & $\langle\langle Q_z \rangle\rangle_{11}$ & $\langle\langle Q_z \rangle\rangle_{10}$ \\
\hline
12.5 & 72.3 & 44.5 & 52.3 & 57.5 & 64.4 \\
70   & 2.2  & 3.4  & 25.1 & 36.2 & 37.1 \\
110  & 0.8  & 0.5  & 8.9  & 17.9 & 15.4 \\
\hline
$\Delta x_{l_{\rm max}}$ [m] & $\langle\langle Q_\varphi \rangle\rangle_{14}$ & $\langle\langle Q_\varphi \rangle\rangle_{13}$ & $\langle\langle Q_\varphi \rangle\rangle_{12}$ & $\langle\langle Q_\varphi \rangle\rangle_{11}$ & $\langle\langle Q_\varphi \rangle\rangle_{10}$ \\
\hline
12.5 & 751.1 & 668.0 & 594.1 & 716.8 & 560.0 \\
70   & 37.6  & 55.4  & 229.0 & 438.2 & 377.0 \\
110  & 20.1  & 21.8  & 106.1 & 237.0 & 141.3 \\
\hline
$\Delta x_{l_{\rm max}}$ [m] & $\langle\langle {\cal R} \rangle\rangle_{14}$ & $\langle\langle {\cal R} \rangle\rangle_{13}$ & $\langle\langle {\cal R} \rangle\rangle_{12}$ & $\langle\langle {\cal R} \rangle\rangle_{11}$ & $\langle\langle {\cal R} \rangle\rangle_{10}$ \\
\hline
12.5 & 0.20 & 0.16 & 0.24 & 0.26 & 0.26 \\
70   & 0.10 & 0.10 & 0.24 & 0.23 & 0.20 \\
110  & 0.06 & 0.05 & 0.13 & 0.18 & 0.19 \\
\hline
$\Delta x_{l_{\rm max}}$ [m] & $\langle\langle \alpha_{\rm mag} \rangle\rangle_{14}$ & $\langle\langle \alpha_{\rm mag} \rangle\rangle_{13}$ & $\langle\langle \alpha_{\rm mag} \rangle\rangle_{12}$ & $\langle\langle \alpha_{\rm mag} \rangle\rangle_{11}$ & $\langle\langle \alpha_{\rm mag} \rangle\rangle_{10}$ \\
\hline
12.5 & -0.03 & 0.34 & 0.52 & 0.49 & 0.40 \\
70   & -0.11 & 0.25 & 0.53 & 0.49 & 0.41 \\
110  & -0.18 & 0.14 & 0.43 & 0.47 & 0.38 \\
\hline
$\Delta x_{l_{\rm max}}$ [m] & $\langle\langle \alpha \rangle\rangle_{14}$ & $\langle\langle \alpha \rangle\rangle_{13}$ & $\langle\langle \alpha \rangle\rangle_{12}$ & $\langle\langle \alpha \rangle\rangle_{11}$ & $\langle\langle \alpha \rangle\rangle_{10}$ \\
\hline
12.5 & 0.0005 & 0.005 & 0.017 & 0.012 & 0.005 \\
70   & 0.0002 & 0.003 & 0.017 & 0.012 & 0.005 \\
110  & 0.0004 & 0.002 & 0.010 & 0.010 & 0.003 \\
\hline\hline
\end{tabular}
\end{table*}

\begin{figure*}[t]
\hspace{-20mm}
\begin{minipage}{0.27\hsize}
\begin{center}
\includegraphics[width=6.8cm,angle=0]{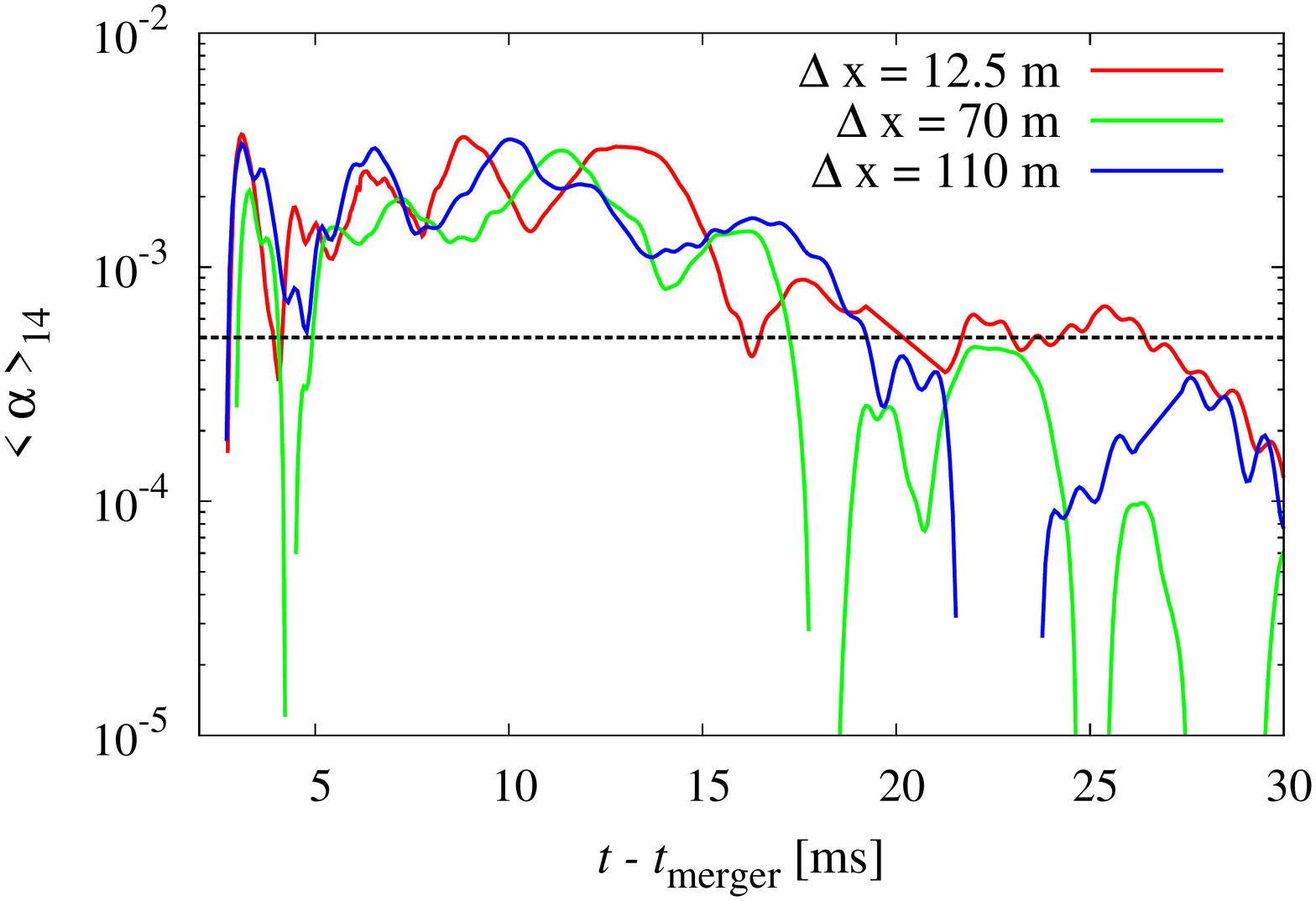}
\end{center}
\end{minipage}
\hspace{9mm}
\begin{minipage}{0.27\hsize}
\begin{center}
\includegraphics[width=6.8cm,angle=0]{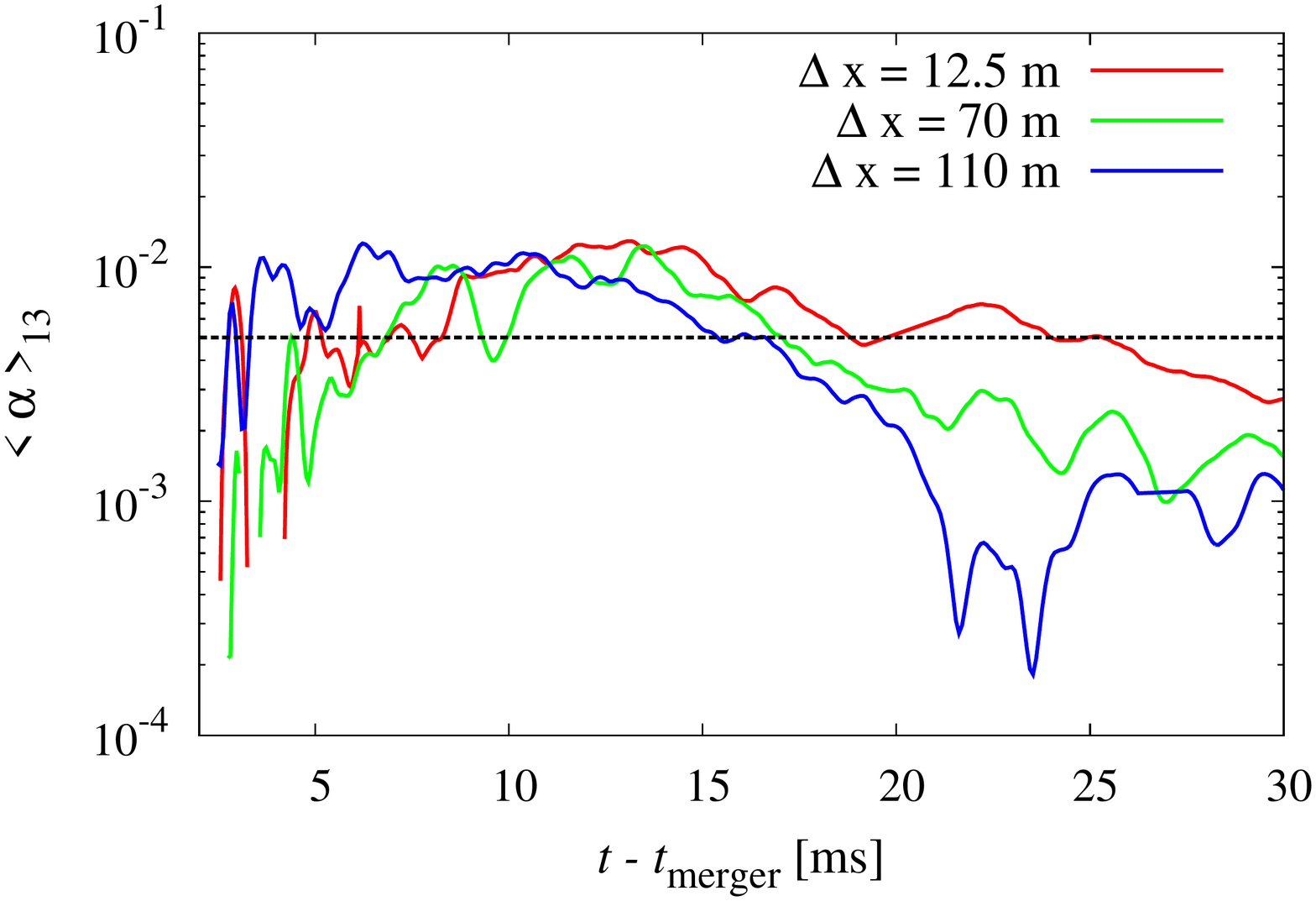}
\end{center}
\end{minipage}
\hspace{9mm}
\begin{minipage}{0.27\hsize}
\begin{center}
\includegraphics[width=6.8cm,angle=0]{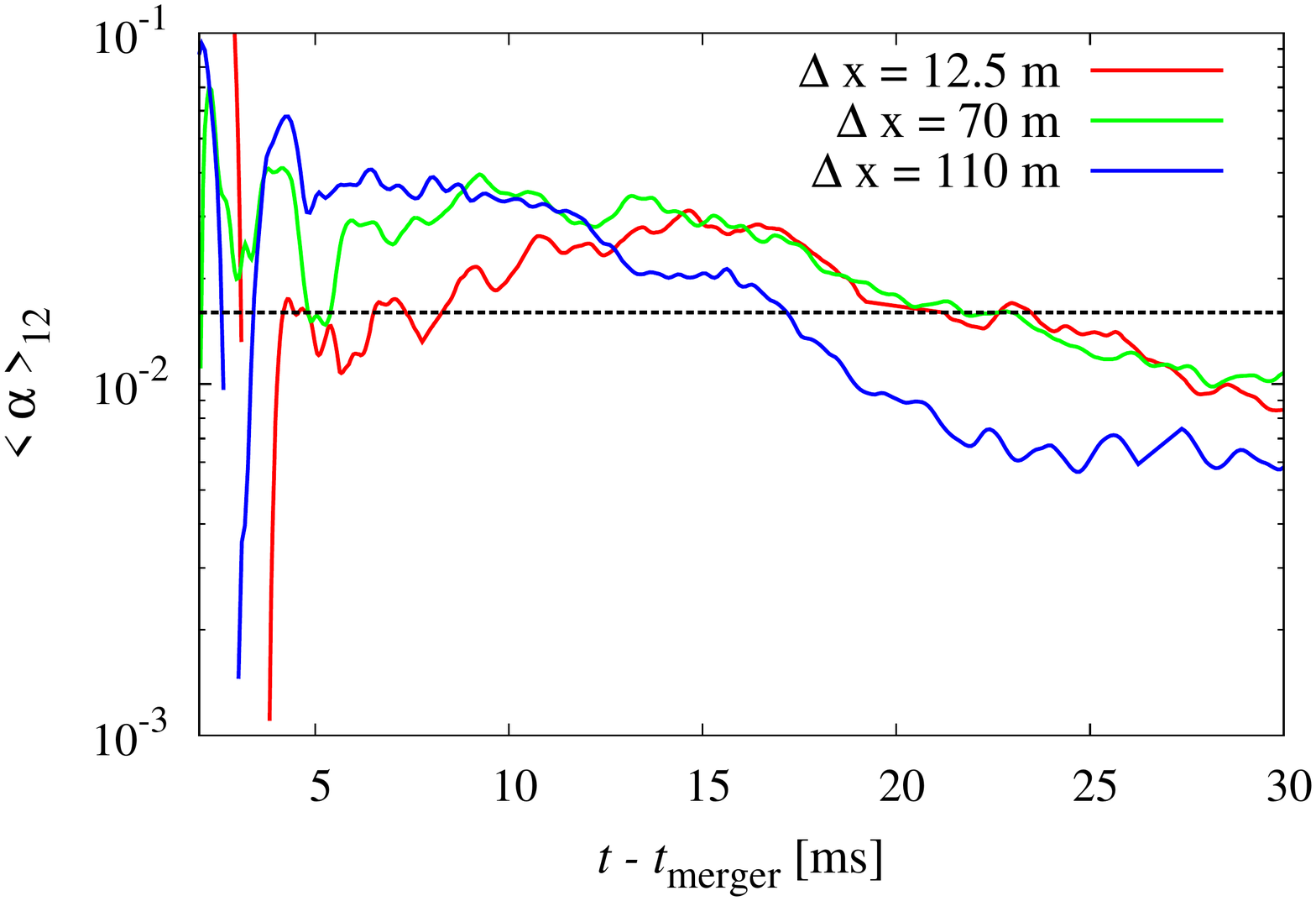}
\end{center}
\end{minipage}\\
\vspace{0mm}
\hspace{-20mm}
\begin{minipage}{0.27\hsize}
\begin{center}
\includegraphics[width=6.8cm,angle=0]{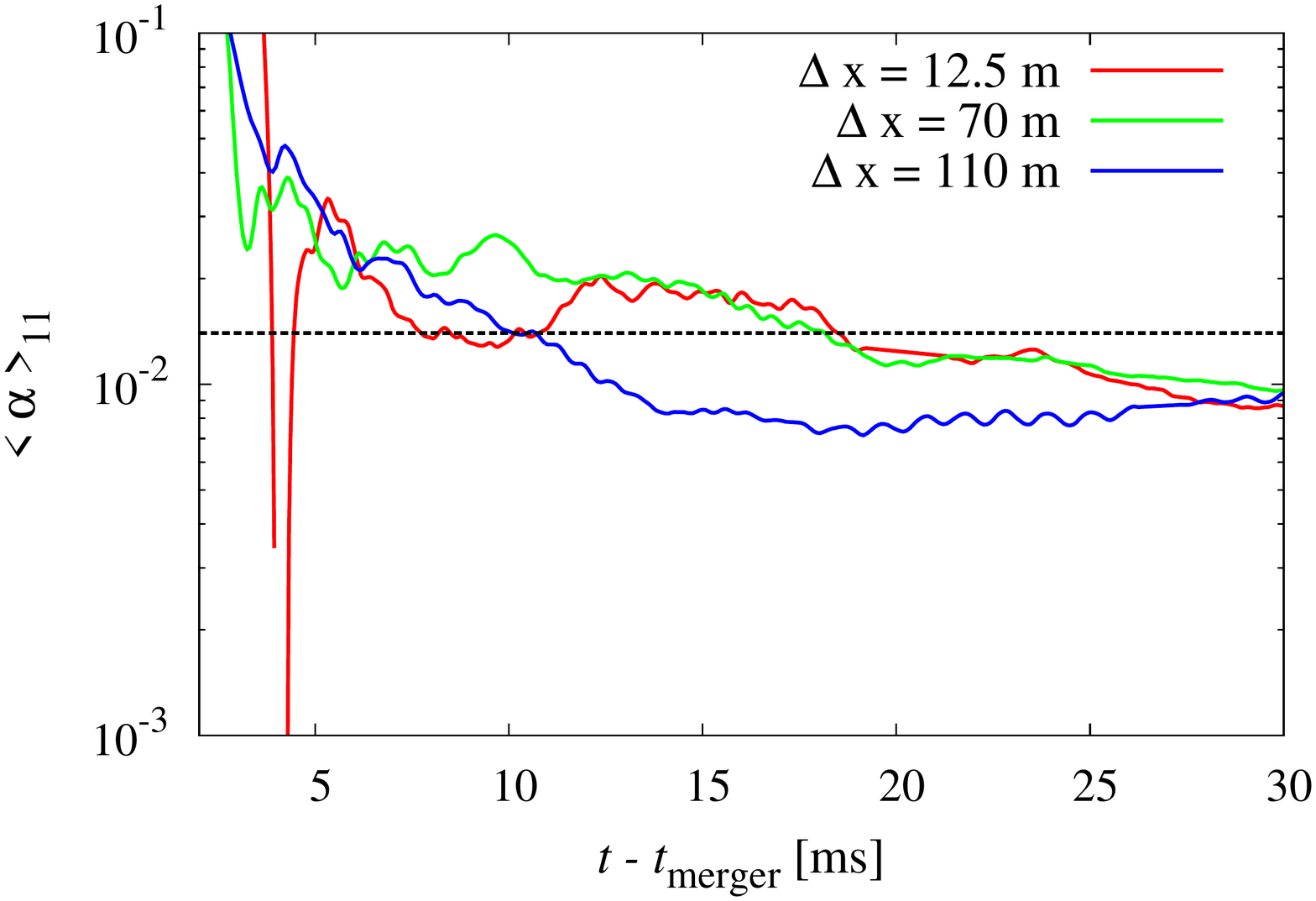}
\end{center}
\end{minipage}
\hspace{9mm}
\begin{minipage}{0.27\hsize}
\begin{center}
\includegraphics[width=6.8cm,angle=0]{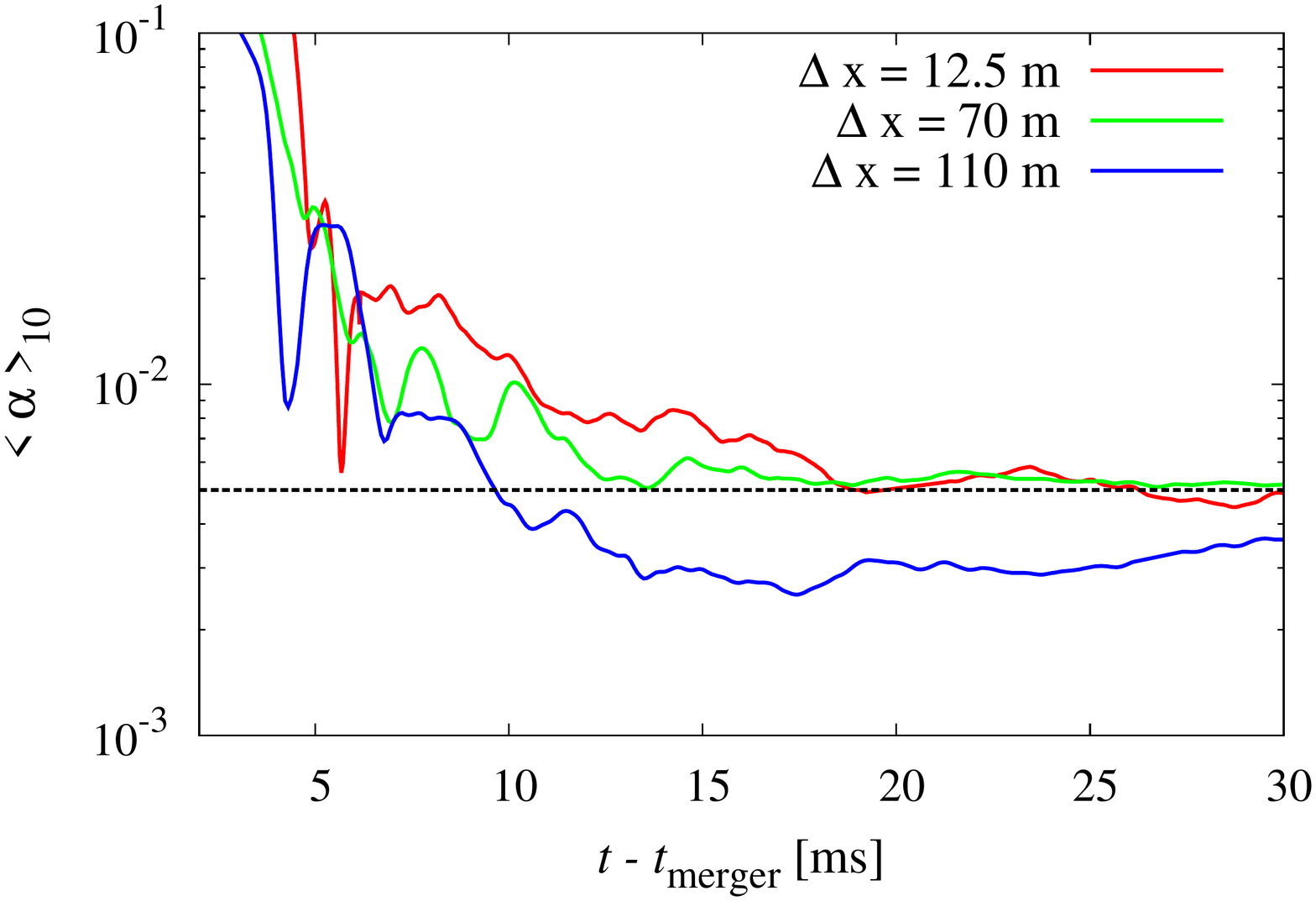}
\end{center}
\end{minipage}
\caption{\label{fig7}
Same as Fig.~\ref{fig6}, but for the $\alpha$ parameter. 
The black-dashed horizontal lines are the time-averaged values for the $12.5$ m run in Table~\ref{tab1}.
}
\end{figure*}

\begin{table*}
\centering
\caption{\label{tab2}Volume- and time-averaged specific angular momentum, the sound speed, 
and viscous timescale for $15$ ms $\le t-t_{\rm merger}\le30$ ms with the highest resolution run. 
The parenthesis of the viscous timescale for $a=13$ and $a=14$ implies that it should be shorter in reality because we underestimate the 
$\alpha$ parameter.}
\begin{tabular}{ccccccc}
\hline\hline
$\langle\langle j \rangle\rangle_{14}~[{\rm cm^2~s^{-1}}]$ & $\langle\langle j \rangle\rangle_{13}~[{\rm cm^2~s^{-1}}]$ & $\langle\langle j \rangle\rangle_{12}~[{\rm cm^2~s^{-1}}]$ & $\langle\langle j \rangle\rangle_{11}~[{\rm cm^2~s^{-1}}]$ & $\langle\langle j \rangle\rangle_{10}~[{\rm cm^2~s^{-1}}]$ \\
\hline
$0.78 \times 10^{16}$ & $2.51 \times 10^{16}$& $3.16 \times 10^{16}$& $3.38 \times 10^{16}$& $3.76 \times 10^{16}$\\
\hline
$\langle\langle c_s \rangle\rangle_{14}$ [c] & $\langle\langle c_s \rangle\rangle_{13}$ [c] & $\langle\langle c_s \rangle\rangle_{12}$ [c] & $\langle\langle c_s \rangle\rangle_{11}$ [c] & $\langle\langle c_s \rangle\rangle_{10}$ [c] \\
\hline
0.33 & 0.13 & 0.14 & 0.10 & 0.06 \\
\hline
$\langle\langle t_{\rm vis} \rangle\rangle_{14}~[\rm s] $ & $\langle\langle t_{\rm vis} \rangle\rangle_{13}~[\rm s] $ & $\langle\langle t_{\rm vis} \rangle\rangle_{12}~[\rm s] $ & $\langle\langle t_{\rm vis} \rangle\rangle_{11}~[\rm s] $ & $\langle\langle t_{\rm vis} \rangle\rangle_{10}~[s] $ \\
\hline
($<$ 0.16) & ($<$ 0.33) & 0.11 & 0.27 & 2.32\\
\hline\hline
\end{tabular}
\end{table*}

\begin{figure*}[t]
\hspace{-40mm}
\begin{minipage}{0.27\hsize}
\begin{center}
\includegraphics[width=10.5cm,angle=0]{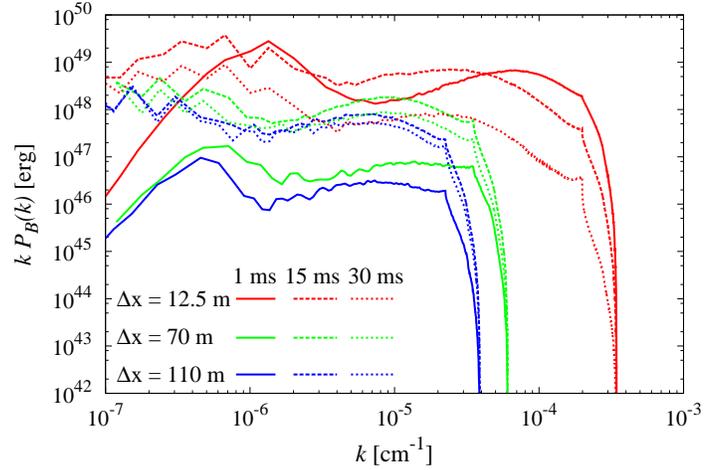}
\end{center}
\end{minipage}
\caption{\label{fig8}
Spectrum of the poloidal magnetic-field energy for all the runs. 
Three snapshots with $t-t_{\rm merger}=1$ ms, $15$ ms, and $30$ ms are shown with the solid, dashed, and 
dotted curves, respectively. 
}
\end{figure*}

\begin{figure*}[t]
\hspace{-50mm}
\begin{minipage}{0.27\hsize}
\begin{center}
\includegraphics[width=7.0cm,angle=0]{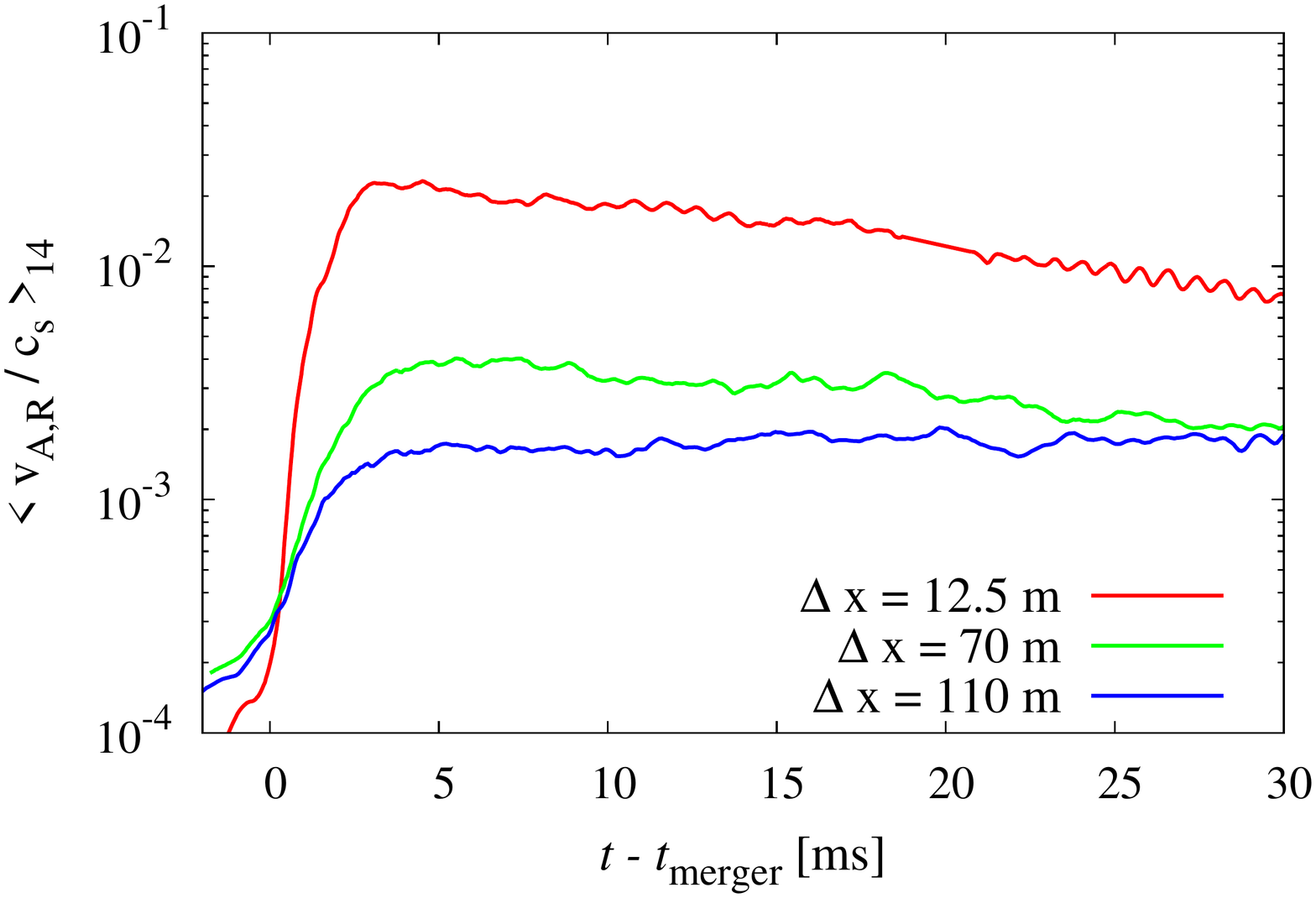}
\end{center}
\end{minipage}
\hspace{20mm}
\begin{minipage}{0.27\hsize}
\begin{center}
\includegraphics[width=7.0cm,angle=0]{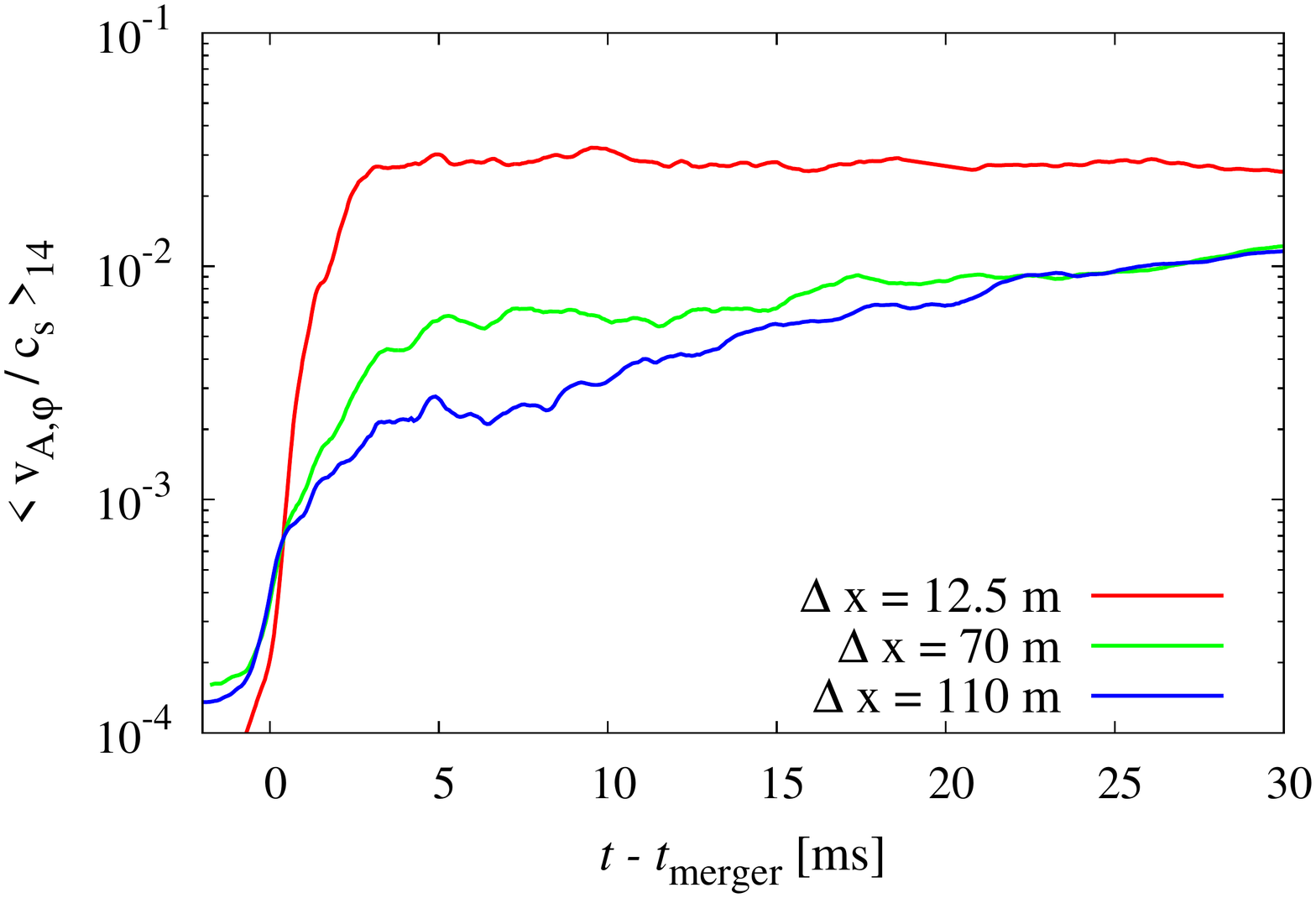}
\end{center}
\end{minipage}\\
\vspace{-10mm}
\hspace{-50mm}
\begin{minipage}{0.27\hsize}
\begin{center}
\includegraphics[width=7.0cm,angle=0]{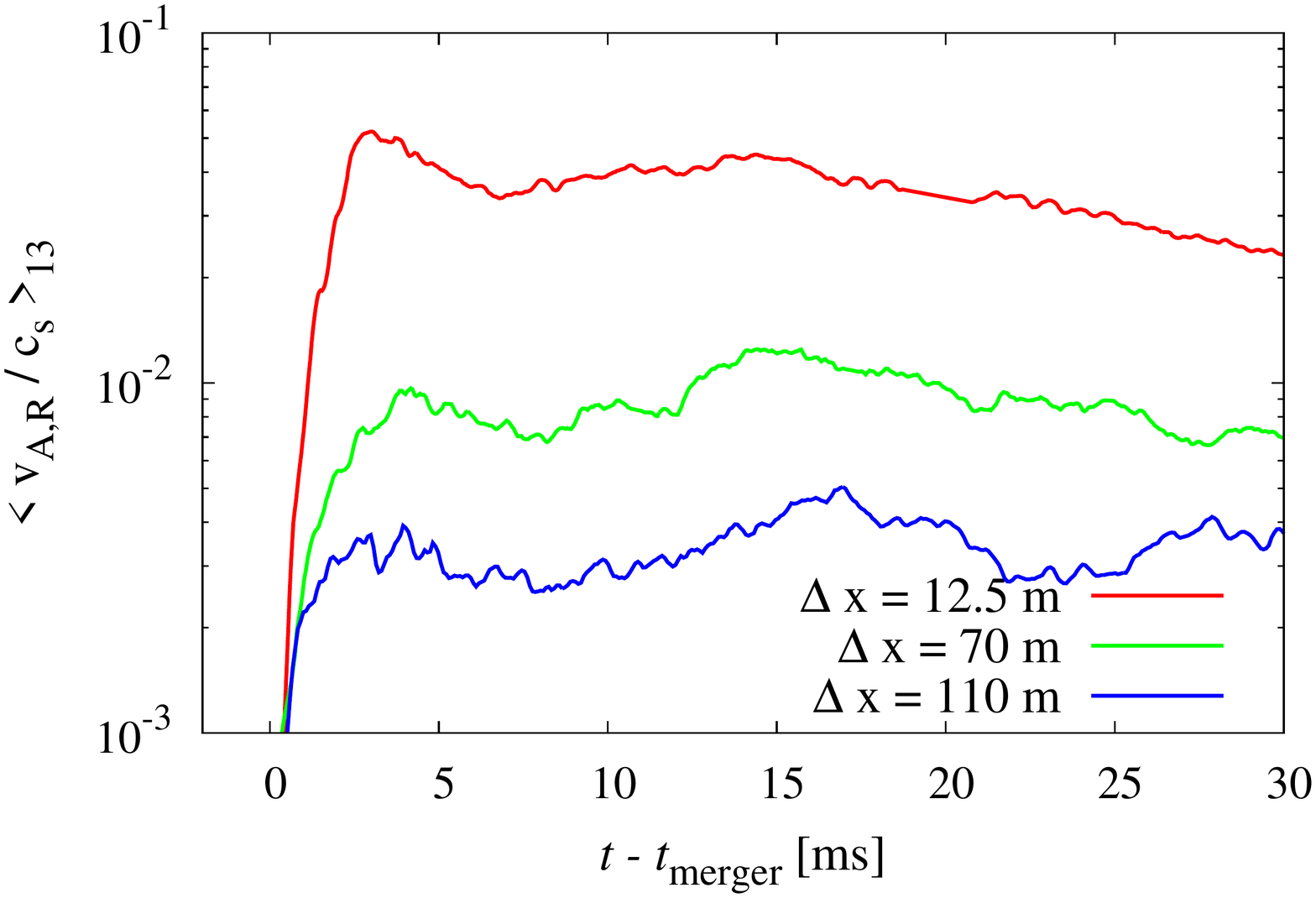}
\end{center}
\end{minipage}
\hspace{20mm}
\begin{minipage}{0.27\hsize}
\begin{center}
\includegraphics[width=7.0cm,angle=0]{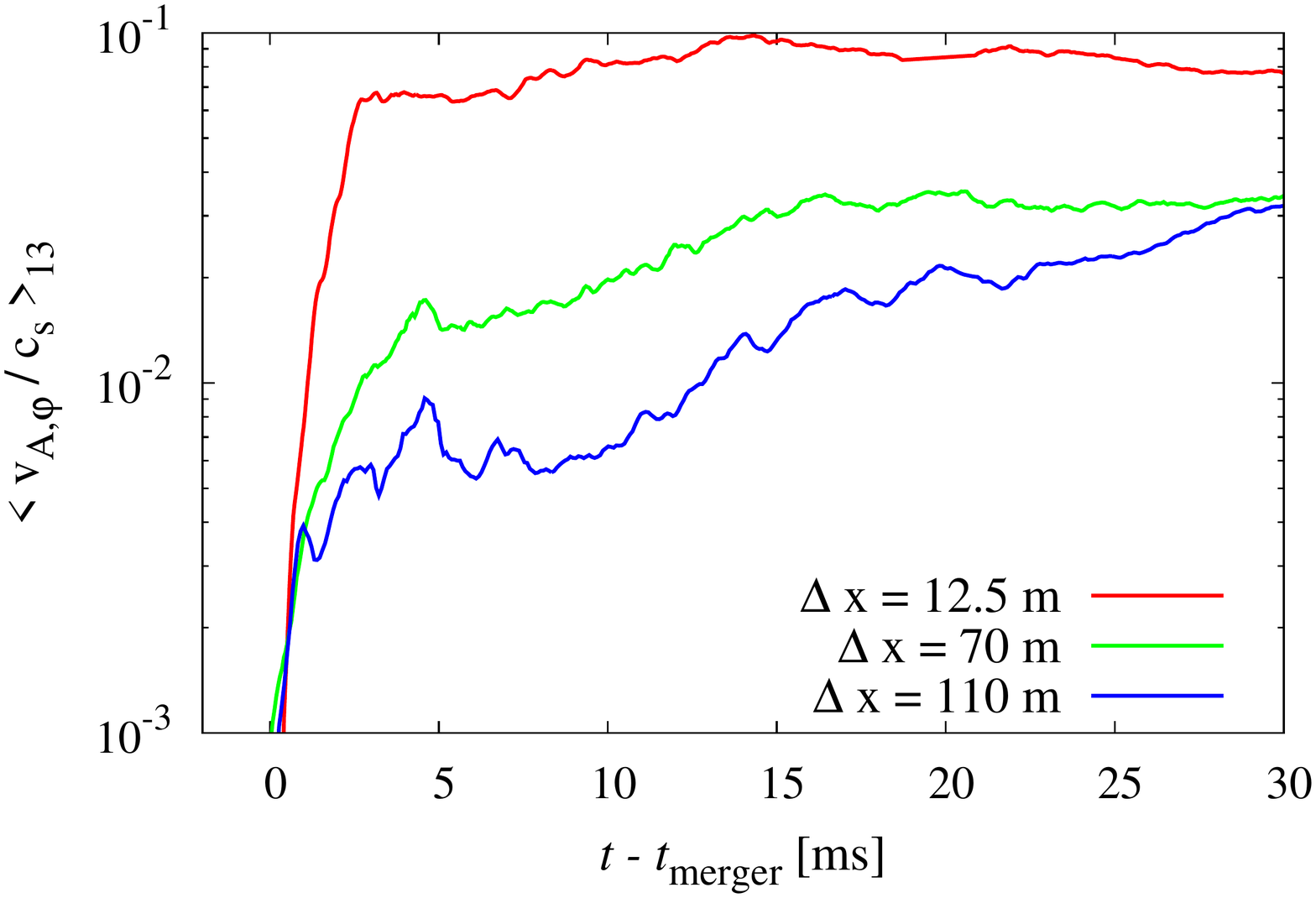}
\end{center}
\end{minipage}\\
\vspace{-10mm}
\hspace{-50mm}
\begin{minipage}{0.27\hsize}
\begin{center}
\includegraphics[width=7.0cm,angle=0]{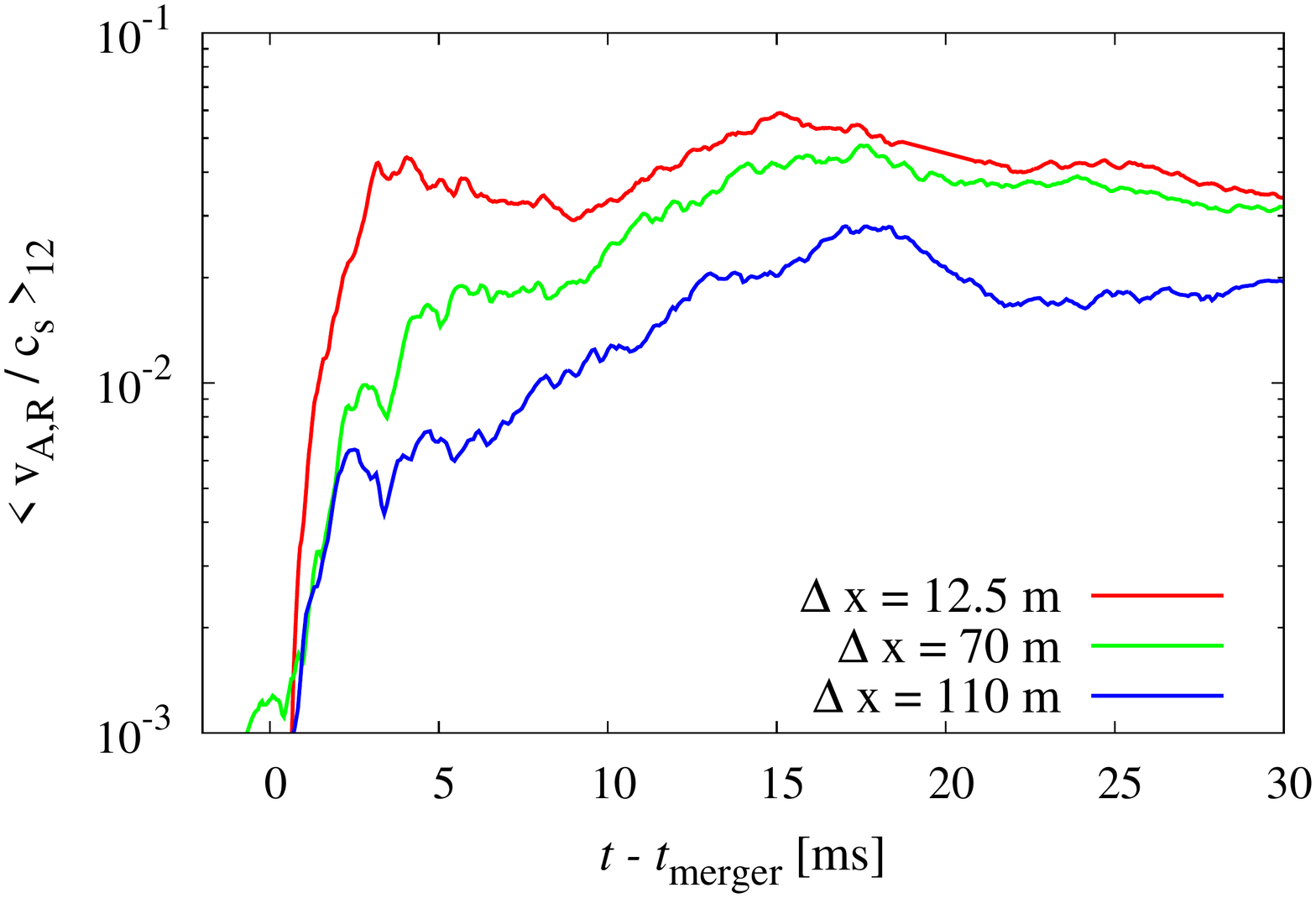}
\end{center}
\end{minipage}
\hspace{20mm}
\begin{minipage}{0.27\hsize}
\begin{center}
\includegraphics[width=7.0cm,angle=0]{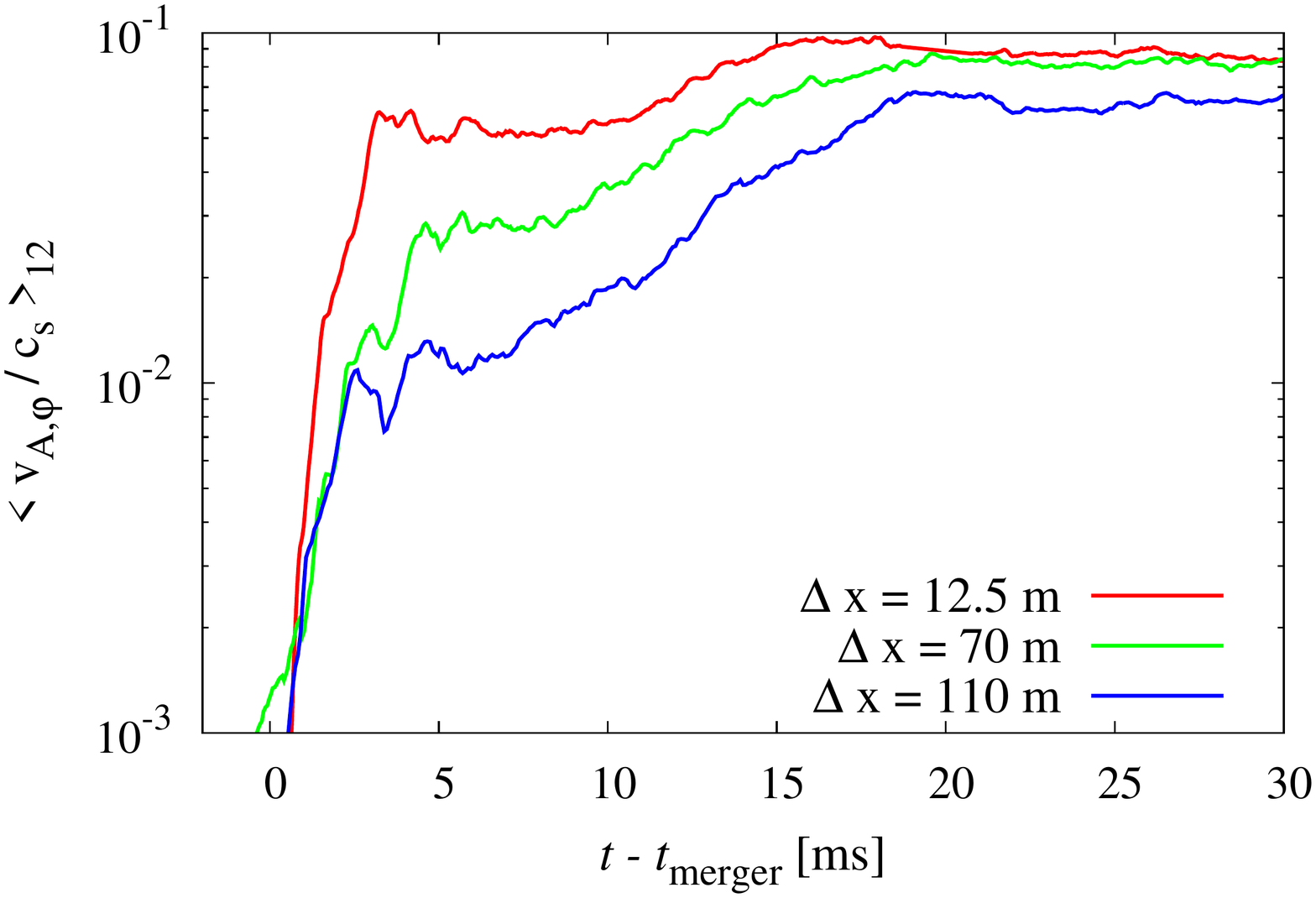}
\end{center}
\end{minipage}\\
\vspace{-10mm}
\hspace{-50mm}
\begin{minipage}{0.27\hsize}
\begin{center}
\includegraphics[width=7.0cm,angle=0]{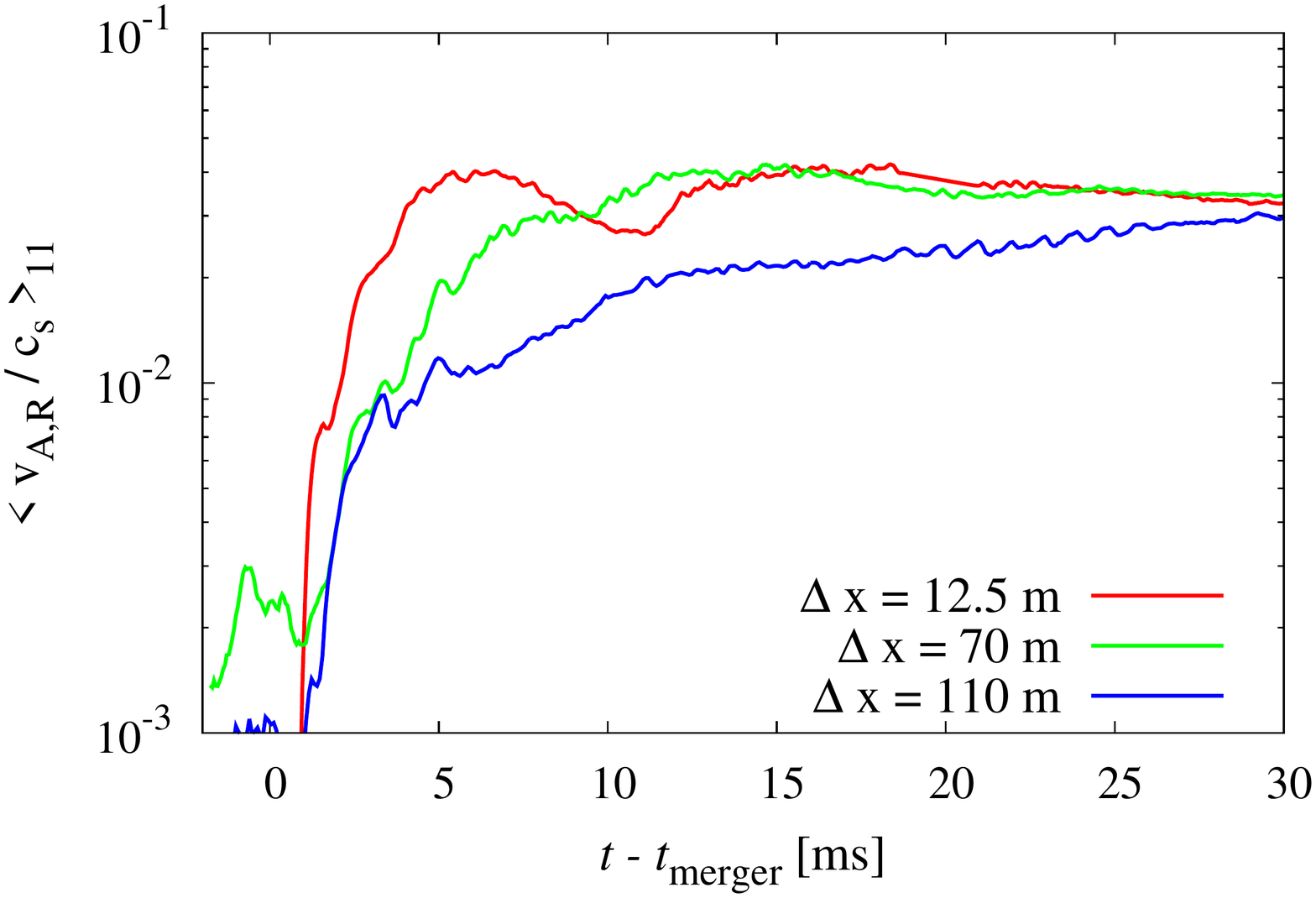}
\end{center}
\end{minipage}
\hspace{20mm}
\begin{minipage}{0.27\hsize}
\begin{center}
\includegraphics[width=7.0cm,angle=0]{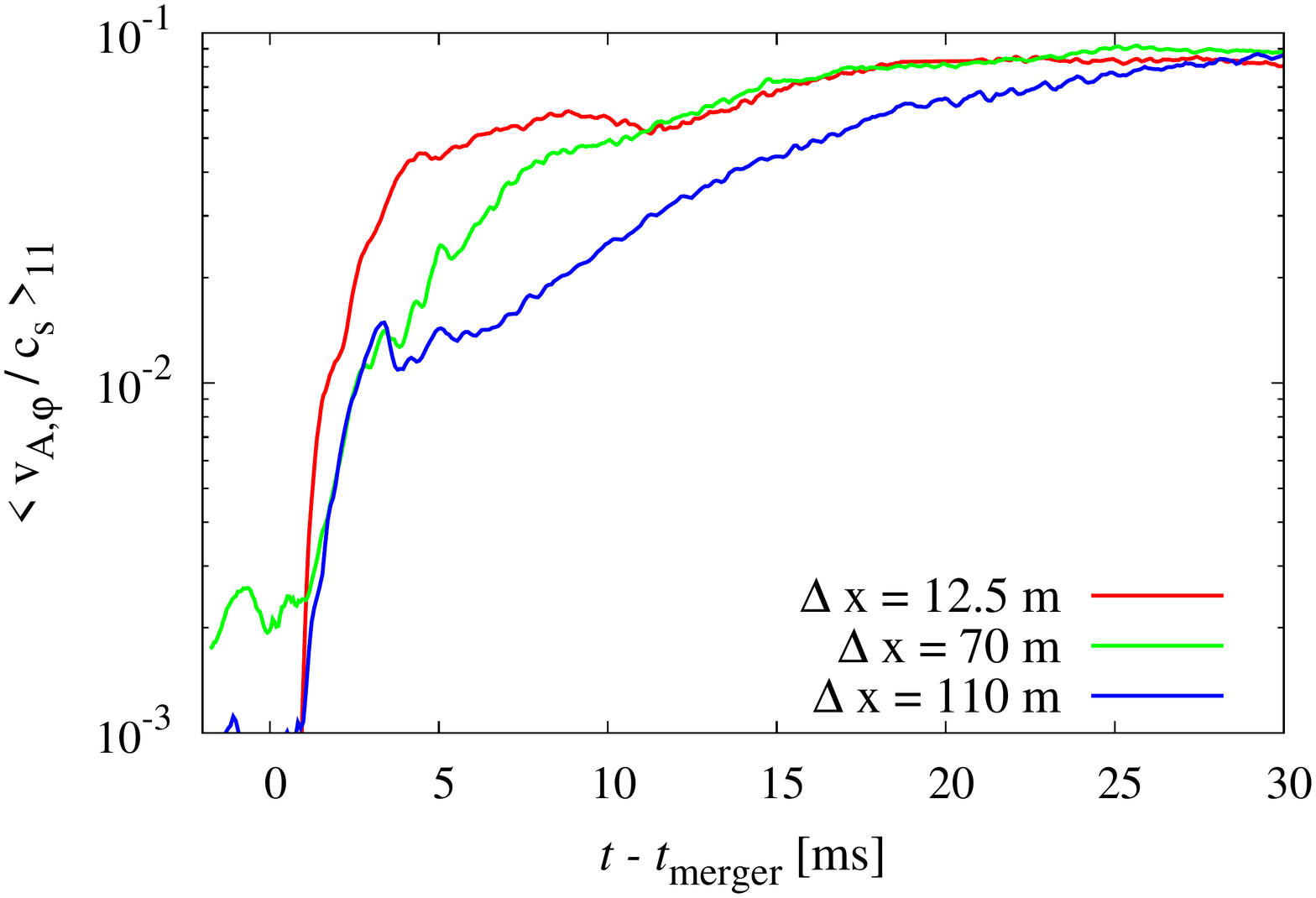}
\end{center}
\end{minipage}\\
\vspace{-10mm}
\hspace{-50mm}
\begin{minipage}{0.27\hsize}
\begin{center}
\includegraphics[width=7.0cm,angle=0]{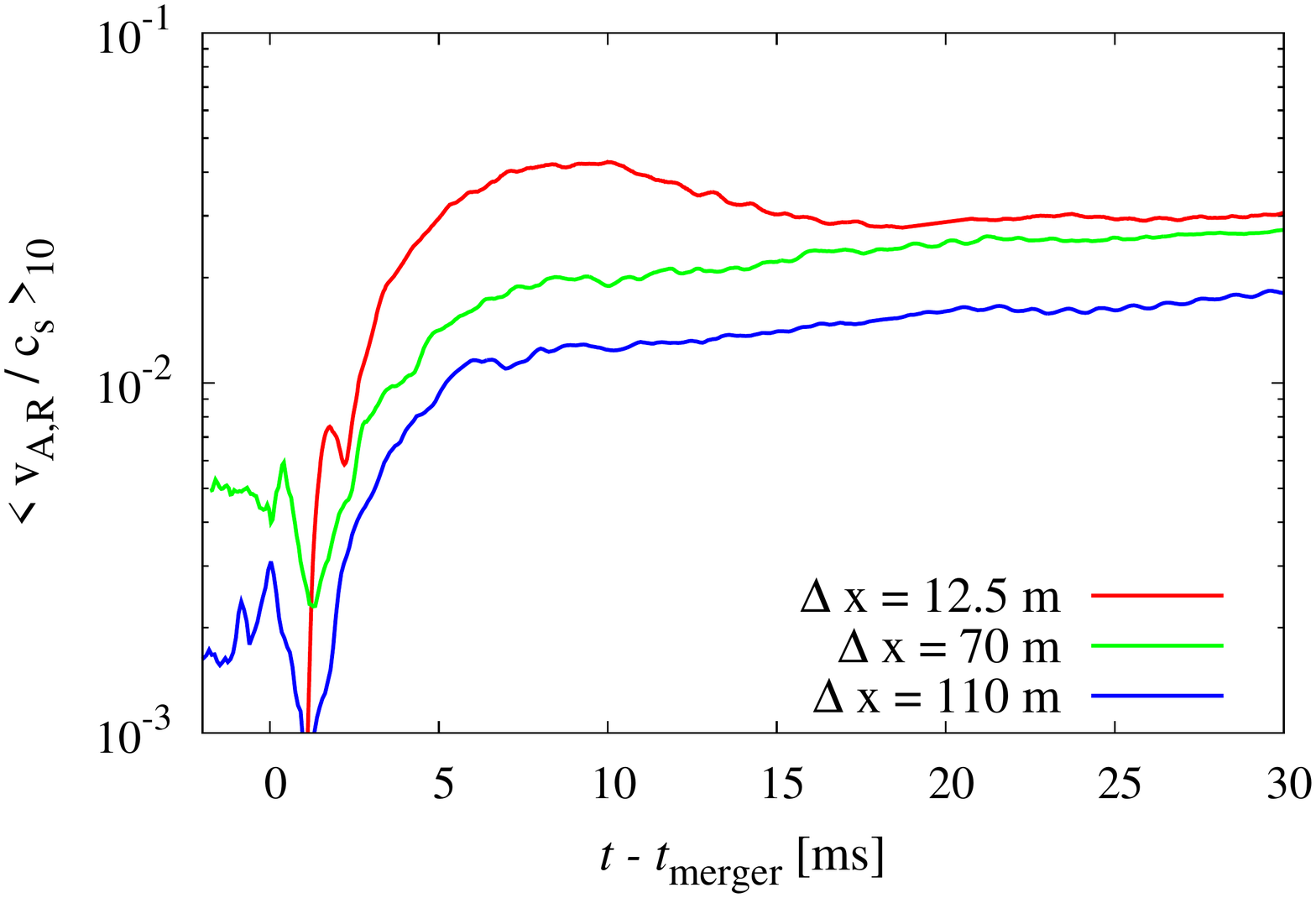}
\end{center}
\end{minipage}
\hspace{20mm}
\begin{minipage}{0.27\hsize}
\begin{center}
\includegraphics[width=7.0cm,angle=0]{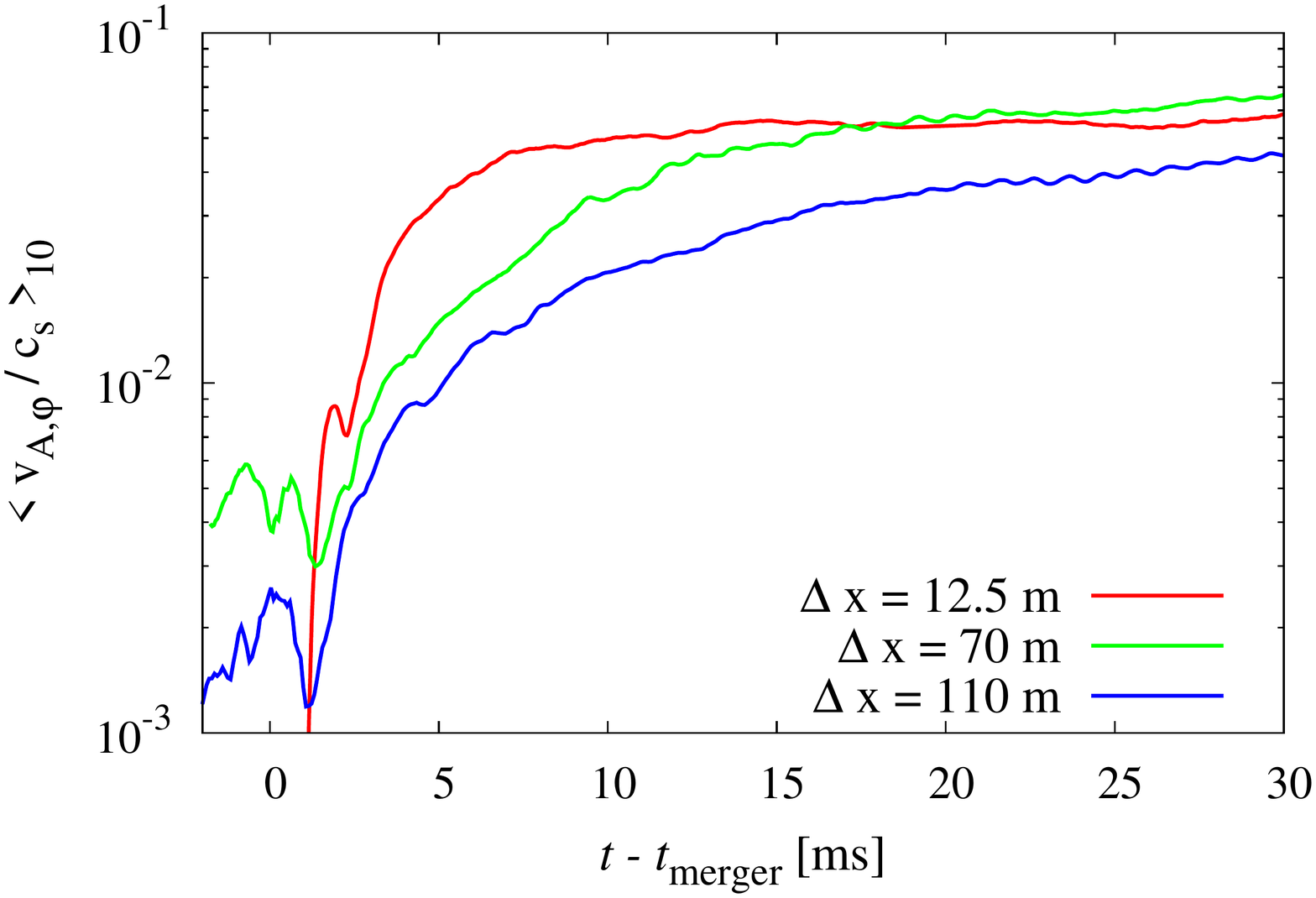}
\end{center}
\end{minipage}\\
\caption{\label{fig9}
Volume-averaged Alfv\'{e}n velocity normalized by the sound speed as functions of time. 
The left and right columns show the radial and azimuthal component, respectively. 
$\langle\cdot\rangle_a$ indicates a volume-average in a 
density range with $a \le\log_{10}[\rho~({\rm g~cm^{-3}})] < a+1$ with $a=10$--$14$. 
}
\end{figure*}

\section{Discussion}\label{sec:dis}

\subsection{Inferred value of the saturated magnetic-field energy and $\alpha$ parameter in the core region}
Because the viscous parameter as well as the convergence metrics in the RMNS core exhibit strong dependence on the grid resolution, 
we were not able to obtain the convergent values of these quantities. In the following, we infer the convergent values for the core region 
by extrapolating from the simulation data 
in the envelope region in which the result does not depend significantly on the grid resolution. 

Figure~\ref{fig9} plots volume-averaged Alfv\'{e}n velocity normalized by the sound speed as a function of time in each density range. 
We show both radial and azimuthal components of the Alfv\'{e}n velocity:
\begin{align}
v_{A,i} \equiv \frac{b_i}{\sqrt{4\pi\rho h + b^2}}c~(i=R~{\rm or}~\varphi).
\end{align} 
For $a \le 12$, $\langle v_{A,R}/c_s \rangle$ and $\langle v_{A,\varphi}/c_s \rangle$ converge at $0.03$--$0.04$ and $0.08$--$0.09$, respectively. 
By contrast, for $a \ge 13$, both quantities are far from the convergence. 
We extrapolate the toroidal magnetic-field strength in the high-density region from the results with $a\le 12$ as
\begin{align}
b_{\varphi} \sim \sqrt{4\pi\rho}v_{A,\varphi} \sim 10^{17} \left(\frac{c_s}{0.3c}\right)\left(\frac{\rho}{10^{15}{\rm g~cm^{-3}}}\right)^{1/2}{\rm G}
\end{align}
with $v_{A,\varphi}\sim 0.1 c_s$. The poloidal magnetic-field strength in the high-density region is expected to be
\begin{align}
b_R \sim \frac{1}{3}b_\varphi \sim 3\times10^{16} \left(\frac{c_s}{0.3c}\right)\left(\frac{\rho}{10^{15}{\rm g~cm^{-3}}}\right)^{1/2}{\rm G}.
\end{align}
This estimation suggests that in reality the toroidal magnetic field in the core region of the RMNS might be further 
amplified by a factor of $\sim 3$--$4$ which results in 
$E_B \sim 10^{51}$ erg. This energy is $\sim 1\%$ of $E_{\rm rot}$ and $E_{\rm int}$. 

\begin{figure*}[t]
\hspace{-30mm}
\begin{minipage}{0.27\hsize}
\begin{center}
\includegraphics[width=9.5cm,angle=0]{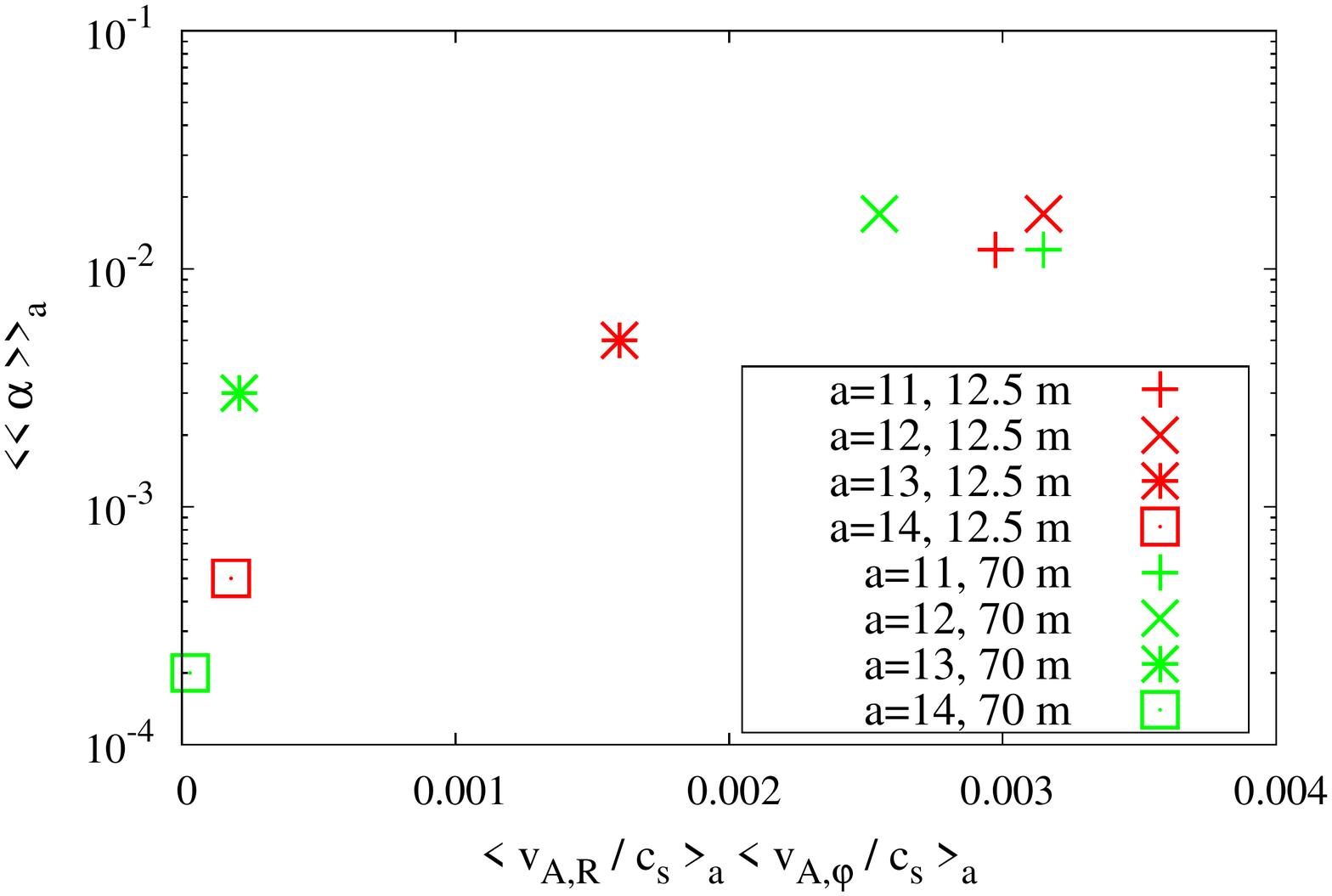}
\end{center}
\end{minipage}
\caption{\label{fig10}
Time- and volume-averaged $\alpha$ parameter as a function of $\langle v_{A,R} / c_s \rangle_a \langle v_{A,\varphi} / c_s \rangle_a$ 
for $a=11$--$14$. We show the results of the $12.5$ m run and the $70$ m run. 
}
\end{figure*}

Figure~\ref{fig10} plots the time- and volume-averaged $\alpha$ parameter 
as a function of ${\cal M} \equiv \langle v_{A,R} /c_s \rangle_a \langle v_{A,\varphi} / c_s \rangle_a$ for $a=11$--14. 
We show the results for the $12.5$ m and $70$ m runs. 
For $a=12$ and $11$, the $\alpha$ parameter converges to $0.01$--$0.02$ 
at ${\cal M} \approx 0.003$. 
If we extrapolate the $\alpha$ parameter for $a=13$ with respect to $\cal M$, it should be $\sim 0.01$--0.02. 
Consequently, we get 
\begin{align}
t_{\rm vis,13} &\approx 83~{\rm ms}\left(\frac{\langle\langle\alpha\rangle\rangle_{13}}{0.02}\right)^{-1}
\left(\frac{\langle\langle j \rangle\rangle_{13}}{2.5\times 10^{16}~{\rm cm^2~s^{-1}}}\right)\nonumber\\
&\times \left(\frac{\langle\langle c_s \rangle\rangle_{13}}{0.13c}\right)^{-2}, \label{eq:tvis}
\end{align}
where $\langle\langle j \rangle\rangle_{a}$ and $\langle\langle c_s \rangle\rangle_{a}$ with $a=13$ 
are time- and volume-averaged specific angular momentum and 
the sound speed in the RMNS for the $12.5$ m run with $15 \le t-t_{\rm merger} \le 30$ ms, respectively (see Table~\ref{tab2}).  

Because the $\alpha$ parameter in the density range with $a=14$ is far from the convergence, it is difficult to extrapolate with respect to $\cal M$. 
Nonetheless, speculating that the $\alpha$ parameter in the density range with $a=14$ would be 0.01--$0.02$ because 
the feature of the magneto-turbulence may be similar inside the envelope and the core, 
the viscous timescale in this high-density region would be 
\begin{align}
t_{\rm vis,14} &\approx 4~{\rm ms}\left(\frac{\langle\langle\alpha\rangle\rangle_{14}}{0.02}\right)^{-1}
\left(\frac{\langle\langle j \rangle\rangle_{14}}{7.8\times 10^{15}~{\rm cm^2~s^{-1}}}\right)\nonumber\\
&\times \left(\frac{\langle\langle c_s \rangle\rangle_{14}}{0.33c}\right)^{-2}, \label{eq:tvis2}
\end{align}
Therefore according to the estimation of the viscous timescale Eqs.~(\ref{eq:tvis}) or (\ref{eq:tvis2}), 
we speculate that the RMNS gradually approaches a rigid-rotation state within this viscous timescale.

\subsection{Magnetic winding and braking in the high-density region}
Our current results show that for the region with $\rho \gtrsim 10^{14}~{\rm g~cm^{-3}}$, the radial gradient of the angular velocity is positive as shown in 
Figs.~\ref{fig2} and \ref{fig3}. The magnetic winding would work in this high-density region even though the MRI might not turn on and 
the magnetic braking timescale is estimated to give 
\begin{align*}
t_{\rm brake} &= \frac{R}{v_{A,R}} \approx 5~{\rm ms} \left(\frac{b_R}{3 \times 10^{16}{\rm G}}\right)^{-1}
\left(\frac{\rho}{10^{15}{\rm~g~cm^{-3}}}\right)^{1/2}\\
&\times \left(\frac{R}{15~{\rm km}}\right).
\end{align*}
Because there is a room for the magnetic-field amplification due to the Kelvin-Helmholtz instability as discussed in Sec.~\ref{subsec:bamp}, 
the braking timescale could be shorter than $5$ ms in reality. 
However note that the coherent poloidal field is assumed to be developed for this estimation. 

For the small-scale randomly-oriented magnetic fields like those found in this study, the braking timescale may be written as 
\begin{align}
t_{\rm brake} &= \frac{R}{v_{A,R}}\left(\frac{R}{\delta R}\right) \approx 750~{\rm ms} \left(\frac{b_R}{3 \times 10^{16}{\rm G}}\right)^{-1}\nonumber\\
&\times \left(\frac{\rho}{10^{15}{\rm~g~cm^{-3}}}\right)^{1/2}
 \left(\frac{R}{15~{\rm km}}\right)\left(\frac{\delta R}{0.1~{\rm km}}\right)^{-1}, 
\end{align} 
where we set a spatial scale of the turbulent magnetic field $\delta R$ as the numerically resolvable scale $\approx 0.1~{\rm km}$, but 
it should be much smaller than this value in reality.

\subsection{Remark on the angular velocity profile}
As shown in Figs.~\ref{fig2} and \ref{fig3}, the radial profile of the angular velocity shows the positive gradient for $R \lesssim 10~{\rm km}$. 
However this may not be the case in reality. In our (insufficiently resolved) simulations, 
the growth timescale of the Kelvin-Helmholtz instability and 
the timescale of the resultant magnetic-field amplification are longer than the rotational period of the RMNS, 
which is $\sim 1~{\rm ms}$. In reality, however, the growth timescale of the Kelvin-Helmholtz instability should be much shorter 
than $1~{\rm ms}$. 
Therefore the Kelvin-Helmholtz vortices and magneto-turbulence could transport the angular momentum within a timescale much shorter 
than the rotational timescale of the RMNS. This suggests that the angular velocity profile which we found in this paper 
could be significantly modified. We have to keep in mind this possibility. 

\section{Summary} 

We performed high-resolution GRMHD simulations for $\approx 30$ ms after merger of a BNS. 
We carried out a detailed analysis of the MHD-driven turbulence and 
evaluated the effective viscosity generated by the MRI-driven turbulence (see also Ref.~\cite{Marshall} 
for the Magnetic-Taylor instability as an angular momentum transport agent).

We obtain the convergent result for the $\alpha$ parameter in the RMNS envelope and torus which have low values of density with $\rho < 10^{13}~{\rm g~cm^{-3}}$. 
For the high-density range with $10^{13}~{\rm g~cm^{-3}} \le \rho \le 10^{14}~{\rm g~cm^{-3}}$, 
we estimate that the MRI-driven turbulence could generate the effective viscous parameter of $\sim 0.01$--$0.02$. 
For the deep inside the RMNS core with $\rho \ge 10^{14}~{\rm g~cm^{-3}}$, the viscous parameter depends significantly on the grid resolution. 
However, we speculate that the Kelvin-Helmholtz instability and 
resultant magneto-turbulence could transport the angular momentum in the core region of RMNS. 
To solve this issue, future ultra high-resolution GRMHD simulations are necessary. 

The final goal of this project is to reveal the long-term evolution process of RMNSs formed after BNS mergers. 
As a next step, we plan to calibrate a viscous hydrodynamical simulation~\cite{Shibata:2017jyf,Radice:2017zta} by a GRMHD simulation. 
Then, we will perform a long-term viscous hydrodynamical simulation to explore the fate of long-lived RMNSs. 

\acknowledgments 
We thank C. Ott and M. C. Werner for giving invaluable comments and reading the manuscript carefully. 
Numerical computation was performed on K computer at AICS
(project numbers hp160211 and hp170230), on Cray XC30 at cfca of
National Astronomical Observatory of Japan, FX10 and Oakforest-PACS at Information
Technology Center of the University of Tokyo, HOKUSAI FX100 at RIKEN,
and on Cray XC40 at Yukawa Institute for Theoretical Physics, Kyoto
University.  This work was supported by Grant-in-Aid for Scientific
Research (16H02183, JP16H06342, JP17H01131, 15K05077, 17H06361, 16K17706, 16H06341, 15H00782, 26400237) 
of JSPS and by a post-K computer project (Priority issue No.~9) of Japanese MEXT.




\begin{thebibliography}{99}
\bibitem{footnote1}
The $\alpha$-parameter for $10^{10}~{\rm g~cm^{-3}} \le \rho \le 10^{11}~{\rm g~cm^{-3}}$ is smaller than 
those in the other regions of the envelope. This might be because the pressure scale height of this density region is 
larger than the spatial scale determined by the foliation in terms of the density, i.e., 
our analysis implicitly sets the maximum size of the eddy generated by the MHD instabilities 
which should be comparable to the pressure scale height.
\end{thebibliography}
\end{document}